\documentclass[12pt]{article}
\usepackage{amssymb}
\usepackage{amsmath}

\allowdisplaybreaks
\begin{document}

\author{C. Bizdadea\thanks{%
e-mail address: bizdadea@central.ucv.ro}, E. M. Cioroianu\thanks{%
e-mail address: manache@central.ucv.ro}, D. Cornea\thanks{%
e-mail address: danubiusmd@yahoo.com}, \and S. O. Saliu\thanks{%
e-mail address: osaliu@central.ucv.ro}, S. C. S\u{a}raru\thanks{%
e-mail address: scsararu@central.ucv.ro} \\
Faculty of Physics, University of Craiova\\
13 A. I. Cuza Str., Craiova 200585, Romania}
\title{No interactions for a collection of \\
spin-two fields intermediated by \\
a massive Rarita-Schwinger field}
\maketitle

\begin{abstract}
The cross-couplings among several massless spin-two fields (described in the
free limit by a sum of Pauli-Fierz actions) in the presence of a massive
Rarita-Schwinger field are investigated in the framework of the deformation
theory based on local BRST cohomology. Under the hypotheses of locality,
smoothness of the interactions in the coupling constant, Poincar\'{e}
invariance, Lorentz covariance, and the preservation of the number of
derivatives on each field, we prove that there are no consistent
cross-interactions among different gravitons with a positively defined
metric in internal space in the presence of a massive Rarita-Schwinger
field. The basic features of the couplings between a single Pauli-Fierz
field and a massive Rarita-Schwinger field are also emphasized.

\end{abstract}

\section{Introduction}

Over the last twenty years there was a sustained effort for constructing
theories involving a multiplet of spin-two fields~\cite%
{cutwald1,wald2,ovrutwald,ancoann}. At the same time, various couplings of a
single massless spin-two field to other fields (including itself) have been
studied in~\cite%
{gupta54,kraich55,wein65,deser70,boul75,fangfr,berburgdam,wald86,hatffeyn,boulcqg,noijhepdirac}%
. In this context the impossibility of cross-interactions among several
Einstein gravitons under certain assumptions has been proved recently in~%
\cite{multi} by means of a cohomological approach based on the Lagrangian
BRST symmetry~\cite{batvilk1,batvilk2,batvilk3,henproc,hencarte}. Moreover,
in~\cite{multi} the impossibility of cross-interactions among different
Einstein gravitons in the presence of a scalar field has also been shown.

The main aim of this paper is to investigate the cross-couplings among
several massless spin-two fields (described in the free limit by a sum of
Pauli-Fierz actions) in the presence of a massive Rarita-Schwinger field.
More precisely, under the hypotheses of locality, smoothness of the
interactions in the coupling constant, Poincar\'{e} invariance, (background)
Lorentz invariance, and the preservation of the number of derivatives on
each field, we prove that there are no consistent cross-interactions among
different gravitons with a positively defined metric in internal space in
the presence of a massive Rarita-Schwinger field. This result is obtained by
using the deformation technique~\cite{def} combined with the local BRST
cohomology~\cite{gen1}. It is well-known the fact that the spin-two field in
metric formulation (Einstein-Hilbert theory) cannot be coupled to a spin-3/2
field. However, as it will be shown below, if we decompose the metric like $%
g_{\mu \nu }=\sigma _{\mu \nu }+\lambda h_{\mu \nu }$, where $\sigma _{\mu
\nu }$ is the flat metric and $\lambda $ is the coupling constant, then we
can indeed couple the massive spin-3/2 field to $h_{\mu \nu }$ in the space
of formal series with the maximum derivative order equal to one in $h_{\mu
\nu }$. Thus, our approach envisages two different aspects. One is related
to the couplings between the spin-two fields and one massive
Rarita-Schwinger field, while the other focuses on proving the impossibility
of cross-interactions among different gravitons via a single massive
Rarita-Schwinger field. In order to make the analysis as clear as possible,
we initially consider the case of the couplings between a single Pauli-Fierz
field~\cite{pf} and a massive Rarita-Schwinger field~\cite{raritaschwinger}.
In this setting we compute the interaction terms to order two in the
coupling constant. Next, we prove the isomorphism between the local BRST
cohomologies corresponding to the Pauli-Fierz theory and respectively to the
linearized version of the vierbein formulation of the spin-two field. Since
the deformation procedure is controlled by the local BRST cohomology of the
free theory (in ghost number zero and one), the previous isomorphism allows
us to translate the results emerging from the Pauli-Fierz formulation into
the vierbein version and conversely. In this manner we obtain that the first
two orders of the interacting Lagrangian resulting from our setting
originate in the development of the full interacting Lagrangian
\begin{eqnarray*}
&&\mathcal{L}^{\left( \mathrm{int}\right) }=\frac{e}{2}\left( -\mathrm{i}%
\bar{\psi}_{\mu }e_{a}^{\;\;\mu }e_{b}^{\;\;\nu }e_{c}^{\;\;\rho }\gamma
^{abc}D_{\nu }\psi _{\rho }+m\bar{\psi}_{\mu }e_{a}^{\;\;\mu }\gamma
^{ab}e_{b}^{\;\;\nu }\psi _{\nu }\right) \\
&&+\lambda \left[ eV\left( X,Y,Z\right) +d_{1}\left( X,Y,Z\right)
e_{a}^{\;\;\nu }\bar{\psi}_{\nu }\gamma ^{a}D_{\mu }\left( e\psi ^{\mu
}\right) \right. \\
&&\left. +ed_{2}\left( X,Y,Z\right) \left( \bar{\psi}^{\mu }\gamma
^{b}+e_{a}^{\;\;\mu }e_{\;\;\rho }^{b}\bar{\psi}^{\rho }\gamma ^{a}\right)
D_{\mu }\left( e_{b}^{\;\;\nu }\psi _{\nu }\right) \right] .
\end{eqnarray*}%
Here, $e_{a}^{\;\;\mu }$ represent the vierbein fields, $e$ is the inverse
of their determinant, $e=\left( \det \left( e_{a}^{\;\;\mu }\right) \right)
^{-1}$, $D_{\mu }$ signifies the full covariant derivative, and $\gamma ^{a}$
stand for the flat Dirac matrices. The fields $\psi _{\nu }$ denote the
(curved) Rarita-Schwinger spinors ($\psi _{\nu }=e_{\;\;\nu }^{a}\psi _{a}$%
). The quantities denoted by $V$, $d_{1}$, and $d_{2}$ are arbitrary
polynomials of $X\equiv \bar{\psi}_{a}\psi ^{a}$, $Y\equiv \bar{\psi}%
_{a}\gamma ^{ab}\psi _{b}$, and $Z=\mathrm{i}\bar{\psi}_{a}\gamma _{5}\psi
^{a}$. Here and in the sequel $\lambda $ is the coupling constant
(deformation parameter). We observe that the first two terms in $\mathcal{L}%
^{\left( \mathrm{int}\right) }$ describe the standard minimal couplings
between the spin-two and massive Rarita-Schwinger fields. The last terms
from $\mathcal{L}^{\left( \mathrm{int}\right) }$, namely those proportional
with $V$, $d_{1}$, or $d_{2}$, produce non-minimal couplings. To our
knowledge, these non-minimal interaction terms are not discussed in the
literature. However, they are consistent with the gauge symmetries of the
Lagrangian $\mathcal{L}_{2}+\mathcal{L}^{\left( \mathrm{int}\right) }$,
where $\mathcal{L}_{2}$ is the full spin-two Lagrangian in the vierbein
formulation. With this result at hand, we start from a finite sum of
Pauli-Fierz actions with a positively defined metric in internal space and a
massive Rarita-Schwinger field, and prove that there are no consistent
cross-interactions between different gravitons in the presence of such a
fermionic matter field.

This paper is organized in seven sections. In Section 2 we construct the
BRST symmetry of a free model with a single Pauli-Fierz field and one
massive Rarita-Schwinger field. Section 3 briefly addresses the deformation
procedure based on BRST symmetry. In Section 4 we compute the first two
orders of the interactions between one graviton and one massive
Rarita-Schwinger spinor. Section 5 presents the Lagrangian formulation of
the interacting theory. Section 6 is devoted to the proof of the fact that
there are no consistent cross-interactions among different gravitons in the
presence of a massive Rarita-Schwinger field. Section 7 exposes the main
conclusions of the paper. The present paper also contains two appendix
sections, in which various notations and conditions are listed and also some
statements from the body of the paper are proved.

\section{Free model: Lagrangian formulation and\protect\linebreak BRST
symmetry}

Our starting point is represented by a free model, whose Lagrangian action
is written like the sum between the action of the linearized version of
Einstein-Hilbert gravity (the Pauli-Fierz action~\cite{pf}) and that of a
massive Rarita-Schwinger field \cite{raritaschwinger}
\begin{eqnarray}
S_{0}^{\mathrm{L}}\left[ h_{\mu \nu },\psi _{\mu }\right] &=&\int
d^{4}x\left( -\frac{1}{2}\left( \partial _{\mu }h_{\nu \rho }\right) \left(
\partial ^{\mu }h^{\nu \rho }\right) +\left( \partial _{\mu }h^{\mu \rho
}\right) \left( \partial ^{\nu }h_{\nu \rho }\right) \right.  \notag \\
&&-\left( \partial _{\mu }h\right) \left( \partial _{\nu }h^{\nu \mu
}\right) +\frac{1}{2}\left( \partial _{\mu }h\right) \left( \partial ^{\mu
}h\right)  \notag \\
&&\left. -\frac{\mathrm{i}}{2}\bar{\psi}_{\mu }\gamma ^{\mu \nu \rho
}\partial _{\nu }\psi _{\rho }+\frac{m}{2}\bar{\psi}_{\mu }\gamma ^{\mu \nu
}\psi _{\nu }\right)  \notag \\
&\equiv &\int d^{4}x\left( \mathcal{L}^{\left( \mathrm{PF}\right) }+\mathcal{%
L}_{0}^{(\mathrm{RS})}\right) =S_{0}^{\mathrm{PF}}\left[ h_{\mu \nu }\right]
+S_{0}^{\mathrm{RS}}\left[ \psi _{\mu }\right] .  \label{fract}
\end{eqnarray}%
Everywhere in this paper we use the flat Minkowski metric of `mostly minus'
signature, $\sigma _{\mu \nu }=\left( +---\right) $. In the above $h$
denotes the trace of the Pauli-Fierz field, $h=\sigma _{\mu \nu }h^{\mu \nu
} $, and the fermionic fields $\psi _{\mu }$ are considered to be real
(Majorana) spinors. We work with a representation of the Clifford algebra
\begin{equation}
\gamma _{\mu }\gamma _{\nu }+\gamma _{\nu }\gamma _{\mu }=2\sigma
_{\mu \nu }\mathbf{1}  \label{gama1}
\end{equation}%
in which all the $\gamma $ matrices are purely imaginary, so we have that%
\begin{equation}
\gamma _{\mu }^{\intercal }=-\gamma _{0}\gamma _{\mu }\gamma _{0},\quad \mu =%
\overline{0,3},  \label{PFRS2}
\end{equation}%
where here and in the sequel the notation $N^{\intercal }$ signifies the
transposed of the matrix $N$. In addition, $\gamma _{0}$ is Hermitian and
antisymmetric, while $\left( \gamma _{i}\right) _{i=\overline{1,3}}$ are
anti-Hermitian and symmetric. The Dirac conjugation is defined as usually
through
\begin{equation}
\bar{\psi}_{\mu }=\left( \psi _{\mu }\right) ^{\dagger }\gamma _{0},
\label{gama2}
\end{equation}%
and the Majorana conjugation via
\begin{equation}
\psi ^{c}=\left( \mathcal{C}\psi \right) ^{\intercal },  \label{gama3}
\end{equation}%
with the corresponding charge conjugation given by
\begin{equation}
\mathcal{C}=-\gamma _{0}.  \label{gama4}
\end{equation}%
(The operation $^{\dagger }$ signifies the Hermitian conjugation.) Action (%
\ref{fract}) possesses an irreducible and Abelian generating set of gauge
transformations
\begin{equation}
\delta _{\epsilon }h_{\mu \nu }=\partial _{(\mu }\epsilon _{\nu )},\quad
\delta _{\epsilon }\psi _{\mu }=0,  \label{PFRS3}
\end{equation}%
with $\epsilon _{\mu }$ bosonic gauge parameters. The parentheses signify
symmetrization; they are never divided by the number of terms: e.g., $%
\partial _{(\mu }\epsilon _{\nu )}=\partial _{\mu }\epsilon _{\nu }+\partial
_{\nu }\epsilon _{\mu }$, and the minimum number of terms is always used.
The same is valid with respect to the notation $\left[ \mu \cdots \nu \right]
$, which means antisymmetrization with respect to the indices between
brackets.

In order to construct the BRST symmetry for (\ref{fract}) we introduce the
fermionic ghosts $\eta _{\mu }$ corresponding to the gauge parameters $%
\epsilon _{\mu }$ and associate antifields with the original fields and
ghosts, respectively denoted by $\left\{ h^{\ast \mu \nu },\psi _{\mu
}^{\ast }\right\} $ and $\left\{ \eta ^{\ast \mu }\right\} $. (The
statistics of the antifields is opposite to that of the correlated
fields/ghosts.) The antifields of the Rarita-Schwinger fields are bosonic,
purely imaginary spinors. Since the gauge generators of the free theory
under study are field independent and irreducible, it follows that the BRST
differential simply decomposes into
\begin{equation}
s=\delta +\gamma ,  \label{PFRS4}
\end{equation}%
where $\delta $ represents the Koszul-Tate differential, graded by the
antighost number $\mathrm{agh}$ ($\mathrm{agh}\left( \delta \right) =-1$),
and $\gamma $ stands for the exterior derivative along the gauge orbits,
whose degree is named pure ghost number $\mathrm{pgh}$ ($\mathrm{pgh}\left(
\gamma \right) =1$). These two degrees do not interfere ($\mathrm{pgh}\left(
\delta \right) =0$, $\mathrm{agh}\left( \gamma \right) =0$). The overall
degree from the BRST complex is known as the ghost number $\mathrm{gh}$ and
is defined like the difference between the pure ghost number and the
antighost number, such that $\mathrm{gh}\left( \delta \right) =\mathrm{gh}%
\left( \gamma \right) =\mathrm{gh}\left( s\right) =1$. If we make the
notations
\begin{equation}
\Phi ^{\alpha _{0}}=\left( h_{\mu \nu },\psi _{\mu }\right) ,\quad \Phi
_{\alpha _{0}}^{\ast }=\left( h^{\ast \mu \nu },\psi _{\mu }^{\ast }\right) ,
\label{PFRS5}
\end{equation}%
then, according to the standard rules of the BRST formalism, the degrees of
the BRST generators are valued like
\begin{eqnarray}
\mathrm{agh}\left( \Phi ^{\alpha _{0}}\right) &=&\mathrm{agh}\left( \eta
_{\mu }\right) =0,\quad \mathrm{agh}\left( \Phi _{\alpha _{0}}^{\ast
}\right) =1,\quad \mathrm{agh}\left( \eta ^{\ast \mu }\right) =2,
\label{PFRS6} \\
\mathrm{pgh}\left( \Phi ^{\alpha _{0}}\right) &=&0,\quad \mathrm{pgh}\left(
\eta _{\mu }\right) =1,\quad \mathrm{pgh}\left( \Phi _{\alpha _{0}}^{\ast
}\right) =\mathrm{pgh}\left( \eta ^{\ast \mu }\right) =0.  \label{PFRS7}
\end{eqnarray}%
The actions of the differentials $\delta $ and $\gamma $ on the generators
from the BRST complex are given by
\begin{eqnarray}
\delta h^{\ast \mu \nu } &=&2H^{\mu \nu },\quad \delta \psi ^{\ast \mu }=m%
\bar{\psi}_{\lambda }\gamma ^{\lambda \mu }-\mathrm{i}\partial _{\rho }\bar{%
\psi}_{\lambda }\gamma ^{\rho \lambda \mu },  \label{PFRS8} \\
\delta \eta ^{\ast \mu } &=&-2\partial _{\nu }h^{\ast \mu \nu },
\label{PFRS9} \\
\delta \Phi ^{\alpha _{0}} &=&0=\delta \eta _{\mu },  \label{PFRS10} \\
\gamma \Phi _{\alpha _{0}}^{\ast } &=&0=\gamma \eta ^{\ast \mu },
\label{PFRS11} \\
\gamma h_{\mu \nu } &=&\partial _{(\mu }\eta _{\nu )},\quad \gamma \psi
_{\mu }=0,\quad \gamma \eta _{\mu }=0,  \label{PFRS12}
\end{eqnarray}%
where $H^{\mu \nu }$ is the linearized Einstein tensor
\begin{equation}
H^{\mu \nu }=K^{\mu \nu }-\frac{1}{2}\sigma ^{\mu \nu }K,  \label{PFRS13}
\end{equation}%
with $K^{\mu \nu }$ and $K$ the linearized Ricci tensor and respectively the
linearized scalar curvature, both obtained from the linearized Riemann
tensor
\begin{eqnarray}
K_{\mu \nu \alpha \beta } &=&-\frac{1}{2}\left( \partial _{\mu }\partial
_{\alpha }h_{\nu \beta }+\partial _{\nu }\partial _{\beta }h_{\mu \alpha
}\right.  \notag \\
&&\left. -\partial _{\nu }\partial _{\alpha }h_{\mu \beta }-\partial _{\mu
}\partial _{\beta }h_{\nu \alpha }\right) ,  \label{PFRS14}
\end{eqnarray}%
via its trace and respectively double trace
\begin{equation}
K_{\mu \alpha }=\sigma ^{\nu \beta }K_{\mu \nu \alpha \beta },\quad K=\sigma
^{\mu \alpha }\sigma ^{\nu \beta }K_{\mu \nu \alpha \beta }.  \label{PFRS15}
\end{equation}

The BRST differential is known to have a canonical action in a structure
named antibracket and denoted by the symbol $\left( ,\right) $ ($s\cdot
=\left( \cdot ,\bar{S}\right) $), which is obtained by decreeing the
fields/ghosts respectively conjugated to the corresponding antifields. The
generator of the BRST symmetry is a bosonic functional of ghost number zero,
which is solution to the classical master equation $\left( \bar{S},\bar{S}%
\right) =0$. The full solution to the classical master equation for the free
model under study reads as
\begin{equation}
\bar{S}=S_{0}^{\mathrm{L}}\left[ h_{\mu \nu },\psi _{\mu }\right] +\int
d^{4}x\,h^{\ast \mu \nu }\partial _{(\mu }\eta _{\nu )}.  \label{PFRS16}
\end{equation}

\section{Deformation of the solution to the master equation: a brief review}

We begin with a \textquotedblleft free\textquotedblright\ gauge theory,
described by a Lagrangian action $S_{0}^{\mathrm{L}}\left[ \Phi ^{\alpha
_{0}}\right] $, invariant under some gauge transformations $\delta
_{\epsilon }\Phi ^{\alpha _{0}}=Z_{\;\;\alpha _{1}}^{\alpha _{0}}\epsilon
^{\alpha _{1}}$, i.e. $\frac{\delta S_{0}^{\mathrm{L}}}{\delta \Phi ^{\alpha
_{0}}}Z_{\;\;\alpha _{1}}^{\alpha _{0}}=0$, and consider the problem of
constructing consistent interactions among the fields $\Phi ^{\alpha _{0}}$
such that the couplings preserve both the field spectrum and the original
number of gauge symmetries. This matter is addressed by means of
reformulating the problem of constructing consistent interactions as a
deformation problem of the solution to the master equation corresponding to
the \textquotedblleft free\textquotedblright\ theory~\cite{def}. Such a
reformulation is possible due to the fact that the solution to the master
equation contains all the information on the gauge structure of the theory.
If an interacting gauge theory can be consistently constructed, then the
solution $\bar{S}$ to the master equation $\left( \bar{S},\bar{S}\right) =0$
associated with the \textquotedblleft free\textquotedblright\ theory can be
deformed into a solution $S$
\begin{eqnarray}
\bar{S}\rightarrow S &=&\bar{S}+\lambda S_{1}+\lambda ^{2}S_{2}+\cdots
\notag \\
&=&\bar{S}+\lambda \int d^{D}x\,a+\lambda ^{2}\int d^{D}x\,b+\cdots ,
\label{PFRS2.2}
\end{eqnarray}%
of the master equation for the deformed theory
\begin{equation}
\left( S,S\right) =0,  \label{PFRS2.3}
\end{equation}%
such that both the ghost and antifield spectra of the initial theory are
preserved. The equation (\ref{PFRS2.3}) splits, according to the various
orders in the coupling constant (deformation parameter) $\lambda $, into a
tower of equations:
\begin{eqnarray}
\left( \bar{S},\bar{S}\right) &=&0,  \label{PFRS2.4} \\
2\left( S_{1},\bar{S}\right) &=&0,  \label{PFRS2.5} \\
2\left( S_{2},\bar{S}\right) +\left( S_{1},S_{1}\right) &=&0,
\label{PFRS2.6} \\
\left( S_{3},\bar{S}\right) +\left( S_{1},S_{2}\right) &=&0,  \label{PFRS2.7}
\\
&&\vdots  \notag
\end{eqnarray}

The equation (\ref{PFRS2.4}) is fulfilled by hypothesis. The next
equation requires that the first-order deformation of the solution
to the master equation, $S_{1}$, is a cocycle of the
\textquotedblleft
free\textquotedblright\ BRST differential $s\cdot =\left( \cdot ,\bar{S}%
\right) $. However, only cohomologically non-trivial solutions to (\ref%
{PFRS2.5}) should be taken into account, as the BRST-exact solutions
can be eliminated by some (in general non-linear) field
redefinitions. This means
that $S_{1}$ pertains to the ghost number zero cohomological space of $s$, $%
H^{0}\left( s\right) $, which is generically non-empty because it is
isomorphic to the space of physical observables of the
\textquotedblleft free\textquotedblright\ theory. It has been shown
(by the triviality of the antibracket map in the cohomology of the
BRST differential) that there are no obstructions in finding
solutions to the remaining equations, namely (\ref{PFRS2.6}),
(\ref{PFRS2.7}), etc. However, the resulting interactions may be
non-local, and there might even appear obstructions if one insists
on their locality. The analysis of these obstructions can be done by
means of standard cohomological techniques.

\section{Consistent interactions between the spin-two field and the massive
Rarita-Schwinger field}

\subsection{Standard material: $H\left( \protect\gamma \right) $ and $%
H\left( \protect\delta |d\right) $}

This section is devoted to the investigation of consistent cross-couplings
that can be introduced between a spin-two field and a massive
Rarita-Schwinger field. This matter is addressed in the context of the
antifield-BRST deformation procedure briefly addressed in the above and
relies on computing the solutions to the equations (\ref{PFRS2.5})--(\ref%
{PFRS2.7}), etc., with the help of the free BRST cohomology.

For obvious reasons, we consider only smooth, local, (background)
Lorentz invariant quantities and, moreover, Poincar\'{e} invariant
quantities (i.e. we do not allow explicit dependence on the
spacetime coordinates). The smoothness of the deformations refers to
the fact that the deformed solution to the master equation
(\ref{PFRS2.2}) is smooth in the coupling constant $\lambda $ and
reduces to the original solution (\ref{PFRS16}) in the free limit
$\lambda =0 $. In addition, we require the conservation of the
number of derivatives
on each field (this condition is frequently met in the literature~\cite%
{multi,boulcqg}). If we make the notation $S_{1}=\int d^{4}x\,a$, with $a$ a
local function, then the equation (\ref{PFRS2.5}), which we have seen that
controls the first-order deformation, takes the local form
\begin{equation}
sa=\partial _{\mu }m^{\mu },\quad \mathrm{gh}\left( a\right) =0,\quad
\varepsilon \left( a\right) =0,  \label{PFRS3.1}
\end{equation}%
for some local $m^{\mu }$, and it shows that the non-integrated density of
the first-order deformation pertains to the local cohomology of the BRST
differential in ghost number zero, $a\in H^{0}\left( s|d\right) $, where $d$
denotes the exterior spacetime differential. The solution to the equation (%
\ref{PFRS3.1}) is unique up to $s$-exact pieces plus divergences
\begin{equation}
a\rightarrow a+sb+\partial _{\mu }n^{\mu },\;\mathrm{gh}\left( b\right)
=-1,\;\varepsilon \left( b\right) =1,\;\mathrm{gh}\left( n^{\mu }\right)
=0,\;\varepsilon \left( n^{\mu }\right) =0.  \label{PFRS3.1a}
\end{equation}%
At the same time, if the general solution of (\ref{PFRS3.1}) is found to be
completely trivial, $a=sb+\partial _{\mu }n^{\mu }$, then it can be made to
vanish $a=0$.

In order to analyze the equation (\ref{PFRS3.1}), we develop $a$ according
to the antighost number
\begin{equation}
a=\sum\limits_{i=0}^{I}a_{i},\quad \mathrm{agh}\left( a_{i}\right) =i,\quad
\mathrm{gh}\left( a_{i}\right) =0,\quad \varepsilon \left( a_{i}\right) =0,
\label{PFRS3.2}
\end{equation}%
and take this decomposition to stop at some finite value $I$ of the
antighost number. The fact that $I$ in (\ref{PFRS3.2}) is finite can
be argued like in~\cite{multi}. Inserting the above expansion into
the equation (\ref{PFRS3.1}) and projecting it on the various values
of the antighost number with the help of the split (\ref{PFRS4}), we
obtain the tower of equations
\begin{eqnarray}
\gamma a_{I} &=&\partial _{\mu }\overset{\left( I\right) }{m}^{\mu },
\label{PFRS3.3} \\
\delta a_{I}+\gamma a_{I-1} &=&\partial _{\mu }\overset{\left( I-1\right) }{m%
}^{\mu },  \label{PFRS3.4} \\
\delta a_{i}+\gamma a_{i-1} &=&\partial _{\mu }\overset{\left( i-1\right) }{m%
}^{\mu },\quad 1\leq i\leq I-1,  \label{PFRS3.5}
\end{eqnarray}%
where $\left( \overset{\left( i\right) }{m}^{\mu }\right) _{i=\overline{0,I}%
} $ are some local currents with $\mathrm{agh}\left( \overset{\left(
i\right) }{m}^{\mu }\right) =i$. Moreover, according to the general result
from~\cite{multi} in the absence of the collection indices, the equation (%
\ref{PFRS3.3}) can be replaced\footnote{%
This is because the presence of the matter fields does not modify the
general results on $H\left( \gamma \right) $ presented in~\cite{multi}.} in
strictly positive antighost numbers by
\begin{equation}
\gamma a_{I}=0,\quad I>0.  \label{PFRS3.6}
\end{equation}%
Due to the second-order nilpotency of $\gamma $ ($\gamma ^{2}=0$), the
solution to the equation (\ref{PFRS3.6}) is clearly unique up to $\gamma $%
-exact contributions
\begin{equation}
a_{I}\rightarrow a_{I}+\gamma b_{I},\quad \mathrm{agh}\left( b_{I}\right)
=I,\quad \mathrm{pgh}\left( b_{I}\right) =I-1,\quad \varepsilon \left(
b_{I}\right) =1.  \label{PFRSr68}
\end{equation}%
Meanwhile, if it turns out that $a_{I}$ reduces to $\gamma $-exact terms
only, $a_{I}=\gamma b_{I}$, then it can be made to vanish, $a_{I}=0$. The
non-triviality of the first-order deformation $a$ is thus translated at its
highest antighost number component into the requirement that $a_{I}\in
H^{I}\left( \gamma \right) $, where $H^{I}\left( \gamma \right) $ denotes
the cohomology of the exterior longitudinal derivative $\gamma $ in pure
ghost number equal to $I$. So, in order to solve the equation (\ref{PFRS3.1}%
) (equivalent with (\ref{PFRS3.6}) and (\ref{PFRS3.4})--(\ref{PFRS3.5})), we
need to compute the cohomology of $\gamma $, $H\left( \gamma \right) $, and,
as it will be made clear below, also the local cohomology of $\delta $ in
pure ghost number zero, $H\left( \delta |d\right) $.

Using the results on the cohomology of the exterior longitudinal
differential for a Pauli-Fierz field~\cite{multi}, as well as the
definitions (\ref{PFRS11}) and (\ref{PFRS12}), we can state that $H\left(
\gamma \right) $ is generated on the one hand by $\Phi _{\alpha _{0}}^{\ast
} $, $\eta _{\mu }^{\ast }$, $\psi _{\mu }$ and $K_{\mu \nu \alpha \beta }$
together with all of their spacetime derivatives and, on the other hand, by
the ghosts $\eta _{\mu }$ and $\partial _{\lbrack \mu }\eta _{\nu ]}$. So,
the most general (and non-trivial), local solution to (\ref{PFRS3.6}) can be
written, up to $\gamma $-exact contributions, as
\begin{equation}
a_{I}=\alpha _{I}\left( \left[ \psi _{\mu }\right] ,\left[ K_{\mu \nu \alpha
\beta }\right] ,\left[ \Phi _{\alpha _{0}}^{\ast }\right] ,\left[ \eta _{\mu
}^{\ast }\right] \right) \omega ^{I}\left( \eta _{\mu },\partial _{\lbrack
\mu }\eta _{\nu ]}\right) ,  \label{PFRS3.10}
\end{equation}%
where the notation $f\left( \left[ q\right] \right) $ means that $f$ depends
on $q$ and its derivatives up to a finite order, while $\omega ^{I}$ denotes
the elements of a basis in the space of polynomials with pure ghost number $%
I $ in the corresponding ghosts and their antisymmetrized first-order
derivatives. The objects $\alpha _{I}$ have the pure ghost number equal to
zero and are required to fulfill the property $\mathrm{agh}\left( \alpha
_{I}\right) =I$ in order to ensure that the ghost number of $a_{I}$ is equal
to zero. Since they have a bounded number of derivatives and a finite
antighost number, $\alpha _{I}$ are actually polynomials in the linearized
Riemann tensor, in the antifields, in all of their derivatives, as well as
in the derivatives of the Rarita-Schwinger fields. The anticommuting
behaviour of the vector-spinors induces that $\alpha _{I}$ are also
polynomials in the undifferentiated Rarita-Schwinger fields, so we conclude
that these elements exhibit a polynomial character in all of their
arguments. Due to their $\gamma $-closeness, $\gamma \alpha _{I}=0$, $\alpha
_{I}$ will be called invariant polynomials. In zero antighost number the
invariant polynomials are polynomials in the linearized Riemann tensor $%
K_{\mu \nu \alpha \beta }$, in the Rarita-Schwinger spinors, as well as in
their derivatives.

Inserting (\ref{PFRS3.10}) in (\ref{PFRS3.4}) we obtain that a necessary
(but not sufficient) condition for the existence of (non-trivial) solutions $%
a_{I-1}$ is that the invariant polynomials $\alpha _{I}$ are (non-trivial)
objects from the local cohomology of the Koszul-Tate differential $H\left(
\delta |d\right) $ in pure ghost number zero and in strictly positive
antighost numbers $I>0$%
\begin{equation}
\delta \alpha _{I}=\partial _{\mu }\overset{\left( I-1\right) }{j}^{\mu
},\quad \mathrm{agh}\left( \overset{\left( I-1\right) }{j}^{\mu }\right)
=I-1,\quad \mathrm{pgh}\left( \overset{\left( I-1\right) }{j}^{\mu }\right)
=0.  \label{PFRS3.11}
\end{equation}
We recall that $H\left( \delta |d\right) $ is completely trivial in both
strictly positive antighost \textit{and} pure ghost numbers (for instance,
see~\cite{gen1}, Theorem 5.4 and~\cite{commun1}). Using the fact that the
Cauchy order of the free theory under study is equal to two together with
the general results from~\cite{gen1}, according to which the local
cohomology of the Koszul-Tate differential in pure ghost number zero is
trivial in antighost numbers strictly greater than its Cauchy order, we can
state that
\begin{equation}
H_{J}\left( \delta |d\right) =0\quad \mathrm{for\;all\;}J>2,
\label{PFRS3.12}
\end{equation}
where $H_{J}\left( \delta |d\right) $ represents the local cohomology of the
Koszul-Tate differential in zero pure ghost number and in antighost number $%
J $. An interesting property of invariant polynomials for the free model
under study is that if an invariant polynomial $\alpha _{J}$, with $\mathrm{%
agh}\left( \alpha _{J}\right) =J\geq 2$, is trivial in $H_{J}\left( \delta
|d\right) $, then it can be taken to be trivial also in $H_{J}^{\mathrm{inv}%
}\left( \delta |d\right) $, i.e.
\begin{equation}
\left( \alpha _{J}=\delta b_{J+1}+\partial _{\mu }\overset{(J)}{c}^{\mu },\;%
\mathrm{agh}\left( \alpha _{J}\right) =J\geq 2\right) \Rightarrow \alpha
_{J}=\delta \beta _{J+1}+\partial _{\mu }\overset{(J)}{\gamma }^{\mu },
\label{PFRS3.12ax}
\end{equation}
with both $\beta _{J+1}$ and $\overset{(J)}{\gamma }^{\mu }$ invariant
polynomials. Here, $H_{J}^{\mathrm{inv}}\left( \delta |d\right) $ denotes
the invariant characteristic cohomology (the local cohomology of the
Koszul-Tate differential in the space of invariant polynomials) in antighost
number $J$. This property is proved in~\cite{multi} in the case of a
collection of Pauli-Fierz fields and remains valid in the case considered
here since the matter fields do not carry gauge symmetries, so we can write
that
\begin{equation}
H_{J}^{\mathrm{inv}}\left( \delta |d\right) =0\quad \mathrm{for\;all\;}J>2.
\label{PFRS3.12x}
\end{equation}
For the same reason, the antifields of the matter fields can bring only
trivial contributions to $H_{J}\left( \delta |d\right) $ and $H_{J}^{\mathrm{%
inv}}\left( \delta |d\right) $ for $J\geq 2$, so the results from~\cite%
{multi} concerning both $H_{2}\left( \delta |d\right) $ in pure ghost number
zero and $H_{2}^{\mathrm{inv}}\left( \delta |d\right) $ remain valid. These
cohomological spaces are still spanned by the undifferentiated antifields
corresponding to the ghosts
\begin{equation}
H_{2}\left( \delta |d\right) \;\mathrm{and}\;H_{2}^{\mathrm{inv}}\left(
\delta |d\right) :\left( \eta ^{*\mu }\right) .  \label{PFRS3.12b}
\end{equation}
In contrast to the groups $\left( H_{J}\left( \delta |d\right) \right)
_{J\geq 2}$ and $\left( H_{J}^{\mathrm{inv}}\left( \delta |d\right) \right)
_{J\geq 2}$, which are finite-dimensional, the cohomology $H_{1}\left(
\delta |d\right) $ in pure ghost number zero, known to be related to global
symmetries and ordinary conservation laws, is infinite-dimensional since the
theory is free. Moreover, $H_{1}\left( \delta |d\right) $ involves
non-trivially the antifields of the matter fields.

The previous results on $H\left( \delta |d\right) $ and $H^{\mathrm{inv}%
}\left( \delta |d\right) $ in strictly positive antighost numbers are
important because they control the obstructions to removing the antifields
from the first-order deformation. More precisely, based on the formulas (\ref%
{PFRS3.11})--(\ref{PFRS3.12x}), one can successively eliminate all the
pieces of antighost number strictly greater that two from the non-integrated
density of the first-order deformation by adding only trivial terms, so one
can take, without loss of non-trivial objects, the condition $I\leq 2$ in
the decomposition (\ref{PFRS3.2}). In addition, the last representative is
of the form (\ref{PFRS3.10}), where the invariant polynomial is necessarily
a non-trivial object from $H_{2}^{\mathrm{inv}}\left( \delta |d\right) $ for
$I=2$, and respectively from $H_{1}\left( \delta |d\right) $ for $I=1$.

\subsection{First-order deformation \label{firstord}}

In the case $I=2$ the non-integrated density of the first-order deformation (%
\ref{PFRS3.2}) becomes
\begin{equation}
a=a_{0}+a_{1}+a_{2}.  \label{PFRS3.12t}
\end{equation}%
We can further decompose $a$ in a natural manner as a sum between three
kinds of deformations
\begin{equation}
a=a^{\left( \mathrm{PF}\right) }+a^{\left( \mathrm{int}\right) }+a^{\left(
\mathrm{RS}\right) },  \label{PFRS3.12a}
\end{equation}%
where $a^{\left( \mathrm{PF}\right) }$ contains only
fields/ghosts/antifields from the Pauli-Fierz sector, $a^{\left( \mathrm{int}%
\right) }$ describes the cross-interactions between the two theories (so it
effectively mixes both sectors), and $a^{\left( \mathrm{RS}\right) }$
involves only the Rarita-Schwinger sector. The component $a^{\left( \mathrm{%
PF}\right) }$ is completely known (for a detailed analysis see~\cite{multi})
and satisfies individually an equation of the type (\ref{PFRS3.1}). It
admits a decomposition similar to (\ref{PFRS3.12t})
\begin{equation}
a^{\left( \mathrm{PF}\right) }=a_{0}^{\left( \mathrm{PF}\right)
}+a_{1}^{\left( \mathrm{PF}\right) }+a_{2}^{\left( \mathrm{PF}\right) },
\label{descPF}
\end{equation}%
where
\begin{eqnarray}
a_{2}^{\left( \mathrm{PF}\right) } &=&\frac{1}{2}\eta ^{\ast \mu }\eta ^{\nu
}\partial _{\left[ \mu \right. }\eta _{\left. \nu \right] },  \label{PFa2} \\
a_{1}^{\left( \mathrm{PF}\right) } &=&h^{\ast \mu \rho }\left( \left(
\partial _{\rho }\eta ^{\nu }\right) h_{\mu \nu }-\eta ^{\nu }\partial
_{\lbrack \mu }h_{\nu ]\rho }\right) ,  \label{PFa1}
\end{eqnarray}%
and $a_{0}^{\left( \mathrm{PF}\right) }$ is the cubic vertex of the
Einstein-Hilbert Lagrangian plus a cosmological term\footnote{%
The terms $a_{2}^{\left( \mathrm{PF}\right) }$ and $a_{1}^{\left( \mathrm{PF}%
\right) }$ given in (\ref{PFa2}) and (\ref{PFa1}) differ from the
corresponding ones in~\cite{multi} by a $\gamma $-exact and respectively a $%
\delta $-exact contribution. However, the difference between our $%
a_{2}^{\left( \mathrm{PF}\right) }+$ $a_{1}^{\left( \mathrm{PF}\right) }$
and the corresponding sum from~\cite{multi} is a $s$-exact modulo $d$
quantity. The associated component of antighost number zero, $a_{0}^{\left(
\mathrm{PF}\right) }$, is nevertheless the same in both formulations. As a
consequence, the object $a^{\left( \mathrm{PF}\right) }$ and the first-order
deformation in~\cite{multi} belong to the same cohomological class from $%
H^{0}\left( s|d\right) $.}. Due to the fact that $a^{\left( \mathrm{int}%
\right) }$ and $a^{\left( \mathrm{RS}\right) }$ involve different kinds of
fields, it follows that $a^{\left( \mathrm{int}\right) }$ and $a^{\left(
\mathrm{RS}\right) }$ are subject to some separate equations
\begin{eqnarray}
sa^{\left( \mathrm{int}\right) } &=&\partial _{\mu }m^{\left( \mathrm{int}%
\right) \mu },  \label{PFRSint} \\
sa^{\left( \mathrm{RS}\right) } &=&\partial _{\mu }m^{\left( \mathrm{RS}%
\right) \mu },  \label{PFRSrs}
\end{eqnarray}%
for some local $m^{\mu }$'s. In the sequel we analyze the general solutions
to these equations.

Since the massive Rarita-Schwinger field does not carry gauge symmetries of
its own, it results that the massive gravitino sector can only occur in
antighost number one and zero, so, without loss of generality, we can take
\begin{equation}
a^{\left( \mathrm{int}\right) }=a_{0}^{\left( \mathrm{int}\right)
}+a_{1}^{\left( \mathrm{int}\right) }  \label{PFRS3.21}
\end{equation}%
in (\ref{PFRSint}), where the components involved in the right-hand side of (%
\ref{PFRS3.21}) are subject to the equations
\begin{eqnarray}
\gamma a_{1}^{\left( \mathrm{int}\right) } &=&0,  \label{PFRS3.14a} \\
\delta a_{1}^{\left( \mathrm{int}\right) }+\gamma a_{0}^{\left( \mathrm{int}%
\right) } &=&\partial _{\mu }\overset{(0)}{m}^{\left( \mathrm{int}\right)
\mu }.  \label{PFRS3.14b}
\end{eqnarray}%
According to (\ref{PFRS3.10}) in pure ghost number one and because $\omega
^{1}$ is spanned by
\begin{equation*}
\omega ^{1}=\left( \eta _{\mu },\partial _{\lbrack \mu }\eta _{\nu ]}\right)
,
\end{equation*}%
we infer that the most general expression of $a_{1}^{\left( \mathrm{int}%
\right) }$ as solution to the equation (\ref{PFRS3.14a}) is\footnote{\label%
{hstar}We remark that in principle we might have added to $a_{1}^{(\mathrm{%
int})}$ a component $\tilde{a}_{1}^{\left( \mathrm{int}\right) }$ linear in
the antifield of the Pauli-Fierz field, $h^{\ast \mu \nu }$. However, such
terms cannot produce a consistent component of the first-order deformation
in antighost number zero, as it is shown in Appendix B.}
\begin{equation}
a_{1}^{\left( \mathrm{int}\right) }=\psi ^{\ast \mu }\left( N_{\;\;\mu
}^{\rho }\eta _{\rho }+N_{\;\;\;\;\mu }^{\rho \lambda }\partial _{\lbrack
\rho }\eta _{\lambda ]}\right) ,  \label{PFRS3.22}
\end{equation}%
where $N_{\;\;\mu }^{\rho }$ and $N_{\;\;\;\;\mu }^{\rho \lambda }$ are
real, odd spinor-like functions, with $N_{\;\;\;\;\mu }^{\rho \lambda }$
antisymmetric in its upper indices. All the objects denoted by $N$ are
gauge-invariant, so they may depend on $\psi _{\mu }$, $K_{\mu \nu \rho
\lambda }$, and their spacetime derivatives. At this stage we recall the
hypothesis on the conservation of the number of derivatives on each field,
which allows us to simplify the solution (\ref{PFRS3.22}) to the equation (%
\ref{PFRS3.14a}) by imposing that the following requirements are
simultaneously satisfied:

i) the interaction vertices present in $a_{0}^{\left( \mathrm{int}\right) }$
as solution to (\ref{PFRS3.14b}), assuming $a_{0}^{\left( \mathrm{int}%
\right) }$ exists, contain at most two derivatives of the fields;

ii) the deformed field equations associated with $a_{0}^{\left( \mathrm{int}%
\right) }$ involve at most the first-order derivatives of the spinor fields
and at most the second-order derivatives of the Pauli-Fierz field.

By applying the differential $\delta $ on (\ref{PFRS3.22}) and using the
definitions (\ref{PFRS8})--(\ref{PFRS12}), we infer that
\begin{equation}
\delta a_{1}^{\left( \mathrm{int}\right) }=\partial _{\mu }m^{\mu }+\gamma
b_{0}+c_{0},  \label{PFRS3.22a}
\end{equation}%
where
\begin{equation}
m^{\mu }=-\mathrm{i}\bar{\psi}_{\beta }\gamma ^{\mu \beta \nu }\left(
N_{\;\;\nu }^{\rho }\eta _{\rho }+N_{\;\;\;\;\mu }^{\rho \lambda }\partial
_{\lbrack \rho }\eta _{\lambda ]}\right) ,  \label{PFRS3.22b}
\end{equation}%
\begin{equation}
b_{0}=\frac{\mathrm{i}}{2}\bar{\psi}_{\beta }\gamma ^{\alpha \beta \mu
}\left( N_{\;\;\mu }^{\rho }h_{\alpha \rho }+2N_{\;\;\;\;\mu }^{\rho \lambda
}\partial _{\lbrack \rho }h_{\lambda ]\alpha }\right) ,  \label{PFRS3.22c}
\end{equation}%
\begin{eqnarray}
c_{0} &=&\left( m\bar{\psi}_{\alpha }\gamma ^{\alpha \mu }N_{\;\;\mu }^{\rho
}+\mathrm{i}\bar{\psi}_{\beta }\gamma ^{\alpha \beta \mu }\partial _{\alpha
}N_{\;\;\mu }^{\rho }\right) \eta _{\rho }  \notag \\
&&+\left( m\bar{\psi}_{\alpha }\gamma ^{\alpha \mu }N_{\;\;\;\;\mu }^{\rho
\lambda }+\mathrm{i}\bar{\psi}_{\beta }\gamma ^{\alpha \beta \mu }\partial
_{\alpha }N_{\;\;\;\;\mu }^{\rho \lambda }\right.  \notag \\
&&\left. +\frac{\mathrm{i}}{2}\bar{\psi}_{\beta }\gamma ^{\rho \beta \mu
}N_{\;\;\mu }^{\lambda }\right) \partial _{\lbrack \rho }\eta _{\lambda ]}.
\label{PFRS3.22d}
\end{eqnarray}%
Taking into account the previous two requirements on the derivative
behaviour of $a_{0}^{\left( \mathrm{int}\right) }$, from (\ref{PFRS3.22c})
we get that the spinor-tensor $N_{\;\;\mu }^{\rho }$ may contain at most one
derivative of the spinor $\psi _{\mu }$, while the spinor-tensor $%
N_{\;\;\;\;\mu }^{\rho \lambda }$ can only depend on the undifferentiated
Rarita-Schwinger field. As a consequence, we have that
\begin{equation}
N_{\;\;\mu }^{\rho }=\bar{N}_{\;\;\;\;\mu }^{\rho \lambda }\psi _{\lambda }+%
\bar{N}_{\;\;\;\;\;\;\mu }^{\rho \lambda \sigma }\partial _{\lambda }\psi
_{\sigma },\quad N_{\;\;\;\;\mu }^{\rho \lambda }=N_{\;\;\;\;\;\;\mu }^{\rho
\lambda \sigma }\psi _{\sigma },  \label{ww1}
\end{equation}%
and hence
\begin{equation}
a_{1}^{\left( \mathrm{int}\right) }=\psi ^{\ast \mu }\left( \bar{N}%
_{\;\;\;\;\mu }^{\rho \lambda }\psi _{\lambda }+\bar{N}_{\;\;\;\;\;\;\mu
}^{\rho \lambda \sigma }\partial _{\lambda }\psi _{\sigma }\right) \eta
_{\rho }+\psi ^{\ast \mu }N_{\;\;\;\;\;\;\mu }^{\rho \lambda \sigma }\psi
_{\sigma }\partial _{\lbrack \rho }\eta _{\lambda ]},  \label{PFRS3.23}
\end{equation}%
where $\bar{N}_{\;\;\;\;\mu }^{\rho \lambda }$, $\bar{N}_{\;\;\;\;\;\;\mu
}^{\rho \lambda \sigma }$, and $N_{\;\;\;\;\;\;\mu }^{\rho \lambda \sigma }$
are real, bosonic $4\times 4$ matrices that may depend only on the
undifferentiated spinor-vector $\psi _{\mu }$. Inserting (\ref{ww1}) in the
formulas (\ref{PFRS3.22c})--(\ref{PFRS3.22d}), we get
\begin{eqnarray}
b_{0} &=&\frac{\mathrm{i}}{2}\bar{\psi}_{\beta }\gamma ^{\alpha \beta \mu
}\left( \left( \bar{N}_{\;\;\;\;\mu }^{\rho \lambda }\psi _{\lambda }+\bar{N}%
_{\;\;\;\;\;\;\mu }^{\rho \lambda \sigma }\partial _{\lambda }\psi _{\sigma
}\right) h_{\alpha \rho }\right.  \notag \\
&&\left. +2N_{\;\;\;\;\;\;\mu }^{\rho \lambda \sigma }\psi _{\sigma
}\partial _{\lbrack \rho }h_{\lambda ]\alpha }\right) ,  \label{ww2}
\end{eqnarray}%
\begin{eqnarray}
c_{0} &=&\left( m\bar{\psi}_{\alpha }\gamma ^{\alpha \mu }\left( \bar{N}%
_{\;\;\;\;\mu }^{\rho \lambda }\psi _{\lambda }+\bar{N}_{\;\;\;\;\;\;\mu
}^{\rho \lambda \sigma }\partial _{\lambda }\psi _{\sigma }\right) \right.
\notag \\
&&\left. +\mathrm{i}\bar{\psi}_{\beta }\gamma ^{\alpha \beta \mu }\partial
_{\alpha }\left( \bar{N}_{\;\;\;\;\mu }^{\rho \lambda }\psi _{\lambda }+\bar{%
N}_{\;\;\;\;\;\;\mu }^{\rho \lambda \sigma }\partial _{\lambda }\psi
_{\sigma }\right) \right) \eta _{\rho }  \notag \\
&&+\left( m\bar{\psi}_{\alpha }\gamma ^{\alpha \mu }N_{\;\;\;\;\;\;\mu
}^{\rho \lambda \sigma }\psi _{\sigma }+\mathrm{i}\bar{\psi}_{\beta }\gamma
^{\alpha \beta \mu }\partial _{\alpha }\left( N_{\;\;\;\;\;\;\mu }^{\rho
\lambda \sigma }\psi _{\sigma }\right) \right.  \notag \\
&&\left. +\frac{\mathrm{i}}{2}\bar{\psi}_{\beta }\gamma ^{\rho \beta \mu
}\left( \bar{N}_{\;\;\;\;\mu }^{\lambda \sigma }\psi _{\sigma }+\bar{N}%
_{\;\;\;\;\;\;\mu }^{\lambda \alpha \sigma }\partial _{\alpha }\psi _{\sigma
}\right) \right) \partial _{\lbrack \rho }\eta _{\lambda ]}.  \label{ww3}
\end{eqnarray}%
The condition that $\delta a_{1}^{\left( \mathrm{int}\right) }$ should be
written like in (\ref{PFRS3.14b}) restricts $c_{0}$ expressed in (\ref{ww3})
to be a $\gamma $-exact modulo $d$ quantity, i.e.
\begin{equation}
c_{0}=\gamma m+\partial _{\mu }n^{\mu }.  \label{ww4}
\end{equation}%
At this stage it is useful to split $c_{0}$ like
\begin{equation}
c_{0}=\sum\limits_{k=0}^{2}\left( c_{0}\right) _{k},  \label{ww5}
\end{equation}%
where $\left( c_{0}\right) _{k}$ denotes the piece from $c_{0}$ with $k$%
-derivatives. According to this decomposition, it follows that each $\left(
c_{0}\right) _{k}$ should be written in a $\gamma $-exact modulo $d$ form,
such that (\ref{PFRS3.14b}) is indeed satisfied. Using (\ref{ww3}), we
obtain that
\begin{equation}
\left( c_{0}\right) _{0}=m\bar{\psi}_{\alpha }\gamma ^{\alpha \mu }\bar{N}%
_{\;\;\;\;\mu }^{\rho \lambda }\psi _{\lambda }\eta _{\rho }.  \label{ww6}
\end{equation}%
As the right-hand side of (\ref{ww6}) is derivative-free, it follows that
these terms neither reduce to a total derivative nor can be expressed in a $%
\gamma $-exact form, so they must vanish
\begin{equation}
\bar{\psi}_{\alpha }\gamma ^{\alpha \mu }\bar{N}_{\;\;\;\;\mu }^{\rho
\lambda }\psi _{\lambda }=0.  \label{ww7}
\end{equation}%
Simple computation exhibit that (\ref{ww7}) is checked if
\begin{equation}
\gamma ^{0}\gamma ^{\alpha \mu }\bar{N}_{\;\;\;\;\mu }^{\rho \lambda
}=\left( \gamma ^{0}\gamma ^{\lambda \mu }\bar{N}_{\;\;\;\;\mu }^{\rho
\alpha }\right) ^{\intercal },  \label{gama5}
\end{equation}%
whose general solution is expressed by
\begin{equation}
\bar{N}_{\;\;\;\;\mu }^{\rho \lambda }=c_{1}\delta _{\mu }^{\rho }\gamma
^{\lambda }+c_{2}\delta _{\mu }^{\lambda }\gamma ^{\rho }+c_{3}\sigma ^{\rho
\lambda }\gamma _{\mu }+\frac{1}{2}\left( c_{1}+2c_{2}+3c_{3}\right) \gamma
_{\;\;\;\;\mu }^{\rho \lambda },  \label{ww8}
\end{equation}%
with $c_{1}$, $c_{2}$, and $c_{3}$ some arbitrary functions depending on $%
\psi _{\mu }$. As it has been shown in Appendix B, the functions $c_{1}$, $%
c_{2}$, and $c_{3}$ from (\ref{ww8}) can be made to vanish by adding some
trivial, $s$-exact terms and by conveniently redefining the functions $\bar{N%
}_{\;\;\;\;\;\;\mu }^{\rho \lambda \sigma }$. In consequence, we can take
\begin{equation}
\bar{N}_{\;\;\;\;\mu }^{\rho \lambda }=0.  \label{ww9}
\end{equation}%
The equation (\ref{ww4}) for $k=1$ becomes
\begin{equation}
m\bar{\psi}_{\alpha }\gamma ^{\alpha \mu }\bar{N}_{\;\;\;\;\;\;\mu }^{\rho
\lambda \sigma }\left( \partial _{\lambda }\psi _{\sigma }\right) \eta
_{\rho }+m\bar{\psi}_{\alpha }\gamma ^{\alpha \mu }N_{\;\;\;\;\;\;\mu
}^{\rho \lambda \sigma }\psi _{\sigma }\partial _{\lbrack \rho }\eta
_{\lambda ]}=\gamma m_{0}+\partial _{\mu }n_{0}^{\mu },  \label{ww10}
\end{equation}%
where $\gamma m_{0}=\left( \partial m_{0}/\partial h_{\rho \lambda }\right)
\partial _{\left( \rho \right. }\eta _{\left. \lambda \right) }$. By taking
the Euler-Lagrange derivatives of the relation (\ref{ww10}) with respect to $%
\eta _{\nu }$ we obtain that the quantity $m\bar{\psi}_{\alpha }\gamma
^{\alpha \mu }\bar{N}_{\;\;\;\;\;\;\mu }^{\rho \lambda \sigma }\left(
\partial _{\lambda }\psi _{\sigma }\right) $ should reduce to a total
derivative
\begin{equation}
m\bar{\psi}_{\alpha }\gamma ^{\alpha \mu }\bar{N}_{\;\;\;\;\;\;\mu }^{\rho
\lambda \sigma }\left( \partial _{\lambda }\psi _{\sigma }\right) =\partial
_{\lambda }M^{\rho \lambda }.  \label{ww11}
\end{equation}%
The left-hand side of (\ref{ww11}) is a full divergence if the following
conditions
\begin{eqnarray}
\partial _{\lambda }\bar{N}_{\;\;\;\;\;\;\mu }^{\rho \lambda \sigma } &=&0,
\label{ww12} \\
\gamma ^{0}\gamma ^{\alpha \mu }\bar{N}_{\;\;\;\;\;\;\mu }^{\rho \lambda
\sigma } &=&-\left( \gamma ^{0}\gamma ^{\sigma \mu }\bar{N}_{\;\;\;\;\;\;\mu
}^{\rho \lambda \alpha }\right) ^{\intercal }  \label{ww13}
\end{eqnarray}%
are simultaneously satisfied. The general solution to (\ref{ww12})--(\ref%
{ww13}) takes the form
\begin{eqnarray}
\bar{N}_{\;\;\;\;\;\;\mu }^{\rho \lambda \sigma } &=&k_{1}\left( \sigma
^{\lambda \sigma }\left( \delta _{\mu }^{\rho }+\frac{1}{2}\gamma _{\;\;\mu
}^{\rho }\right) +\sigma ^{\rho \sigma }\left( \delta _{\mu }^{\lambda }+%
\frac{1}{2}\gamma _{\;\;\mu }^{\lambda }\right) \right)  \notag \\
&&+k_{2}\sigma ^{\rho \lambda }\left( \delta _{\mu }^{\sigma }+\frac{1}{2}%
\gamma _{\;\;\mu }^{\sigma }\right) +k_{4}\sigma ^{\rho \lambda }\delta
_{\mu }^{\sigma }  \notag \\
&&+k_{3}\left( \sigma ^{\lambda \sigma }\delta _{\mu }^{\rho }-\sigma ^{\rho
\sigma }\delta _{\mu }^{\lambda }-\delta _{\mu }^{\sigma }\gamma ^{\rho
\lambda }+\gamma _{\;\;\;\;\;\;\mu }^{\rho \lambda \sigma }\right.  \notag \\
&&\left. -\frac{1}{2}\left( \delta _{\mu }^{\lambda }\gamma ^{\rho \sigma
}-\delta _{\mu }^{\rho }\gamma ^{\lambda \sigma }\right) +\frac{1}{2}\left(
\sigma ^{\sigma \lambda }\gamma _{\;\;\mu }^{\rho }-\sigma ^{\rho \sigma
}\gamma _{\;\;\mu }^{\lambda }\right) \right)  \notag \\
&=&\bar{N}_{1\;\;\;\;\;\;\mu }^{\rho \lambda \sigma }+\bar{N}%
_{2\;\;\;\;\;\;\mu }^{\rho \lambda \sigma },  \label{ww14}
\end{eqnarray}%
with
\begin{eqnarray}
\bar{N}_{1\;\;\;\;\;\;\mu }^{\rho \lambda \sigma } &=&k_{1}\left( \sigma
^{\lambda \sigma }\left( \delta _{\mu }^{\rho }+\frac{1}{2}\gamma _{\;\;\mu
}^{\rho }\right) +\sigma ^{\rho \sigma }\left( \delta _{\mu }^{\lambda }+%
\frac{1}{2}\gamma _{\;\;\mu }^{\lambda }\right) \right)  \notag \\
&&+k_{2}\sigma ^{\rho \lambda }\left( \delta _{\mu }^{\sigma }+\frac{1}{2}%
\gamma _{\;\;\mu }^{\sigma }\right) +k_{4}\sigma ^{\rho \lambda }\delta
_{\mu }^{\sigma },  \label{ww15}
\end{eqnarray}%
and $\left( k_{i}\right) _{i=\overline{1,4}}$ some arbitrary constants.
Under these circumstances (if the equations (\ref{ww12})--(\ref{ww13}) are
verified), we find that
\begin{eqnarray}
&&m\bar{\psi}_{\alpha }\gamma ^{\alpha \mu }\bar{N}_{\;\;\;\;\;\;\mu }^{\rho
\lambda \sigma }\left( \partial _{\lambda }\psi _{\sigma }\right) \eta
_{\rho }+m\bar{\psi}_{\alpha }\gamma ^{\alpha \mu }N_{\;\;\;\;\;\;\mu
}^{\rho \lambda \sigma }\psi _{\sigma }\partial _{\lbrack \rho }\eta
_{\lambda ]}  \notag \\
&=&\gamma \left( -\frac{1}{4}m\bar{\psi}_{\alpha }\gamma ^{\alpha \mu }\bar{N%
}_{1\;\;\;\;\;\;\mu }^{\rho \lambda \sigma }\psi _{\sigma }h_{\rho \lambda
}\right) +\partial _{\lambda }\left( \frac{1}{2}m\bar{\psi}_{\alpha }\gamma
^{\alpha \mu }\bar{N}_{\;\;\;\;\;\;\mu }^{\rho \lambda \sigma }\psi _{\sigma
}\eta _{\rho }\right)  \notag \\
&&+m\bar{\psi}_{\alpha }\gamma ^{\alpha \mu }\left( N_{\;\;\;\;\;\;\mu
}^{\rho \lambda \sigma }+\frac{1}{4}\bar{N}_{2\;\;\;\;\;\;\mu }^{\rho
\lambda \sigma }\right) \psi _{\sigma }\partial _{\lbrack \rho }\eta
_{\lambda ]}.  \label{ww16}
\end{eqnarray}%
By comparing the last equation to (\ref{ww10}) we observe that the last term
from the right-hand side of (\ref{ww16}) must be $\gamma $-exact modulo $d$.
This takes place if
\begin{equation}
\bar{\psi}_{\alpha }\gamma ^{\alpha \mu }\left( N_{\;\;\;\;\;\;\mu }^{\rho
\lambda \sigma }+\frac{1}{4}\bar{N}_{2\;\;\;\;\;\;\mu }^{\rho \lambda \sigma
}\right) \psi _{\sigma }=0,  \label{ww17}
\end{equation}%
from which we further deduce
\begin{equation}
N_{\;\;\;\;\;\;\mu }^{\rho \lambda \sigma }=-\frac{1}{4}\bar{N}%
_{2\;\;\;\;\;\;\mu }^{\rho \lambda \sigma }+\hat{N}_{\;\;\;\;\;\;\mu }^{\rho
\lambda \sigma },  \label{ww18}
\end{equation}%
where $\hat{N}_{\;\;\;\;\;\;\mu }^{\rho \lambda \sigma }$ is solution to the
equation
\begin{equation}
\bar{\psi}_{\alpha }\gamma ^{\alpha \mu }\hat{N}_{\;\;\;\;\;\;\mu }^{\rho
\lambda \sigma }\psi _{\sigma }=0.  \label{ww19}
\end{equation}%
It is simple to see that (\ref{ww19}) holds if
\begin{equation}
\gamma ^{0}\gamma ^{\alpha \mu }\hat{N}_{\;\;\;\;\;\;\mu }^{\rho \lambda
\sigma }=\left( \gamma ^{0}\gamma ^{\sigma \mu }\hat{N}_{\;\;\;\;\;\;\mu
}^{\rho \lambda \alpha }\right) ^{\intercal },  \label{ww21}
\end{equation}%
whose general solution is given by
\begin{eqnarray}
\hat{N}_{\;\;\;\;\;\;\mu }^{\rho \lambda \sigma } &=&\bar{k}_{1}\left(
\sigma ^{\lambda \sigma }\delta _{\mu }^{\rho }-\sigma ^{\rho \sigma }\delta
_{\mu }^{\lambda }\right) +\bar{k}_{2}\delta _{\mu }^{\sigma }\gamma ^{\rho
\lambda }  \notag \\
&&+\bar{k}_{3}\left( \delta _{\mu }^{\lambda }\gamma ^{\rho \sigma }-\delta
_{\mu }^{\rho }\gamma ^{\lambda \sigma }\right) +\bar{k}_{4}\gamma
_{\;\;\;\;\;\;\mu }^{\rho \lambda \sigma }  \notag \\
&&+\frac{1}{2}\left( \bar{k}_{1}-2\bar{k}_{2}+\bar{k}_{4}\right) \left(
\sigma ^{\sigma \lambda }\gamma _{\;\;\mu }^{\rho }-\sigma ^{\rho \sigma
}\gamma _{\;\;\mu }^{\lambda }\right) ,  \label{ww20}
\end{eqnarray}%
with $\left( \bar{k}_{i}\right) _{i=\overline{1,4}}$ some arbitrary
functions depending on $\psi _{\mu }$.

Next, we analyze the solution to the equation (\ref{ww4}) for $k=2$. It
takes the concrete form
\begin{eqnarray}
&&\frac{\mathrm{i}}{2}\bar{\psi}_{\beta }\left( \gamma ^{\alpha \beta \mu }%
\bar{N}_{\;\;\;\;\;\;\mu }^{\rho \lambda \sigma }+\gamma ^{\lambda \beta \mu
}\bar{N}_{\;\;\;\;\;\;\mu }^{\rho \alpha \sigma }\right) \left( \partial
_{\alpha }\partial _{\lambda }\psi _{\sigma }\right) \eta _{\rho }  \notag \\
&&+\mathrm{i}\bar{\psi}_{\beta }\gamma ^{\alpha \beta \mu }\partial _{\alpha
}\left( N_{\;\;\;\;\;\;\mu }^{\rho \lambda \sigma }\psi _{\sigma }\right)
\partial _{\lbrack \rho }\eta _{\lambda ]}  \notag \\
&&+\frac{\mathrm{i}}{2}\bar{\psi}_{\beta }\gamma ^{\rho \beta \mu }\bar{N}%
_{\;\;\;\;\;\;\mu }^{\lambda \alpha \sigma }\left( \partial _{\alpha }\psi
_{\sigma }\right) \partial _{\lbrack \rho }\eta _{\lambda ]}  \notag \\
&=&\gamma m_{1}+\partial _{\mu }n_{1}^{\mu },  \label{ww22}
\end{eqnarray}%
with $\bar{N}_{\;\;\;\;\;\;\mu }^{\rho \lambda \sigma }$ and $%
N_{\;\;\;\;\;\;\mu }^{\rho \lambda \sigma }$ determined previously. By
taking the Euler-Lagrange derivatives of (\ref{ww22}) with respect to $\eta
_{\nu }$ and by using the result that $\gamma m_{1}=\left( \delta
m_{1}/\delta h_{\rho \lambda }\right) \partial _{\left( \rho \right. }\eta
_{\left. \lambda \right) }+\partial _{\lambda }v^{\lambda }$, with $\delta
m_{1}/\delta h_{\rho \lambda }$ the variational derivative of $m_{1}$ with
respect to $h_{\rho \lambda }$, it follows that
\begin{equation}
\frac{\mathrm{i}}{2}\bar{\psi}_{\beta }\left( \gamma ^{\alpha \beta \mu }%
\bar{N}_{\;\;\;\;\;\;\mu }^{\rho \lambda \sigma }+\gamma ^{\lambda \beta \mu
}\bar{N}_{\;\;\;\;\;\;\mu }^{\rho \alpha \sigma }\right) \left( \partial
_{\alpha }\partial _{\lambda }\psi _{\sigma }\right) =\partial _{\lambda
}P^{\rho \lambda },  \label{ww23}
\end{equation}%
for some $P^{\rho \lambda }$. The left-hand side of the last equation is
written as a full divergence if
\begin{equation}
\left( \partial _{\lambda }\bar{\psi}_{\beta }\right) \left( \gamma ^{\alpha
\beta \mu }\bar{N}_{\;\;\;\;\;\;\mu }^{\rho \lambda \sigma }+\gamma
^{\lambda \beta \mu }\bar{N}_{\;\;\;\;\;\;\mu }^{\rho \alpha \sigma }\right)
\left( \partial _{\alpha }\psi _{\sigma }\right) =0,  \label{ww24}
\end{equation}%
which further produces
\begin{equation}
k_{1}=k_{2}=k_{3}=0,  \label{ww25}
\end{equation}%
such that we have
\begin{eqnarray}
&&\frac{\mathrm{i}}{2}\bar{\psi}_{\beta }\left( \gamma ^{\alpha \beta \mu }%
\bar{N}_{\;\;\;\;\;\;\mu }^{\rho \lambda \sigma }+\gamma ^{\lambda \beta \mu
}\bar{N}_{\;\;\;\;\;\;\mu }^{\rho \alpha \sigma }\right) \left( \partial
_{\alpha }\partial _{\lambda }\psi _{\sigma }\right) \eta _{\rho }  \notag \\
&=&-\frac{\mathrm{i}k_{4}}{4}\gamma \left( \bar{\psi}_{\beta }\left( \gamma
^{\alpha \beta \sigma }\left( \partial _{\alpha }\psi _{\sigma }\right)
h+\gamma ^{\lambda \beta \sigma }\left( \partial ^{\rho }\psi _{\sigma
}\right) h_{\lambda \rho }\right) \right)  \notag \\
&&+\frac{\mathrm{i}k_{4}}{8}\bar{\psi}_{\beta }\left( \sigma ^{\alpha \rho
}\gamma ^{\lambda \beta \sigma }-\sigma ^{\alpha \lambda }\gamma ^{\rho
\beta \sigma }\right) \left( \partial _{\alpha }\psi _{\sigma }\right)
\partial _{\lbrack \rho }\eta _{\lambda ]}+\partial _{\lambda }u^{\lambda }.
\label{ww26}
\end{eqnarray}%
On the other hand, it is easy to see that
\begin{eqnarray}
&&\mathrm{i}\bar{\psi}_{\beta }\gamma ^{\alpha \beta \mu }\partial _{\alpha
}\left( N_{\;\;\;\;\;\;\mu }^{\rho \lambda \sigma }\psi _{\sigma }\right)
\partial _{\lbrack \rho }\eta _{\lambda ]}  \notag \\
&=&-\gamma \left( \mathrm{i}\bar{\psi}_{\beta }\gamma ^{\alpha \beta \mu }%
\hat{N}_{\;\;\;\;\;\;\mu }^{\rho \lambda \sigma }\psi _{\sigma }\partial
_{\lbrack \rho }h_{\lambda ]\alpha }\right) +\partial _{\lambda }\bar{u}%
^{\lambda }  \notag \\
&&-\mathrm{i}\left( \partial _{\alpha }\bar{\psi}_{\beta }\right) \gamma
^{\alpha \beta \mu }\hat{N}_{\;\;\;\;\;\;\mu }^{\rho \lambda \sigma }\psi
_{\sigma }\partial _{\lbrack \rho }\eta _{\lambda ]}.  \label{ww27}
\end{eqnarray}%
Inserting (\ref{ww26})--(\ref{ww27}) in (\ref{ww22}) and taking into account
the result (\ref{ww25}), the equation (\ref{ww22}) reduces to%
\begin{eqnarray}
&&-\mathrm{i}\left( \partial _{\alpha }\bar{\psi}_{\beta }\right) \gamma
^{\alpha \beta \mu }\hat{N}_{\;\;\;\;\;\;\mu }^{\rho \lambda \sigma }\psi
_{\sigma }\partial _{\lbrack \rho }\eta _{\lambda ]}  \notag \\
&&-\frac{\mathrm{i}k_{4}}{8}\bar{\psi}_{\beta }\left( \sigma ^{\alpha \rho
}\gamma ^{\lambda \beta \sigma }-\sigma ^{\alpha \lambda }\gamma ^{\rho
\beta \sigma }\right) \left( \partial _{\alpha }\psi _{\sigma }\right)
\partial _{\lbrack \rho }\eta _{\lambda ]}  \notag \\
&=&\gamma \bar{m}_{1}+\partial _{\mu }\bar{n}_{1}^{\mu }.  \label{ww28}
\end{eqnarray}%
Now, we decompose $\gamma ^{\alpha \beta \mu }\hat{N}_{\;\;\;\;\;\;\mu
}^{\rho \lambda \sigma }$ like
\begin{equation}
\gamma ^{\alpha \beta \mu }\hat{N}_{\;\;\;\;\;\;\mu }^{\rho \lambda \sigma
}=\left( \gamma ^{\alpha \beta \mu }\hat{N}_{\;\;\;\;\;\;\mu }^{\rho \lambda
\sigma }\right) _{1}+\left( \gamma ^{\alpha \beta \mu }\hat{N}%
_{\;\;\;\;\;\;\mu }^{\rho \lambda \sigma }\right) _{2},  \label{ww29}
\end{equation}%
with
\begin{eqnarray}
\left( \gamma ^{\alpha \beta \mu }\hat{N}_{\;\;\;\;\;\;\mu }^{\rho \lambda
\sigma }\right) _{1} &=&\frac{1}{2}\left( \frac{1}{2}\bar{k}_{1}+\bar{k}%
_{2}-2\bar{k}_{3}-\frac{1}{2}\bar{k}_{4}\right) \times  \notag \\
&&\left( \sigma ^{\lambda \sigma }\gamma ^{\alpha \beta \rho }+\sigma
^{\lambda \beta }\gamma ^{\alpha \sigma \rho }-\sigma ^{\rho \sigma }\gamma
^{\alpha \beta \lambda }-\sigma ^{\rho \beta }\gamma ^{\alpha \sigma \lambda
}\right)  \notag \\
&&+\bar{k}_{3}\left( 2\sigma ^{\sigma \beta }\gamma ^{\rho \alpha \lambda
}-\sigma ^{\sigma \alpha }\gamma ^{\rho \beta \lambda }-\sigma ^{\beta
\alpha }\gamma ^{\rho \sigma \lambda }\right)  \notag \\
&&+\left( \bar{k}_{1}-\bar{k}_{2}+\bar{k}_{3}+\bar{k}_{4}\right) \left(
\sigma ^{\sigma \rho }\sigma ^{\lambda \beta }-\sigma ^{\sigma \lambda
}\sigma ^{\rho \beta }\right) \gamma ^{\alpha }  \notag \\
&&+\frac{1}{2}\left( \bar{k}_{1}+\bar{k}_{3}\right) \left( \left( \sigma
^{\sigma \lambda }\sigma ^{\alpha \rho }-\sigma ^{\sigma \rho }\sigma
^{\lambda \alpha }\right) \gamma ^{\beta }\right.  \notag \\
&&\left. +\left( \sigma ^{\beta \rho }\sigma ^{\alpha \lambda }-\sigma
^{\beta \lambda }\sigma ^{\alpha \rho }\right) \gamma ^{\sigma }\right) +%
\frac{1}{2}\left( \bar{k}_{3}+\bar{k}_{4}\right) \left( \left( \sigma
^{\beta \rho }\sigma ^{\alpha \sigma }\right. \right.  \notag \\
&&\left. \left. -\sigma ^{\sigma \rho }\sigma ^{\alpha \beta }\right) \gamma
^{\lambda }+\left( \sigma ^{\sigma \lambda }\sigma ^{\alpha \beta }-\sigma
^{\beta \lambda }\sigma ^{\alpha \sigma }\right) \gamma ^{\rho }\right) ,
\label{ww32}
\end{eqnarray}%
\begin{eqnarray}
\left( \gamma ^{\alpha \beta \mu }\hat{N}_{\;\;\;\;\;\;\mu }^{\rho \lambda
\sigma }\right) _{2} &=&\frac{1}{2}\left( \frac{1}{2}\bar{k}_{1}-\bar{k}_{2}-%
\frac{1}{2}\bar{k}_{4}\right) \times  \notag \\
&&\left( \sigma ^{\lambda \sigma }\gamma ^{\alpha \beta \rho }-\sigma
^{\lambda \beta }\gamma ^{\alpha \sigma \rho }-\sigma ^{\rho \sigma }\gamma
^{\alpha \beta \lambda }+\sigma ^{\rho \beta }\gamma ^{\alpha \sigma \lambda
}\right)  \notag \\
&&+\left( \bar{k}_{2}-\bar{k}_{3}\right) \left( \sigma ^{\rho \alpha }\gamma
^{\beta \sigma \lambda }-\sigma ^{\lambda \alpha }\gamma ^{\beta \sigma \rho
}\right) +\bar{k}_{3}\left( \sigma ^{\beta \alpha }\gamma ^{\rho \sigma
\lambda }\right.  \notag \\
&&\left. -\sigma ^{\sigma \alpha }\gamma ^{\rho \beta \lambda }\right) +%
\frac{1}{2}\left( \bar{k}_{1}-2\bar{k}_{2}+\bar{k}_{3}+2\bar{k}_{4}\right)
\left( \left( \sigma ^{\sigma \lambda }\sigma ^{\alpha \rho }\right. \right.
\notag \\
&&\left. \left. -\sigma ^{\sigma \rho }\sigma ^{\lambda \alpha }\right)
\gamma ^{\beta }+\left( \sigma ^{\beta \lambda }\sigma ^{\alpha \rho
}-\sigma ^{\beta \rho }\sigma ^{\alpha \lambda }\right) \gamma ^{\sigma
}\right)  \notag \\
&&+\frac{1}{2}\left( \bar{k}_{3}+\bar{k}_{4}\right) \left( 2\sigma ^{\beta
\sigma }\left( \sigma ^{\alpha \lambda }\gamma ^{\rho }-\sigma ^{\alpha \rho
}\gamma ^{\lambda }\right) +\left( \sigma ^{\beta \rho }\sigma ^{\alpha
\sigma }\right. \right.  \notag \\
&&\left. \left. +\sigma ^{\sigma \rho }\sigma ^{\alpha \beta }\right) \gamma
^{\lambda }-\left( \sigma ^{\alpha \sigma }\sigma ^{\beta \lambda }+\sigma
^{\alpha \beta }\sigma ^{\sigma \lambda }\right) \gamma ^{\rho }\right) .
\label{ww33}
\end{eqnarray}%
By direct computation it can be shown that the two components of $\gamma
^{\alpha \beta \mu }\hat{N}_{\;\;\;\;\;\;\mu }^{\rho \lambda \sigma }$
satisfy the properties
\begin{eqnarray}
\gamma ^{0}\left( \gamma ^{\alpha \beta \mu }\hat{N}_{\;\;\;\;\;\;\mu
}^{\rho \lambda \sigma }\right) _{1} &=&-\left( \gamma ^{0}\left( \gamma
^{\alpha \sigma \mu }\hat{N}_{\;\;\;\;\;\;\mu }^{\rho \lambda \beta }\right)
_{1}\right) ^{\intercal },  \label{ww30} \\
\gamma ^{0}\left( \gamma ^{\alpha \beta \mu }\hat{N}_{\;\;\;\;\;\;\mu
}^{\rho \lambda \sigma }\right) _{2} &=&\left( \gamma ^{0}\left( \gamma
^{\alpha \sigma \mu }\hat{N}_{\;\;\;\;\;\;\mu }^{\rho \lambda \beta }\right)
_{2}\right) ^{\intercal }.  \label{ww31}
\end{eqnarray}%
By means of the formulas (\ref{ww30})--(\ref{ww31}) we can write
\begin{eqnarray}
&&-\mathrm{i}\left( \partial _{\alpha }\bar{\psi}_{\beta }\right) \gamma
^{\alpha \beta \mu }\hat{N}_{\;\;\;\;\;\;\mu }^{\rho \lambda \sigma }\psi
_{\sigma }\partial _{\lbrack \rho }\eta _{\lambda ]}  \notag \\
&=&\gamma \left( \frac{\mathrm{i}}{2}\bar{\psi}_{\beta }\left( \gamma
^{\alpha \beta \mu }\hat{N}_{\;\;\;\;\;\;\mu }^{\rho \lambda \sigma }\right)
_{1}\psi _{\sigma }\partial _{\lbrack \rho }h_{\lambda ]\alpha }\right)
\notag \\
&&+\mathrm{i}\bar{\psi}_{\beta }\left( \gamma ^{\alpha \beta \mu }\hat{N}%
_{\;\;\;\;\;\;\mu }^{\rho \lambda \sigma }\right) _{2}\left( \partial
_{\alpha }\psi _{\sigma }\right) \partial _{\lbrack \rho }\eta _{\lambda ]}
\notag \\
&&+\partial _{\alpha }\left( -\frac{\mathrm{i}}{2}\bar{\psi}_{\beta }\left(
\gamma ^{\alpha \beta \mu }\hat{N}_{\;\;\;\;\;\;\mu }^{\rho \lambda \sigma
}\right) _{1}\psi _{\sigma }\partial _{\lbrack \rho }\eta _{\lambda
]}\right) ,  \label{ww34}
\end{eqnarray}%
such that
\begin{eqnarray}
&&-\mathrm{i}\left( \partial _{\alpha }\bar{\psi}_{\beta }\right) \gamma
^{\alpha \beta \mu }\hat{N}_{\;\;\;\;\;\;\mu }^{\rho \lambda \sigma }\psi
_{\sigma }\partial _{\lbrack \rho }\eta _{\lambda ]}  \notag \\
&&-\frac{\mathrm{i}k_{4}}{8}\bar{\psi}_{\beta }\left( \gamma ^{\lambda \beta
\sigma }\sigma ^{\alpha \rho }-\gamma ^{\rho \beta \sigma }\sigma ^{\alpha
\lambda }\right) \left( \partial _{\alpha }\psi _{\sigma }\right) \partial
_{\lbrack \rho }\eta _{\lambda ]}  \notag \\
&=&\gamma \left( \frac{\mathrm{i}}{2}\bar{\psi}_{\beta }\left( \gamma
^{\alpha \beta \mu }\hat{N}_{\;\;\;\;\;\;\mu }^{\rho \lambda \sigma }\right)
_{1}\psi _{\sigma }\partial _{\lbrack \rho }h_{\lambda ]\alpha }\right)
\notag \\
&&+\partial _{\alpha }\left( -\frac{\mathrm{i}}{2}\bar{\psi}_{\beta }\left(
\gamma ^{\alpha \beta \mu }\hat{N}_{\;\;\;\;\;\;\mu }^{\rho \lambda \sigma
}\right) _{1}\psi _{\sigma }\partial _{\lbrack \rho }\eta _{\lambda ]}\right)
\notag \\
&&-\mathrm{i}\bar{\psi}_{\beta }\left( \frac{k_{4}}{8}\left( \sigma ^{\alpha
\rho }\gamma ^{\lambda \beta \sigma }-\sigma ^{\alpha \lambda }\gamma ^{\rho
\beta \sigma }\right) \right.  \notag \\
&&\left. -\left( \gamma ^{\alpha \beta \mu }\hat{N}_{\;\;\;\;\;\;\mu }^{\rho
\lambda \sigma }\right) _{2}\right) \left( \partial _{\alpha }\psi _{\sigma
}\right) \partial _{\lbrack \rho }\eta _{\lambda ]}.  \label{ww35}
\end{eqnarray}%
Comparing (\ref{ww35}) with (\ref{ww28}) it results that the last term in (%
\ref{ww35}) has to be $\gamma $-exact modulo $d$. This holds if
\begin{equation}
\mathrm{i}\bar{\psi}_{\beta }\left( \left( \frac{k_{4}}{8}\left( \sigma
^{\alpha \rho }\gamma ^{\lambda \beta \sigma }-\sigma ^{\alpha \lambda
}\gamma ^{\rho \beta \sigma }\right) -\left( \gamma ^{\alpha \beta \mu }\hat{%
N}_{\;\;\;\;\;\;\mu }^{\rho \lambda \sigma }\right) _{2}\right) \left(
\partial _{\alpha }\psi _{\sigma }\right) \right) =\partial _{\alpha }\theta
^{\alpha }  \label{ww36}
\end{equation}%
for some $\theta ^{\alpha }$ or, in other words, if
\begin{equation}
M^{\alpha \beta \rho \lambda \sigma }=\gamma ^{0}\left( \frac{k_{4}}{8}%
\left( \sigma ^{\alpha \rho }\gamma ^{\lambda \beta \sigma }-\sigma ^{\alpha
\lambda }\gamma ^{\rho \beta \sigma }\right) -\left( \gamma ^{\alpha \beta
\mu }\hat{N}_{\;\;\;\;\;\;\mu }^{\rho \lambda \sigma }\right) _{2}\right)
\label{ww37}
\end{equation}%
fulfills the condition
\begin{equation}
M^{\alpha \beta \rho \lambda \sigma }=-\left( M^{\alpha \sigma \rho \lambda
\beta }\right) ^{\intercal }.  \label{ww38}
\end{equation}%
With the help of (\ref{ww31}) we obtain the relations
\begin{equation}
M^{\alpha \beta \rho \lambda \sigma }=\left( M^{\alpha \sigma \rho \lambda
\beta }\right) ^{\intercal },  \label{ww39}
\end{equation}%
which indicate that (\ref{ww38}) cannot be satisfied, and hence neither (\ref%
{ww36}). As a consequence, the term $-\mathrm{i}\bar{\psi}_{\beta }M^{\alpha
\beta \rho \lambda \sigma }\left( \partial _{\alpha }\psi _{\sigma }\right)
\partial _{\lbrack \rho }\eta _{\lambda ]}$ from (\ref{ww35}) must be
canceled, which implies
\begin{equation}
M^{\alpha \beta \rho \lambda \sigma }=0.  \label{ww40}
\end{equation}%
The solution to the above equation reads as
\begin{equation}
\bar{k}_{1}=\frac{1}{4}k_{4},\quad \bar{k}_{2}=\frac{1}{8}k_{4},\quad \bar{k}%
_{3}=0,\quad \bar{k}_{4}=0.  \label{ww41}
\end{equation}%
Redenoting $k_{4}$ by $k$, we finally find the relations
\begin{equation}
\bar{N}_{\;\;\;\;\;\;\mu }^{\rho \lambda \sigma }=k\sigma ^{\rho \lambda
}\delta _{\mu }^{\sigma },\;N_{\;\;\;\;\;\;\mu }^{\rho \lambda \sigma }=\hat{%
N}_{\;\;\;\;\;\;\mu }^{\rho \lambda \sigma }=\frac{1}{4}k\left( \sigma
^{\lambda \sigma }\delta _{\mu }^{\rho }-\sigma ^{\rho \sigma }\delta _{\mu
}^{\lambda }+\frac{1}{2}\delta _{\mu }^{\sigma }\gamma ^{\rho \lambda
}\right) .  \label{ww42}
\end{equation}%
Replacing (\ref{ww9}) and (\ref{ww42}) in (\ref{PFRS3.23}), we get that
\begin{eqnarray}
a_{1}^{(\mathrm{int})} &=&k\psi ^{\ast \mu }\left( \partial ^{\nu }\psi
_{\mu }\right) \eta _{\nu }+\frac{k}{2}\psi ^{\ast \mu }\psi ^{\nu }\partial
_{\lbrack \mu }\eta _{\nu ]}  \notag \\
&&+\frac{k}{8}\psi ^{\ast \rho }\gamma ^{\mu \nu }\psi _{\rho }\partial
_{\lbrack \mu }\eta _{\nu ]}.  \label{PFRSa1}
\end{eqnarray}%
Meanwhile, if we insert (\ref{ww42}) in (\ref{ww2}), (\ref{ww16}), (\ref%
{ww26})--(\ref{ww27}), and (\ref{ww35}) and the resulting expressions in (%
\ref{PFRS3.22a}), we deduce that the component of antighost number
zero from the first-order deformation is given by
\begin{eqnarray}
&&a_{0}^{(\mathrm{int})}=\frac{k}{2}\left( \sigma ^{\rho \lambda }\mathcal{L}%
_{0}^{(\mathrm{RS})}-\frac{\mathrm{i}}{2}\bar{\psi}_{\mu }\gamma ^{\mu \nu
\rho }\partial ^{\lambda }\psi _{\nu }\right) h_{\rho \lambda }  \notag \\
&&+\frac{\mathrm{i}k}{4}\left( \frac{1}{2}\bar{\psi}^{\mu }\gamma ^{\rho
}\psi ^{\nu }+\sigma ^{\mu \rho }\bar{\psi}^{\nu }\gamma ^{\sigma }\psi
_{\sigma }+\bar{\psi}_{\sigma }\gamma ^{\sigma \rho \mu }\psi ^{\nu }\right)
\partial _{\lbrack \mu }h_{\nu ]\rho }+\bar{a}_{0}^{(\mathrm{int})},
\label{PFRSa0}
\end{eqnarray}%
where $\bar{a}_{0}^{\left( \mathrm{int}\right) }$ represents the general,
local solution to the homogeneous equation
\begin{equation}
\gamma \bar{a}_{0}^{\left( \mathrm{int}\right) }=\partial _{\mu }\bar{m}%
^{\left( \mathrm{int}\right) \mu },  \label{PFRSom}
\end{equation}%
with some local $\bar{m}^{\left( \mathrm{int}\right) \mu }$.

Such solutions correspond to $\bar{a}_{1}^{\left( \mathrm{int}\right) }=0$
and thus they cannot deform either the gauge algebra or the gauge
transformations, but simply the Lagrangian at order one in the coupling
constant. There are two main types of solutions to (\ref{PFRSom}). The first
one corresponds to $\bar{m}^{\left( \mathrm{int}\right) \mu }=0$ and is
given by gauge-invariant, non-integrated densities constructed from the
original fields and their spacetime derivatives. According to (\ref{PFRS3.10}%
) for both pure ghost and antighost numbers equal to zero, they are
given by $\bar{a}_{0}^{\prime \left( \mathrm{int}\right)
}=\bar{a}_{0}^{\prime \left( \mathrm{int}\right) }\left( \left[ \psi
_{\mu }\right] ,\left[ K_{\mu \nu \alpha \beta }\right] \right) $,
up to the conditions that they effectively describe cross-couplings
between the two types of fields and cannot be written in a
divergence-like form. Unfortunately, this type of solutions must
depend on the linearized Riemann tensor (and possibly of its
derivatives) in order to provide cross-couplings, and thus would
lead to terms with at least two derivatives of the Rarita-Schwinger
spinors in the deformed field equations. So, by virtue of the
derivative order assumption, they must be discarded by setting
$\bar{a}_{0}^{\prime \left( \mathrm{int}\right) }=0$. The second
kind of solutions is associated with $\bar{m}^{\left(
\mathrm{int}\right) \mu }\neq 0 $ in (\ref{PFRSom}) and will be
approached below.

We split the solution to the equation (\ref{PFRSom}) for $\bar{m}^{\left(
\mathrm{int}\right) \mu }\neq 0$ along the number of derivatives present in
the interaction vertices
\begin{equation}
\bar{a}_{0}^{\left( \mathrm{int}\right) }=\sum\limits_{i=0}^{2}\overset{(i)}{%
\omega },  \label{desca0}
\end{equation}%
where $\overset{(i)}{\omega }$ contains $i$ derivatives of the fields. The
decomposition (\ref{desca0}) yields a similar splitting with respect to the
equation (\ref{PFRSom}), which becomes equivalent to three independent
equations
\begin{equation}
\gamma \overset{(i)}{\omega }=\partial ^{\mu }\overset{(i)}{m}_{\mu },\quad
i=\overline{0,2}.  \label{ecech}
\end{equation}

Let us solve (\ref{ecech}) for $i=0$. With the help of the definitions of $%
\gamma $ acting on the generators from the BRST complex we get
\begin{equation}
\gamma \overset{(0)}{\omega }=-2\left( \partial _{\nu }\frac{\partial
\overset{(0)}{\omega }}{\partial h_{\mu \nu }}\right) \eta _{\mu }+\partial
_{\mu }\pi ^{\mu }.  \label{om1}
\end{equation}
Thus, $\overset{(0)}{\omega }$ is solution to (\ref{ecech}) for $i=0$ if and
only if
\begin{equation}
\partial _{\nu }\frac{\partial \overset{(0)}{\omega }}{\partial h_{\mu \nu }}%
=0.  \label{om1a}
\end{equation}
Since $\overset{(0)}{\omega }$ has no derivatives, the equation (\ref{om1a})
implies that $\partial \overset{(0)}{\omega }/\partial h_{\rho \mu }$ must
be constant. As the only constant and symmetric tensor in four spacetime
dimensions is the flat metric, we can write
\begin{equation}
\frac{\partial \overset{(0)}{\omega }}{\partial h_{\mu \nu }}=p\sigma ^{\mu
\nu },  \label{om1b}
\end{equation}
with $p$ a real constant. Integrating (\ref{om1b}), it results that the
solution to the equation (\ref{ecech}) for $i=0$ reads as
\begin{equation*}
\overset{(0)}{\omega }=ph+F\left( \psi _{\mu }\right) ,
\end{equation*}
but since it provides no cross-interactions, we can take
\begin{equation}
\overset{(0)}{\omega }=0.  \label{om1c}
\end{equation}

Next, we pass to the equation (\ref{ecech}) for $i=1$. We obtain
that
\begin{equation}
\gamma \overset{(1)}{\omega }=-2\left( \partial _{\nu }\frac{\delta
\overset{(1)}{\omega }}{\delta h_{\mu \nu }}\right) \eta _{\mu
}+\partial _{\mu }\beta ^{\mu },  \label{om2}
\end{equation}%
so $\overset{(1)}{\omega }$ checks (104) for $i=1$ if and only if
\begin{equation}
\partial _{\nu }\frac{\delta \overset{(1)}{\omega }}{\delta h_{\mu \nu }}%
=0.  \label{om2a}
\end{equation}%
Because $\overset{(1)}{\omega }$ includes just one spacetime derivative, the
solution to (\ref{om2a}) is
\begin{equation}
\frac{\delta \overset{(1)}{\omega }}{\delta h_{\mu \nu }}=\partial _{\rho
}D^{\rho \mu \nu },  \label{om2b}
\end{equation}%
where $D^{\rho \mu \nu }$ depends only on the undifferentiated fields and is
antisymmetric in its first two indices
\begin{equation}
D^{\rho \mu \nu }=-D^{\mu \rho \nu }.  \label{om2c}
\end{equation}%
Since $D^{\rho \mu \nu }$ is derivative-free and $h_{\mu \nu }$ is
symmetric, (\ref{om2b}) implies that $D^{\rho \mu \nu }$ must be symmetric
in its last two indices
\begin{equation}
D^{\rho \mu \nu }=D^{\rho \nu \mu }.  \label{om2d}
\end{equation}%
The properties (\ref{om2c}) and (\ref{om2d}) further lead to
\begin{eqnarray}
D^{\rho \mu \nu } &=&-D^{\mu \rho \nu }=-D^{\mu \nu \rho }=D^{\nu \mu \rho }
\notag \\
&=&D^{\nu \rho \mu }=-D^{\rho \nu \mu }=-D^{\rho \mu \nu },  \label{om2e}
\end{eqnarray}%
so $D^{\rho \mu \nu }=0$. Consequently, (\ref{om2b}) reduces to
\begin{equation}
\frac{\delta \overset{(1)}{\omega }}{\delta h_{\mu \nu }}=0,  \label{om2f}
\end{equation}%
whose solution is expressed by
\begin{equation}
\overset{(1)}{\omega }=L\left( \left[ \psi _{\mu }\right] \right) +\partial
_{\mu }G^{\mu }\left( \psi _{\mu },h_{\alpha \beta }\right)  \label{om2g}
\end{equation}%
and is not suitable as the first term provides no cross-interactions, while
the second is trivial, so we have that
\begin{equation}
\overset{(1)}{\omega }=0.  \label{om2h}
\end{equation}

In the end, we solve (\ref{ecech}) for $i=2$. From the relation
\begin{equation}
\gamma \overset{(2)}{\omega }=-2\left( \partial _{\nu }\frac{\delta \overset{%
(2)}{\omega }}{\delta h_{\mu \nu }}\right) \eta _{\mu }+\partial _{\mu }\xi
^{\mu },  \label{om3}
\end{equation}%
we observe that $\overset{(2)}{\omega }$ verifies (\ref{ecech}) for $i=2$ if
and only if
\begin{equation}
\partial _{\nu }\frac{\delta \overset{(2)}{\omega }}{\delta h_{\mu \nu }}=0.
\label{om3a}
\end{equation}%
The solution to the last equation reads as
\begin{equation}
\frac{\delta \overset{(2)}{\omega }}{\delta h_{\mu \nu }}=\partial _{\alpha
}\partial _{\beta }U^{\mu \alpha \nu \beta },  \label{ww73}
\end{equation}%
where $U^{\mu \alpha \nu \beta }$ displays the symmetry properties of the
Riemann tensor and involves only the undifferentiated fields $\psi _{\mu }$
and $h_{\mu \nu }$. At this stage it is useful to introduce a derivation in
the algebra of the fields $h_{\mu \nu }$ and of their derivatives that
counts the powers of the fields and their derivatives, defined by
\begin{equation}
N=\sum\limits_{k\geq 0}\left( \partial _{\mu _{1}\cdots \mu _{k}}h_{\mu \nu
}\right) \frac{\partial }{\partial \left( \partial _{\mu _{1}\cdots \mu
_{k}}h_{\mu \nu }\right) }.  \label{ww74}
\end{equation}%
Then, it is easy to see that for every nonintegrated density $\chi $, we
have that
\begin{equation}
N\chi =h_{\mu \nu }\frac{\delta \chi }{\delta h_{\mu \nu }}+\partial _{\mu
}s^{\mu }.  \label{ww75}
\end{equation}%
If $\chi ^{\left( l\right) }$ is a homogeneous polynomial of order $l>0$ in
the fields and their derivatives, then $N\chi ^{\left( l\right) }=l\chi
^{\left( l\right) }$. Using (\ref{ww73}), and (\ref{ww75}), we find that
\begin{equation}
N\overset{(2)}{\omega }=-\frac{1}{2}K_{\mu \alpha \nu \beta }U^{\mu
\alpha \nu \beta }+\partial _{\mu }v^{\mu }. \label{ww76a}
\end{equation}%
We expand $\overset{(2)}{\omega }$ like
\begin{equation}
\overset{(2)}{\omega }=\sum\limits_{l>0}\overset{(2)}{\omega }^{\left(
l\right) },  \label{ww77}
\end{equation}%
where $N\overset{(2)}{\omega }^{\left( l\right) }=l\overset{(2)}{\omega }%
^{\left( l\right) }$, such that
\begin{equation}
N\overset{(2)}{\omega }=\sum\limits_{l>0}l\overset{(2)}{\omega }^{\left(
l\right) }.  \label{ww78}
\end{equation}%
Comparing (\ref{ww76a}) with (\ref{ww78}), we reach the conclusion that the
decomposition (\ref{ww77}) induces a similar decomposition with respect to $%
U^{\mu \alpha \nu \beta }$, i.e.
\begin{equation}
U^{\mu \alpha \nu \beta }=\sum\limits_{l>0}U_{\left( l-1\right) }^{\mu
\alpha \nu \beta }.  \label{ww79}
\end{equation}%
Substituting (\ref{ww79}) into (\ref{ww76a}) and comparing the resulting
expression with (\ref{ww78}), we obtain that
\begin{equation}
\overset{(2)}{\omega }^{\left( l\right) }=-\frac{1}{2l}K_{\mu \alpha
\nu \beta }U_{\left( l-1\right) }^{\mu \alpha \nu \beta }+\partial
_{\mu }\bar{v}_{(l)}^{\mu }.  \label{prform}
\end{equation}%
Introducing (\ref{prform}) in (\ref{ww77}), we arrive at
\begin{equation}
\overset{(2)}{\omega }=-\frac{1}{2}K_{\mu \alpha \nu \beta }\bar{U}%
^{\mu \alpha \nu \beta }+\partial _{\mu }\bar{v}^{\mu },
\label{ww81}
\end{equation}%
where
\begin{equation}
\bar{U}^{\mu \alpha \nu \beta }=\sum\limits_{l>0}\frac{1}{l}U_{\left(
l-1\right) }^{\mu \alpha \nu \beta }.  \label{ww82}
\end{equation}%
Even if consistent, an $\overset{(2)}{\omega }$ of the type
(\ref{ww81}) would produce field equations with two spacetime
derivatives acting on the Rarita-Schwinger spinors, which breaks the
hypothesis on the derivative order of the interacting theory, so we
must take
\begin{equation}
\overset{(2)}{\omega }=0.  \label{om3h}
\end{equation}%
The results (\ref{om1c}), (\ref{om2h}), and (\ref{om3h}) enable us to take,
without loss of generality
\begin{equation}
\bar{a}_{0}^{\left( \mathrm{int}\right) }=0  \label{om5}
\end{equation}%
in (\ref{PFRSa0}).

Finally, we analyze the component $a^{\left( \mathrm{RS}\right) }$ from (\ref%
{PFRS3.12a}). As the massive Rarita-Schwinger action from (\ref{fract}) has
no non-trivial gauge invariance, it follows that $a^{\left( \mathrm{RS}%
\right) }$ can only reduce to its component of antighost number zero
\begin{equation}
a^{\left( \mathrm{RS}\right) }=a_{0}^{\left( \mathrm{RS}\right) }\left( %
\left[ \psi _{\mu }\right] \right) ,  \label{om6}
\end{equation}%
which is automatically solution to the equation $sa^{\left( \mathrm{RS}%
\right) }\equiv \gamma a_{0}^{\left( \mathrm{RS}\right) }=0$. It comes from $%
a_{1}^{\left( \mathrm{RS}\right) }=0$ and does not deform the gauge
transformations (\ref{PFRS5}), but merely modifies the massive spin-$3/2$
action. The condition that $a_{0}^{\left( \mathrm{RS}\right) }$ is of
maximum derivative order equal to one is translated into
\begin{equation}
a_{0}^{\left( \mathrm{RS}\right) }=V\left( \psi _{\mu }\right) +V^{\alpha
\beta }\left( \psi _{\mu }\right) \partial _{\alpha }\psi _{\beta },
\label{om7}
\end{equation}%
where $V$ and $V^{\alpha \beta }$ are polynomials in the undifferentiated
spinor fields (since they anticommute). The first polynomial is a scalar
(bosonic and real), while the tensor $V^{\alpha \beta }$ is fermionic and
anti-Majorana spinor-like.

The general conclusion of this subsection is that the first-order
deformation associated with the Pauli-Field theory plus the massive
Rarita-Schwinger field can be written like
\begin{equation}
S_{1}=S_{1}^{\left( \mathrm{PF}\right) }+S_{1}^{\left( \mathrm{int}\right) },
\label{om8}
\end{equation}
with
\begin{equation}
S_{1}^{\left( \mathrm{PF}\right) }=\int d^{4}x\left( a_{0}^{\left( \mathrm{PF%
}\right) }+a_{1}^{\left( \mathrm{PF}\right) }+a_{2}^{\left( \mathrm{PF}%
\right) }\right) ,  \label{om9}
\end{equation}
and
\begin{equation}
S_{1}^{\left( \mathrm{int}\right) }=\int d^{4}x\left( a_{0}^{\left( \mathrm{%
int}\right) }+a_{1}^{\left( \mathrm{int}\right) }+a_{0}^{\left( \mathrm{RS}%
\right) }\right) .  \label{om10}
\end{equation}
The first two components of (\ref{om10}) are expressed by (\ref{PFRSa1}) and (%
\ref{PFRSa0}) with $\bar{a}_{0}^{(\mathrm{int})}=0$, while $a_{0}^{\left(
\mathrm{RS}\right) }$ is given by (\ref{om7}). This is the most general form
that complies with all the hypotheses that must be satisfied by the
deformations, including that related to the derivative order of the deformed
Lagrangian.

\subsection{Second-order deformation\label{secorddef}}

In this subsection we are interested in determining the complete expression
of the second-order deformation for the solution to the master equation,
which is known to be subject to the equation (\ref{PFRS2.6}). Proceeding in
the same manner like during the first-order deformation procedure, we can
write the second-order deformation of the solution to the master equation
like the sum between the Pauli-Fierz and the interacting parts
\begin{equation}
S_{2}=S_{2}^{\left( \mathrm{PF}\right) }+S_{2}^{\left( \mathrm{int}\right) }.
\label{sec1}
\end{equation}%
The piece $S_{2}^{\left( \mathrm{PF}\right) }$ describes the second-order
deformation in the Pauli-Fierz sector and we will not insist on it since we
are merely interested in the cross-couplings. The term $S_{2}^{\left(
\mathrm{int}\right) }$ results as solution to the equation
\begin{equation}
\frac{1}{2}\left( S_{1},S_{1}\right) ^{\left( \mathrm{int}\right)
}+sS_{2}^{\left( \mathrm{int}\right) }=0,  \label{sec2}
\end{equation}%
where
\begin{equation}
\left( S_{1},S_{1}\right) ^{\left( \mathrm{int}\right) }=\left(
S_{1}^{\left( \mathrm{int}\right) },S_{1}^{\left( \mathrm{int}\right)
}\right) +2\left( S_{1}^{\left( \mathrm{PF}\right) },S_{1}^{\left( \mathrm{%
int}\right) }\right)  \label{sec3}
\end{equation}%
and $S_{1}^{\left( \mathrm{int}\right) }$ is presented in (\ref{om10}). If
we denote by $\Delta ^{\left( \mathrm{int}\right) }$ and $b^{\left( \mathrm{%
int}\right) }$ the non-integrated densities of $\left( S_{1},S_{1}\right)
^{\left( \mathrm{int}\right) }$ and respectively of $S_{2}^{\left( \mathrm{%
int}\right) }$, the local form of (\ref{sec2}) becomes
\begin{equation}
\Delta ^{\left( \mathrm{int}\right) }=-2sb^{\left( \mathrm{int}\right)
}+\partial _{\mu }n^{\mu },  \label{sec4}
\end{equation}%
with
\begin{equation}
\mathrm{gh}\left( \Delta ^{\left( \mathrm{int}\right) }\right) =1,\quad
\mathrm{gh}\left( b^{\left( \mathrm{int}\right) }\right) =0,\quad \mathrm{gh}%
\left( n^{\mu }\right) =1,  \label{sec5}
\end{equation}%
for some local current $n^{\mu }$. Direct computation shows that
$\Delta ^{\left( \mathrm{int}\right) }$ decomposes like
\begin{equation}
\Delta ^{\left( \mathrm{int}\right) }=\Delta _{0}^{\left( \mathrm{int}%
\right) }+\Delta _{1}^{\left( \mathrm{int}\right) },\quad \mathrm{agh}\left(
\Delta _{I}^{\left( \mathrm{int}\right) }\right) =I,\quad I=0,1,
\label{sec6}
\end{equation}%
with
\begin{eqnarray}
&&\Delta _{1}^{\left( \mathrm{int}\right) }=\gamma \left( k\left( -\frac{1}{4%
}\left( \psi ^{\ast \lbrack \mu }\psi ^{\sigma ]}+\frac{1}{2}\psi ^{\ast
\rho }\gamma ^{\mu \sigma }\psi _{\rho }\right) \partial _{\lbrack \sigma
}\eta _{\lambda ]}\sigma ^{\nu \lambda }\right. \right.  \notag \\
&&\left. +k\psi ^{\ast \sigma }\left( \partial ^{\mu }\psi _{\sigma }\right)
\eta ^{\nu }\right) h_{\mu \nu }  \notag \\
&&\left. +\frac{k\left( 2-k\right) }{2}\left( \psi ^{\ast \mu }\psi ^{\nu }+%
\frac{1}{4}\psi ^{\ast \sigma }\gamma ^{\mu \nu }\psi _{\sigma }\right) \eta
^{\rho }\partial _{\lbrack \mu }h_{\nu ]\rho }\right)  \notag \\
&&+k\left( 1-k\right) \left( \psi ^{\ast \mu }\left( \partial ^{\nu }\psi
_{\mu }\right) \eta ^{\rho }\partial _{\lbrack \nu }\eta _{\rho ]}+\frac{1}{4%
}\left( \psi ^{\ast \lbrack \mu }\psi ^{\nu ]}\right. \right.  \notag \\
&&\left. \left. +\frac{1}{2}\psi ^{\ast \sigma }\gamma ^{\mu \nu }\psi
_{\sigma }\right) \partial _{\lbrack \mu }\eta _{\rho ]}\partial _{\lbrack
\nu }\eta _{\lambda ]}\sigma ^{\rho \lambda }\right) ,  \label{sec7}
\end{eqnarray}%
and
\begin{eqnarray}
&&\Delta _{0}^{\left( \mathrm{int}\right) }=\gamma \left( \frac{k}{4}%
\mathcal{L}_{0}^{(\mathrm{RS})}h_{\mu \nu }h^{\mu \nu }\right)  \notag \\
&&+k\left( -\mathcal{L}_{0}^{(\mathrm{RS})}\eta ^{\mu }+\frac{\mathrm{i}k}{2}%
\eta _{\sigma }\partial ^{\sigma }\left( \bar{\psi}^{\mu }\gamma ^{\rho
}\psi _{\rho }\right) +\frac{\mathrm{i}k}{4}\bar{\psi}^{\mu }\gamma ^{\rho
}\psi ^{\sigma }\partial _{\lbrack \rho }\eta _{\sigma ]}\right.  \notag \\
&&\left. +\frac{\mathrm{i}k}{4}\bar{\psi}_{\sigma }\gamma ^{\rho }\psi
_{\rho }\partial ^{\lbrack \mu }\eta ^{\sigma ]}+\frac{\mathrm{i}k}{16}\bar{%
\psi}^{\mu }\left[ \gamma ^{\rho },\gamma ^{\alpha \beta }\right] \psi
_{\rho }\partial _{\lbrack \alpha }\eta _{\beta ]}\right) \left( \partial
^{\nu }h_{\mu \nu }-\partial _{\mu }h\right)  \notag \\
&&+\frac{\mathrm{i}k^{2}}{4}\left( \eta _{\sigma }\partial ^{\sigma }\left(
\bar{\psi}^{\mu }\gamma ^{\alpha }\psi ^{\nu }-2\bar{\psi}_{\beta }\gamma
^{\alpha \beta \mu }\psi ^{\nu }\right) +\bar{\psi}^{\mu }\gamma ^{\alpha
}\psi _{\sigma }\partial ^{\lbrack \nu }\eta ^{\sigma ]}\right.  \notag \\
&&-\bar{\psi}_{\beta }\gamma ^{\alpha \beta \mu }\psi _{\sigma }\partial
^{\lbrack \nu }\eta ^{\sigma ]}-\bar{\psi}^{\sigma }\gamma ^{\alpha \beta
\mu }\psi ^{\nu }\partial _{\lbrack \beta }\eta _{\sigma ]}+\left( \frac{1}{8%
}\bar{\psi}^{\mu }\left[ \gamma ^{\alpha },\gamma ^{\rho \lambda }\right]
\psi ^{\nu }\right.  \notag \\
&&\left. \left. -\frac{1}{4}\bar{\psi}_{\beta }\left[ \gamma ^{\alpha \beta
\mu },\gamma ^{\rho \lambda }\right] \psi ^{\nu }\right) \partial _{\lbrack
\rho }\eta _{\lambda ]}\right) \partial _{\lbrack \mu }h_{\nu ]\alpha }
\notag \\
&&+k^{2}\left( \eta _{\sigma }\partial ^{\sigma }\mathcal{L}_{0}^{(\mathrm{RS%
})}-\frac{\mathrm{i}}{2}\bar{\psi}_{\mu }\gamma ^{\mu \nu \rho }\left(
\partial ^{\sigma }\psi _{\rho }\right) \partial _{\nu }\eta _{\sigma }+%
\frac{m}{2}\bar{\psi}_{\mu }\gamma ^{\mu \nu }\psi ^{\sigma }\partial
_{\lbrack \nu }\eta _{\sigma ]}\right.  \notag \\
&&-\frac{\mathrm{i}}{2}\bar{\psi}_{\mu }\gamma ^{\mu \nu \rho }\partial
_{\nu }\left( \psi ^{\sigma }\partial _{\lbrack \rho }\eta _{\sigma
]}\right) -\frac{\mathrm{i}}{2}\bar{\psi}^{\sigma }\gamma ^{\mu \nu \rho
}\left( \partial _{\nu }\psi _{\rho }\right) \partial _{\lbrack \mu }\eta
_{\sigma ]}  \notag \\
&&+\frac{m}{16}\left( \bar{\psi}_{\mu }\left[ \gamma ^{\mu \nu },\gamma
^{\alpha \beta }\right] \psi _{\nu }-\mathrm{i}\bar{\psi}_{\mu }\left[
\gamma ^{\mu \nu \rho },\gamma ^{\alpha \beta }\right] \partial _{\nu }\psi
_{\rho }\right) \partial _{\lbrack \alpha }\eta _{\beta ]}  \notag \\
&&\left. -\frac{\mathrm{i}}{16}\bar{\psi}_{\mu }\gamma ^{\mu \nu \rho
}\gamma ^{\alpha \beta }\psi _{\rho }\partial _{\nu }\left( \partial
_{\lbrack \alpha }\eta _{\beta ]}\right) \right) h-\frac{\mathrm{i}k^{2}}{2}%
\left( \eta _{\sigma }\partial ^{\sigma }\left( \bar{\psi}_{\mu }\gamma
^{\mu \nu \rho }\partial ^{\lambda }\psi _{\nu }\right) \right.  \notag \\
&&+\bar{\psi}_{\mu }\gamma ^{\mu \nu \rho }\left( \partial ^{\sigma }\psi
_{\nu }\right) \partial ^{\lambda }\eta _{\sigma }+\frac{1}{2}\bar{\psi}%
_{\mu }\gamma ^{\mu \nu \rho }\left( \partial ^{\lambda }\psi ^{\sigma
}\partial _{\lbrack \nu }\eta _{\sigma ]}\right)  \notag \\
&&+\frac{1}{2}\bar{\psi}^{\sigma }\gamma ^{\mu \nu \rho }\left( \partial
^{\lambda }\psi _{\nu }\right) \partial _{\lbrack \mu }\eta _{\sigma ]}+%
\frac{1}{8}\bar{\psi}_{\mu }\left[ \gamma ^{\mu \nu \rho },\gamma ^{\alpha
\beta }\right] \left( \partial ^{\lambda }\psi _{\nu }\right) \partial
_{\lbrack \alpha }\eta _{\beta ]}  \notag \\
&&\left. +\frac{1}{8}\bar{\psi}_{\mu }\gamma ^{\mu \nu \rho }\gamma ^{\alpha
\beta }\psi _{\nu }\partial ^{\lambda }\left( \partial _{\lbrack \alpha
}\eta _{\beta ]}\right) \right) h_{\rho \lambda }-\frac{\mathrm{i}k}{4}\bar{%
\psi}_{\mu }\gamma ^{\mu \nu (\rho }\partial ^{\lambda )}\psi _{\nu }\times
\notag \\
&&\times \left( h_{\lambda \sigma }\partial _{\rho }\eta ^{\sigma }-\eta
^{\sigma }\left( \partial _{\rho }h_{\lambda \sigma }-\partial _{\sigma
}h_{\rho \lambda }\right) \right) +\frac{\mathrm{i}k}{4}\bar{\psi}^{\mu
}\gamma ^{(\rho }\psi ^{\lambda )}\times  \notag \\
&&\times \partial _{\mu }\left( h_{\lambda \sigma }\partial _{\rho }\eta
^{\sigma }-\eta ^{\sigma }\left( \partial _{\rho }h_{\lambda \sigma
}-\partial _{\sigma }h_{\rho \lambda }\right) \right) +\frac{\mathrm{i}k}{4}%
\bar{\psi}^{\mu }\gamma ^{\rho }\psi _{\rho }\times  \notag \\
&&\times \partial ^{\nu }\left( h_{\sigma (\mu }\partial _{\nu )}\eta
^{\sigma }-\eta ^{\sigma }\left( \partial _{(\mu }h_{\nu )\sigma }-2\partial
_{\sigma }h_{\mu \nu }\right) \right) -\frac{\mathrm{i}k}{2}\bar{\psi}^{\mu
}\gamma ^{\rho }\psi _{\rho }\times  \notag \\
&&\times \partial _{\mu }\left( h^{\alpha \beta }\partial _{\alpha }\eta
_{\beta }-\eta ^{\alpha }\left( \partial ^{\beta }h_{\alpha \beta }-\partial
_{\alpha }h\right) \right) -\frac{\mathrm{i}k}{4}\bar{\psi}_{\mu }\gamma
^{\mu \nu (\rho }\psi ^{\lambda )}\times  \notag \\
&&\times \partial _{\nu }\left( h_{\lambda \sigma }\partial _{\rho }\eta
^{\sigma }-\eta ^{\sigma }\left( \partial _{\rho }h_{\lambda \sigma
}-\partial _{\sigma }h_{\rho \lambda }\right) \right)  \notag \\
&&+2k\left( \partial ^{\mu }V+V^{\alpha \beta }\partial _{\alpha }\psi
_{\beta }\right) \eta _{\mu }+2kV^{\mu \nu }\left( \partial ^{\sigma }\psi
_{\nu }\right) \partial _{\mu }\eta _{\sigma }+k\frac{\partial ^{R}V}{%
\partial \psi _{\mu }}\psi ^{\nu }\partial _{\lbrack \mu }\eta _{\nu ]}
\notag \\
&&+kV^{\mu \nu }\partial _{\mu }\left( \psi ^{\sigma }\partial _{\lbrack \nu
}\eta _{\sigma ]}\right) +k\bar{\psi}^{\sigma }\frac{\partial ^{L}V^{\mu \nu
}}{\partial \bar{\psi}_{\rho }}\left( \partial _{\mu }\psi _{\nu }\right)
\partial _{\lbrack \rho }\eta _{\sigma ]}  \notag \\
&&+\frac{k}{4}\left( \frac{\partial ^{R}V}{\partial \psi _{\rho }}\gamma
^{\alpha \beta }\psi ^{\rho }-\bar{\psi}_{\rho }\gamma ^{\alpha \beta }\frac{%
\partial ^{L}V^{\mu \nu }}{\partial \bar{\psi}_{\rho }}\partial _{\mu }\psi
_{\nu }\right) \partial _{\lbrack \alpha }\eta _{\beta ]}+  \notag \\
&&+\frac{k}{4}V^{\mu \nu }\gamma ^{\alpha \beta }\partial _{\mu }\left( \psi
_{\nu }\partial _{\lbrack \alpha }\eta _{\beta ]}\right) .  \label{sec8}
\end{eqnarray}

Since the first-order deformation in the interacting sector starts in
antighost number one, we can take, without loss of generality, the
corresponding second-order deformation to start in antighost number two
\begin{eqnarray}
b^{\left( \mathrm{int}\right) } &=&b_{0}^{\left( \mathrm{int}\right)
}+b_{1}^{\left( \mathrm{int}\right) }+b_{2}^{\left( \mathrm{int}\right)
},\quad \mathrm{agh}\left( b_{I}^{\left( \mathrm{int}\right) }\right)
=I,\quad I=0,1,2,  \label{sec9} \\
n^{\mu } &=&n_{0}^{\mu }+n_{1}^{\mu }+n_{2}^{\mu },\quad \mathrm{agh}\left(
n_{I}^{\mu }\right) =I,\quad I=0,1,2.  \label{sec10}
\end{eqnarray}%
By projecting the equation (\ref{sec4}) on various antighost numbers, we
obtain
\begin{eqnarray}
\gamma b_{2}^{\left( \mathrm{int}\right) } &=&\partial _{\mu }\left( \frac{1%
}{2}n_{2}^{\mu }\right) ,  \label{sec11} \\
\Delta _{1}^{\left( \mathrm{int}\right) } &=&-2\left( \delta b_{2}^{\left(
\mathrm{int}\right) }+\gamma b_{1}^{\left( \mathrm{int}\right) }\right)
+\partial _{\mu }n_{1}^{\mu },  \label{sec12} \\
\Delta _{0}^{\left( \mathrm{int}\right) } &=&-2\left( \delta b_{1}^{\left(
\mathrm{int}\right) }+\gamma b_{0}^{\left( \mathrm{int}\right) }\right)
+\partial _{\mu }n_{0}^{\mu }.  \label{sec13}
\end{eqnarray}%
The equation (\ref{sec11}) can always be replaced, by adding trivial terms,
with
\begin{equation}
\gamma b_{2}^{\left( \mathrm{int}\right) }=0.  \label{sec14}
\end{equation}%
Looking at $\Delta _{1}^{\left( \mathrm{int}\right) }$ given in (\ref{sec7}%
), it results that it can be written like in (\ref{sec12}) if
\begin{eqnarray}
\chi  &=&k\left( 1-k\right) \left( \psi ^{\ast \mu }\left( \partial ^{\nu
}\psi _{\mu }\right) \eta ^{\rho }\partial _{\lbrack \nu }\eta _{\rho ]}+%
\frac{1}{4}\left( \psi ^{\ast \lbrack \mu }\psi ^{\nu ]}\right. \right.
\notag \\
&&\left. \left. +\frac{1}{2}\psi ^{\ast \sigma }\gamma ^{\mu \nu }\psi
_{\sigma }\right) \partial _{\lbrack \mu }\eta _{\rho ]}\partial _{\lbrack
\nu }\eta _{\lambda ]}\sigma ^{\rho \lambda }\right)   \label{sec15}
\end{eqnarray}%
can be expressed like
\begin{equation}
\chi =\delta \varphi +\gamma \omega +\partial _{\alpha }l^{\alpha }.
\label{sec16}
\end{equation}%
Supposing that (\ref{sec16}) holds and applying $\delta $ on it, we infer
that
\begin{equation}
\delta \chi =\gamma \left( -\delta \omega \right) +\partial _{\alpha }\left(
\delta l^{\alpha }\right) .  \label{sec17}
\end{equation}%
On the other hand, using the concrete expression of $\chi $, we have that
\begin{eqnarray}
&&\delta \chi =\gamma \left( \frac{k(1-k)}{2}\delta \left( \psi ^{\ast \rho
}\psi _{\rho }\eta _{\nu }\left( \partial _{\mu }h^{\mu \nu }-\partial ^{\nu
}h\right) \right) \right)   \notag \\
&&+\partial ^{\mu }\left( \frac{1}{2}k(1-k)\delta \left( \psi ^{\ast \rho
}\psi _{\rho }\eta ^{\nu }\partial _{\lbrack \mu }\eta _{\nu ]}\right)
\right)   \notag \\
&&+\gamma \left( \frac{\mathrm{i}}{4}k(1-k)\left( \left( \bar{\psi}_{\beta
}\gamma ^{\alpha \beta \sigma }\left( \partial ^{\mu }\psi _{\sigma }\right)
h_{\alpha }^{\rho }-\left( \bar{\psi}_{\beta }\gamma ^{\alpha \beta \lbrack
\mu }\psi ^{\nu ]}\right. \right. \right. \right.   \notag \\
&&\left. \left. -\bar{\psi}^{\mu }\gamma ^{\alpha }\psi ^{\nu }-\sigma
^{\alpha \lbrack \mu }\bar{\psi}^{\nu ]}\gamma ^{\sigma }\psi _{\sigma
}\right) \sigma ^{\rho \lambda }\partial _{\lbrack \nu }h_{\lambda ]\alpha
}\right) \partial _{\lbrack \mu }\eta _{\rho ]}  \notag \\
&&\left. \left. -2\bar{\psi}_{\beta }\gamma ^{\alpha \beta \mu }\left(
\partial ^{\nu }\psi _{\mu }\right) \eta ^{\rho }\partial _{\lbrack \nu
}h_{\rho ]\alpha }\right) \right)   \notag \\
&&+\partial _{\alpha }\left( \frac{\mathrm{i}}{2}k(1-k)\left( \bar{\psi}%
_{\beta }\gamma ^{\alpha \beta \sigma }\left( \partial ^{\mu }\psi _{\sigma
}\right) \eta ^{\rho }-\frac{1}{4}\left( \bar{\psi}_{\beta }\gamma ^{\alpha
\beta \lbrack \mu }\psi ^{\nu ]}\right. \right. \right.   \notag \\
&&\left. \left. \left. -\bar{\psi}^{\mu }\gamma ^{\alpha }\psi ^{\nu
}-\sigma ^{\alpha \lbrack \mu }\bar{\psi}^{\nu ]}\gamma ^{\sigma }\psi
_{\sigma }\right) \sigma ^{\rho \lambda }\partial _{\lbrack \nu }\eta
_{\lambda ]}\right) \partial _{\lbrack \mu }\eta _{\rho ]}\right) .
\label{sec18}
\end{eqnarray}%
The right-hand side of (\ref{sec18}) can be written like in the right-hand
side of (\ref{sec17}) if the following conditions are simultaneously
satisfied
\begin{eqnarray}
&&\delta \omega ^{\prime }=\left( \bar{\psi}_{\beta }\gamma ^{\alpha \beta
\sigma }\left( \partial ^{\mu }\psi _{\sigma }\right) h_{\alpha }^{\rho
}-\left( \bar{\psi}_{\beta }\gamma ^{\alpha \beta \lbrack \mu }\psi ^{\nu
]}\right. \right.   \notag \\
&&\left. \left. -\bar{\psi}^{\mu }\gamma ^{\alpha }\psi ^{\nu }-\sigma
^{\alpha \lbrack \mu }\bar{\psi}^{\nu ]}\gamma ^{\sigma }\psi _{\sigma
}\right) \sigma ^{\rho \lambda }\partial _{\lbrack \nu }h_{\lambda ]\alpha
}\right) \partial _{\lbrack \mu }\eta _{\rho ]}  \notag \\
&&-2\bar{\psi}_{\beta }\gamma ^{\alpha \beta \mu }\left( \partial ^{\nu
}\psi _{\mu }\right) \eta ^{\rho }\partial _{\lbrack \nu }h_{\rho ]\alpha },
\label{sec19}
\end{eqnarray}%
\begin{eqnarray}
&&\delta l^{\prime \alpha }=\left( \bar{\psi}_{\beta }\gamma ^{\alpha \beta
\sigma }\left( \partial ^{\mu }\psi _{\sigma }\right) \eta ^{\rho }-\frac{1}{%
4}\left( \bar{\psi}_{\beta }\gamma ^{\alpha \beta \lbrack \mu }\psi ^{\nu
]}\right. \right.   \notag \\
&&\left. \left. -\bar{\psi}^{\mu }\gamma ^{\alpha }\psi ^{\nu }-\sigma
^{\alpha \lbrack \mu }\bar{\psi}^{\nu ]}\gamma ^{\sigma }\psi _{\sigma
}\right) \sigma ^{\rho \lambda }\partial _{\lbrack \nu }\eta _{\lambda
]}\right) \partial _{\lbrack \mu }\eta _{\rho ]}.  \label{sec20}
\end{eqnarray}%
Since none of the quantities $h_{\mu \beta }$, $\partial ^{\left[ \alpha
\right. }h^{\beta ]\lambda }$, $\eta _{\beta }$, or $\partial ^{\left[
\alpha \right. }\eta ^{\beta ]}$ are $\delta $-exact, the last relations
hold if the equations
\begin{eqnarray}
\bar{\psi}_{\beta }\gamma ^{\alpha \beta \sigma }\left( \partial _{\mu }\psi
_{\sigma }\right)  &=&\delta \Omega _{\;\;\mu }^{\alpha },  \label{sec21a} \\
\bar{\psi}_{\beta }\gamma ^{\alpha \beta \lbrack \mu }\psi ^{\nu ]}-\bar{\psi%
}^{\mu }\gamma ^{\alpha }\psi ^{\nu }-\sigma ^{\alpha \lbrack \mu }\bar{\psi}%
^{\nu ]}\gamma ^{\sigma }\psi _{\sigma } &=&\delta \Gamma ^{\mu \nu \alpha }
\label{sec21b}
\end{eqnarray}%
take place simultaneously. Assuming that both the equations (\ref{sec21a})
and (\ref{sec21b}) are valid, they further give
\begin{eqnarray}
\partial _{\alpha }\left( \bar{\psi}_{\beta }\gamma ^{\alpha \beta \sigma
}\left( \partial _{\mu }\psi _{\sigma }\right) \right)  &=&\delta \left(
\partial _{\alpha }\Omega _{\;\;\mu }^{\alpha }\right) ,  \label{sec22a} \\
\partial _{\alpha }\left( \bar{\psi}_{\beta }\gamma ^{\alpha \beta \lbrack
\mu }\psi ^{\nu ]}-\bar{\psi}^{\mu }\gamma ^{\alpha }\psi ^{\nu }-\sigma
^{\alpha \lbrack \mu }\bar{\psi}^{\nu ]}\gamma ^{\sigma }\psi _{\sigma
}\right)  &=&\delta \left( \partial _{\alpha }\Gamma ^{\mu \nu \alpha
}\right) .  \label{sec22b}
\end{eqnarray}%
On the other hand, by direct computation we obtain that
\begin{equation}
\partial _{\alpha }\left( \bar{\psi}_{\beta }\gamma ^{\alpha \beta \sigma
}\left( \partial _{\mu }\psi _{\sigma }\right) \right) =\delta \left( -%
\mathrm{i}\left( \psi ^{\ast \sigma }\left( \partial _{\mu }\psi _{\sigma
}\right) -\bar{\psi}^{\sigma }\left( \partial _{\mu }\bar{\psi}_{\sigma
}^{\ast }\right) \right) \right) ,  \label{sec23a}
\end{equation}%
\begin{eqnarray}
&&\partial _{\alpha }\left( \bar{\psi}_{\beta }\gamma ^{\alpha \beta \lbrack
\mu }\psi ^{\nu ]}-\bar{\psi}^{\mu }\gamma ^{\alpha }\psi ^{\nu }-\sigma
^{\alpha \lbrack \mu }\bar{\psi}^{\nu ]}\gamma ^{\sigma }\psi _{\sigma
}\right) =\delta \left( -\mathrm{i}\psi ^{\ast \sigma }\gamma ^{\mu \nu
}\psi _{\sigma }\right.   \notag \\
&&\left. -2\mathrm{i}\psi ^{\ast \lbrack \mu }\psi ^{\nu ]}\right) -\bar{\psi%
}_{\alpha }\gamma ^{\alpha \beta \lbrack \mu }\partial ^{\nu ]}\psi _{\beta
},  \label{sec23b}
\end{eqnarray}%
so the right-hand sides of (\ref{sec23a})--(\ref{sec23b}) cannot be written
like in the right-hand sides of (\ref{sec22a})--(\ref{sec22b}). This means
that the relations (\ref{sec21a})--(\ref{sec21b}) are not valid, and
therefore neither are (\ref{sec19})--(\ref{sec20}). As a consequence, $\chi $
must vanish, and hence we must set
\begin{equation}
k\left( 1-k\right) =0.  \label{sec24}
\end{equation}%
Using (\ref{sec24}), we conclude that
\begin{equation}
k=1.  \label{sec25}
\end{equation}%
Inserting (\ref{sec25}) in (\ref{sec7}), we obtain that
\begin{eqnarray}
&&\Delta _{1}^{\left( \mathrm{int}\right) }=\gamma \left( \left( -\frac{1}{4}%
\left( \psi ^{\ast \lbrack \mu }\psi ^{\sigma ]}+\frac{1}{2}\psi ^{\ast \rho
}\gamma ^{\mu \sigma }\psi _{\rho }\right) \partial _{\lbrack \sigma }\eta
_{\lambda ]}\sigma ^{\nu \lambda }\right. \right.   \notag \\
&&\left. +\psi ^{\ast \sigma }\left( \partial ^{\mu }\psi _{\sigma }\right)
\eta ^{\nu }\right) h_{\mu \nu }  \notag \\
&&\left. +\frac{1}{2}\left( \psi ^{\ast \mu }\psi ^{\nu }+\frac{1}{4}\psi
^{\ast \sigma }\gamma ^{\mu \nu }\psi _{\sigma }\right) \eta ^{\rho
}\partial _{\lbrack \mu }h_{\nu ]\rho }\right) .  \label{sec26}
\end{eqnarray}%
Comparing (\ref{sec26}) with (\ref{sec12}), we find that
\begin{equation}
b_{2}^{\left( \mathrm{int}\right) }=0,  \label{sec27}
\end{equation}%
\begin{eqnarray}
&&b_{1}^{\left( \mathrm{int}\right) }=\frac{1}{8}\left( \psi ^{\ast \lbrack
\mu }\psi ^{\sigma ]}+\frac{1}{2}\psi ^{\ast \rho }\gamma ^{\mu \sigma }\psi
_{\rho }\right) h_{\mu }^{\lambda }\partial _{\lbrack \sigma }\eta _{\lambda
]}  \notag \\
&&-\frac{1}{2}\psi ^{\ast \sigma }\left( \partial ^{\mu }\psi _{\sigma
}\right) \eta ^{\nu }h_{\mu \nu }  \notag \\
&&-\frac{1}{4}\left( \psi ^{\ast \mu }\psi ^{\nu }+\frac{1}{4}\psi ^{\ast
\sigma }\gamma ^{\mu \nu }\psi _{\sigma }\right) \eta ^{\rho }\partial
_{\lbrack \mu }h_{\nu ]\rho }.  \label{sec28}
\end{eqnarray}%
Substituting (\ref{sec25}) in (\ref{sec8}) and using (\ref{sec28}), we
deduce
\begin{eqnarray}
&&\Delta _{0}^{\left( \mathrm{int}\right) }+2\delta b_{1}^{\left( \mathrm{int%
}\right) }=\partial _{\mu }n_{0}^{\mu }+\gamma \left( -\frac{1}{4}\mathcal{L}%
_{0}^{(\mathrm{RS})}\left( h^{2}-2h_{\mu \nu }h^{\mu \nu }\right) \right.
\notag \\
&&+\frac{\mathrm{i}}{8}\bar{\psi}^{\mu }\gamma ^{\lambda }\psi ^{\nu }\left(
\left( h_{\rho }^{\lambda }-h\delta _{\rho }^{\lambda }\right) \partial
_{\lbrack \mu }h_{\nu ]\lambda }+h_{\nu }^{\sigma }\left( 2\partial
_{\lbrack \mu }h_{\sigma ]\lambda }+\partial _{\lambda }h_{\mu \sigma
}\right) \right)   \notag \\
&&+\frac{\mathrm{i}}{4}\bar{\psi}^{\mu }\gamma ^{\sigma }\psi _{\sigma
}\left( h\left( \partial _{\mu }h-\partial ^{\nu }h_{\mu \nu }\right)
+h_{\mu }^{\rho }\left( \partial ^{\lambda }h_{\rho \lambda }-\partial
_{\rho }h\right) \right.   \notag \\
&&\left. -2h^{\alpha \beta }\partial _{\mu }h_{\alpha \beta }+\frac{3}{2}%
h^{\rho \lambda }\partial _{\rho }h_{\mu \lambda }+\frac{1}{2}h_{\mu \nu
}\partial _{\rho }h^{\rho \nu }\right)   \notag \\
&&+\frac{\mathrm{i}}{4}\bar{\psi}_{\mu }\gamma ^{\mu \nu \beta }\left(
\partial ^{\alpha }\psi _{\nu }\right) \left( hh_{\alpha \beta }-\frac{3}{2}%
h_{\alpha \sigma }h_{\beta }^{\sigma }\right) -\left( V+V^{\mu \nu }\partial
_{\mu }\psi _{\nu }\right) h  \notag \\
&&+\frac{\mathrm{i}}{8}\bar{\psi}_{\beta }\gamma ^{\beta \mu \alpha
}\psi ^{\nu }\left( \left( h\delta _{\nu }^{\rho }-\frac{1}{2}h_{\nu
}^{\rho }\right) \partial _{\lbrack \mu }h_{\alpha ]\rho }+h_{\mu
}^{\rho }\left( 3\partial _{\alpha }h_{\nu \rho }-2\partial _{\rho
}h_{\alpha \nu }\right)
\right)   \notag \\
&&\left. +V^{\mu \nu }\left( h_{\mu \sigma }\partial ^{\sigma }\psi _{\nu
}+\psi ^{\sigma }\partial _{\lbrack \nu }h_{\sigma ]\mu }+\frac{1}{4}\gamma
^{\alpha \beta }\psi _{\nu }\partial _{\lbrack \alpha }h_{\beta ]\mu
}\right) \right) +\Pi ^{\mu \nu }\partial _{\lbrack \mu }\eta _{\nu ]},
\label{sec29}
\end{eqnarray}%
where
\begin{eqnarray}
&&\Pi ^{\mu \nu }=V^{\mu \rho }\partial ^{\nu }\psi _{\rho }+\frac{\partial
^{R}V}{\partial \psi _{\mu }}\psi ^{\nu }+V^{\rho \mu }\partial _{\rho }\psi
^{\nu }+\bar{\psi}^{\nu }\frac{\partial ^{L}V^{\rho \lambda }}{\partial \bar{%
\psi}_{\mu }}\partial _{\rho }\psi _{\lambda }  \notag \\
&&+\frac{1}{4}\left( \frac{\partial ^{R}V}{\partial \psi _{\rho }}\gamma
^{\mu \nu }\psi _{\rho }+V^{\rho \lambda }\gamma ^{\mu \nu }\partial _{\rho
}\psi _{\lambda }-\bar{\psi}_{\theta }\gamma ^{\mu \nu }\frac{\partial
^{L}V^{\rho \lambda }}{\partial \bar{\psi}_{\theta }}\partial _{\rho }\psi
_{\lambda }\right) .  \label{sec30}
\end{eqnarray}%
We observe that (\ref{sec29}) can be written like in (\ref{sec13}) if and
only if
\begin{equation}
\Pi ^{\mu \nu }-\Pi ^{\nu \mu }=\partial _{\rho }U^{\rho \mu \nu }.
\label{sec31}
\end{equation}%
The right-hand side of (\ref{sec30}) splits according to the number of
derivatives into
\begin{equation}
\Pi ^{\mu \nu }=\Pi _{0}^{\mu \nu }+\Pi _{1}^{\mu \nu },  \label{pi}
\end{equation}%
where we made the notations
\begin{equation}
\Pi _{0}^{\mu \nu }=\frac{\partial ^{R}V}{\partial \psi _{\mu }}\psi ^{\nu }+%
\frac{1}{4}\frac{\partial ^{R}V}{\partial \psi _{\rho }}\gamma ^{\mu \nu
}\psi _{\rho },  \label{pi0}
\end{equation}%
\begin{eqnarray}
&&\Pi _{1}^{\mu \nu }=V^{\mu \rho }\partial ^{\nu }\psi _{\rho }+V^{\rho \mu
}\partial _{\rho }\psi ^{\nu }+\bar{\psi}^{\nu }\frac{\partial ^{L}V^{\rho
\lambda }}{\partial \bar{\psi}_{\mu }}\partial _{\rho }\psi _{\lambda }
\notag \\
&&+\frac{1}{4}\left( V^{\rho \lambda }\gamma ^{\mu \nu }\partial _{\rho
}\psi _{\lambda }-\bar{\psi}_{\theta }\gamma ^{\mu \nu }\frac{\partial
^{L}V^{\rho \lambda }}{\partial \bar{\psi}_{\theta }}\partial _{\rho }\psi
_{\lambda }\right) .  \label{pi1}
\end{eqnarray}%
As $\Pi _{0}^{\mu \nu }$ has no derivatives, it cannot bring to (\ref{sec31}%
) a divergence-like contribution, and $\Pi _{1}^{\mu \nu }$ contains just
one derivative, so in principle it may lead to a total derivative, as
required by (\ref{sec31}). As a consequence, from (\ref{sec31}) projected on
the number of derivatives equal to zero we find that $\Pi _{0}^{\mu \nu }$
is subject to the equation
\begin{equation}
\Pi _{0}^{\mu \nu }-\Pi _{0}^{\nu \mu }=0,  \label{eqpi0}
\end{equation}%
which is, via (\ref{pi0}), equivalent to
\begin{equation}
\frac{\partial ^{R}V}{\partial \psi _{\mu }}\psi ^{\nu }-\frac{\partial ^{R}V%
}{\partial \psi _{\nu }}\psi ^{\mu }=-\frac{1}{2}\frac{\partial ^{R}V}{%
\partial \psi _{\rho }}\gamma ^{\mu \nu }\psi _{\rho }.  \label{sec32}
\end{equation}%
If we generically represent $\partial ^{R}V/\partial \psi _{\mu }$ under the
form%
\begin{equation}
\frac{\partial ^{R}V}{\partial \psi _{\mu }}=\bar{\psi}_{\alpha }M^{\alpha
\mu }\left( \psi ^{\nu }\right) ,  \label{ecmat2}
\end{equation}%
then the equation (\ref{sec32}) requires that
\begin{equation}
\gamma ^{0}V^{\mu \nu \alpha \sigma }=\left( \gamma ^{0}V^{\mu \nu \sigma
\alpha }\right) ^{\intercal },  \label{ecmat1}
\end{equation}%
where
\begin{equation}
V^{\mu \nu \alpha \sigma }=M^{\alpha \mu }\sigma ^{\nu \sigma }-M^{\alpha
\nu }\sigma ^{\mu \sigma }+\frac{1}{2}M^{\alpha \sigma }\gamma ^{\mu \nu
}=-V^{\nu \mu \alpha \sigma }.  \label{qq1}
\end{equation}%
If we decompose $V^{\mu \nu \alpha \sigma }$ like
\begin{eqnarray}
V^{\mu \nu \alpha \sigma } &=&V_{0}^{\mu \nu \alpha \sigma }\mathbf{1}+%
V_{1\;\;\;\;\;\;\;\;\tau }^{\mu \nu \alpha \sigma }\gamma ^{\tau
}+V_{2\;\;\;\;\;\;\;\;\tau \gamma }^{\mu \nu \alpha \sigma }\gamma ^{\tau
\gamma }  \notag \\
&&+V_{3\;\;\;\;\;\;\;\;\tau \gamma \rho }^{\mu \nu \alpha \sigma }\gamma
^{\tau \gamma \rho }+V_{4\;\;\;\;\;\;\;\;\tau \gamma \rho \lambda }^{\mu \nu
\alpha \sigma }\gamma ^{\tau \gamma \rho \lambda },  \label{qq2}
\end{eqnarray}%
then the condition (\ref{ecmat1}) implies the relations
\begin{eqnarray}
V_{0}^{\mu \nu \alpha \sigma } &=&-V_{0}^{\mu \nu \sigma \alpha
},\;V_{1\;\;\;\;\;\;\;\;\tau }^{\mu \nu \alpha \sigma
}=V_{1\;\;\;\;\;\;\;\;\tau }^{\mu \nu \sigma \alpha
},\;V_{2\;\;\;\;\;\;\;\;\tau \gamma }^{\mu \nu \alpha \sigma
}=V_{2\;\;\;\;\;\;\;\;\tau \gamma }^{\mu \nu \sigma \alpha },  \label{qq4} \\
V_{3\;\;\;\;\;\;\;\;\tau \gamma \rho }^{\mu \nu \alpha \sigma }
&=&-V_{3\;\;\;\;\;\;\;\;\tau \gamma \rho }^{\mu \nu \sigma \alpha
},\;V_{4\;\;\;\;\;\;\;\;\tau \gamma \rho \lambda }^{\mu \nu \alpha \sigma
}=-V_{4\;\;\;\;\;\;\;\;\tau \gamma \rho \lambda }^{\mu \nu \sigma \alpha }.
\label{qq3}
\end{eqnarray}%
In a similar manner, if we expand $M^{\alpha \mu }$ along the basis in the
space of constant, $4\times 4$ complex matrices
\begin{equation}
M^{\alpha \mu }=M_{0}^{\alpha \mu }\mathbf{1+}M_{1\;\;\;\;\tau }^{\alpha \mu
}\gamma ^{\tau }+M_{2\;\;\;\;\tau \gamma }^{\alpha \mu }\gamma ^{\tau \gamma
}+M_{3\;\;\;\;\tau \gamma \rho }^{\alpha \mu }\gamma ^{\tau \gamma \rho
}+M_{4\;\;\;\;\tau \gamma \rho \lambda }^{\alpha \mu }\gamma ^{\tau \gamma
\rho \lambda },  \label{qq5}
\end{equation}%
substitute (\ref{qq5}) in (\ref{qq1}), and take into account the relations (%
\ref{qq4})--(\ref{qq3}), then we finally find that
\begin{eqnarray}
M_{0}^{\alpha \mu } &=&m_{0}\left( \psi ^{\nu }\right) \sigma ^{\alpha \mu
},\quad M_{1\;\;\;\;\tau }^{\alpha \mu }=0,\quad M_{2\;\;\;\;\tau \gamma
}^{\alpha \mu }=m_{2}\left( \psi ^{\nu }\right) \delta _{\lbrack \tau
}^{\alpha }\delta _{\gamma ]}^{\mu },  \label{qq6} \\
M_{3\;\;\;\;\tau \gamma \rho }^{\alpha \mu } &=&0,\quad M_{4\;\;\;\;\tau
\gamma \rho \lambda }^{\alpha \mu }=m_{4}\left( \psi ^{\nu }\right)
\varepsilon _{\tau \gamma \rho \lambda }\sigma ^{\alpha \mu },  \label{qq7}
\end{eqnarray}%
where $m_{0}\left( \psi ^{\nu }\right) $, $m_{2}\left( \psi ^{\nu }\right) $%
, and $m_{4}\left( \psi ^{\nu }\right) $ are arbitrary functions. Replacing
now (\ref{qq6})--(\ref{qq7}) in (\ref{qq5}) and then the resulting
expression in (\ref{ecmat2}), we find that
\begin{eqnarray}
\frac{\partial ^{R}V}{\partial \psi _{\mu }} &=&m_{0}\left( \psi ^{\nu
}\right) \bar{\psi}^{\mu }+2m_{2}\left( \psi ^{\nu }\right) \bar{%
\psi}_{\alpha }\gamma ^{\alpha \mu }+24\mathrm{i}m_{4}\left( \psi ^{\nu
}\right) \bar{\psi}_{\mu }\gamma _{5}  \notag \\
&=&\frac{1}{2}m_{0}\left( \psi ^{\nu }\right) \frac{\partial ^{R}X}{\partial
\psi _{\mu }}+m_{2}\left( \psi ^{\nu }\right) \frac{\partial ^{R}Y}{\partial
\psi _{\mu }}+12m_{4}\left( \psi ^{\nu }\right) \frac{\partial ^{R}Z}{%
\partial \psi _{\mu }},  \label{qq10}
\end{eqnarray}%
with
\begin{equation}
X\equiv \bar{\psi}_{\mu }\psi ^{\mu },\quad Y\equiv
\bar{\psi}_{\alpha }\gamma ^{\alpha \mu }\psi _{\mu },\quad Z\equiv
\mathrm{i}\bar{\psi}_{\mu }\gamma _{5}\psi ^{\mu }.  \label{qq11}
\end{equation}%
The equation (\ref{qq10}) shows that the solution to (\ref{sec32}) is
nothing but an arbitrary polynomial of $X$, $Y$, and $Z$, i.e.
\begin{equation}
V=V\left( X,Y,Z\right) .  \label{sec32a}
\end{equation}%
In order to complete the analysis of the equation (\ref{sec31}), we need to
solve its component of order one in the spacetime derivatives
\begin{equation}
\Pi _{1}^{\mu \nu }-\Pi _{1}^{\nu \mu }=\partial _{\rho }U^{\rho \mu \nu },
\label{eqpi1}
\end{equation}%
with $\Pi _{1}^{\mu \nu }$ given in (\ref{pi1}) and $U^{\rho \mu \nu }$
containing no derivatives. Taking into consideration the formula (\ref{pi1}%
), it follows that the equation (\ref{eqpi1}) restricts $V^{\mu \lambda }$
to satisfy the equation
\begin{eqnarray}
&&V^{\mu \lambda }\sigma ^{\nu \rho }+V^{\rho \mu }\sigma ^{\nu \lambda
}-V^{\nu \lambda }\sigma ^{\mu \rho }-V^{\rho \nu }\sigma ^{\mu \lambda }+%
\bar{\psi}^{\nu }\frac{\partial ^{L}V^{\rho \lambda }}{\partial \bar{\psi}%
_{\mu }}  \notag \\
&&-\bar{\psi}^{\mu }\frac{\partial ^{L}V^{\rho \lambda }}{\partial \bar{\psi}%
_{\nu }}+\frac{1}{2}\left( V^{\rho \lambda }\gamma ^{\mu \nu }-\bar{\psi}%
_{\theta }\gamma ^{\mu \nu }\frac{\partial ^{L}V^{\rho \lambda }}{\partial
\bar{\psi}_{\theta }}\right) =\frac{\partial ^{R}U^{\rho \mu \nu }}{\partial
\psi _{\lambda }}.  \label{qq12}
\end{eqnarray}%
The last equation is fulfilled if there exist some objects $Q^{\mu }$ such
that the following conditions take place simultaneously:
\begin{eqnarray}
&&V^{\mu \lambda }=-\frac{\partial ^{R}Q^{\mu }}{\partial \psi _{\lambda }},
\label{qq13} \\
&&\frac{\partial ^{R}Q^{\rho }}{\partial \psi _{\mu }}\sigma ^{\nu \lambda }-%
\frac{\partial ^{R}Q^{\rho }}{\partial \psi _{\nu }}\sigma ^{\mu \lambda }=0.
\label{qq14}
\end{eqnarray}%
On the other hand, by adding to and subtracting from the left-hand side of (%
\ref{qq12}) the quantity $\left( 1/2\right) \left( \partial ^{R}Q^{\rho
}/\partial \psi _{\lambda }\right) \gamma ^{\mu \nu }=\partial ^{R}\left(
1/2\left( Q^{\rho }\gamma ^{\mu \nu }\right) \right) /\partial \psi
_{\lambda }$, we can state that (\ref{qq12}) is checked if (\ref{qq13}) and
\begin{equation}
\frac{\partial ^{R}Q^{\rho }}{\partial \psi _{\mu }}\sigma ^{\nu \lambda }-%
\frac{\partial ^{R}Q^{\rho }}{\partial \psi _{\nu }}\sigma ^{\mu \lambda }+%
\frac{1}{2}\frac{\partial ^{R}Q^{\rho }}{\partial \psi _{\lambda }}\gamma
^{\mu \nu }=0  \label{qq16}
\end{equation}%
are simultaneously verified. By multiplying (\ref{qq16}) from the right with
$\psi _{\lambda }$ we get the equation
\begin{equation}
\frac{\partial ^{R}Q^{\rho }}{\partial \psi _{\mu }}\psi ^{\nu }-\frac{%
\partial ^{R}Q^{\rho }}{\partial \psi _{\nu }}\psi ^{\mu }+\frac{1}{2}\frac{%
\partial ^{R}Q^{\rho }}{\partial \psi _{\lambda }}\gamma ^{\mu \nu }\psi
^{\lambda }=0,  \label{qq15}
\end{equation}%
which shows that (see (\ref{sec32}) and (\ref{sec32a}))
\begin{equation}
Q^{\rho }=Q^{\rho }\left( X,Y,Z\right) .  \label{qq17}
\end{equation}%
Since $Q^{\mu }$ like in (\ref{qq17}) must provide $V^{\mu \lambda }$ via
taking its right derivative with respect to $\psi _{\lambda }$ (see (\ref%
{qq13})), it results that
\begin{equation}
Q^{\mu }=Q\left( X,Y,Z\right) \gamma ^{\mu },  \label{qq17aaaa}
\end{equation}%
with $Q\left( X,Y,Z\right) $ an arbitrary polynomial. Formulas (\ref{qq13})
and (\ref{qq17aaaa}) together with some appropriate Fierz identities further
yield
\begin{equation}
V^{\mu \nu }=\bar{\psi}_{\rho }P^{\rho \mu \nu }\left( X,Y,Z\right) ,
\label{sec33ab}
\end{equation}%
where
\begin{equation}
P^{\rho \mu \nu }\left( X,Y,Z\right) =\left( P^{\rho \mu \nu }\right)
_{\alpha }\left( X,Y,Z\right) \gamma ^{\alpha }+\left( P^{\rho \mu \nu
}\right) _{\alpha \beta \gamma }\left( X,Y,Z\right) \gamma ^{\alpha \beta
\gamma }.  \label{sec33}
\end{equation}%
The dependence on $X$, $Y$, and $Z$ of the functions $\left( P^{\rho \mu \nu
}\right) _{\alpha }$ and $\left( P^{\rho \mu \nu }\right) _{\alpha \beta
\gamma }$ enables us to conclude that the most general form of these
coefficients reads as
\begin{eqnarray}
\left( P^{\rho \mu \nu }\right) _{\alpha }\left( X,Y,Z\right)
&=&d_{1}\delta _{\alpha }^{\rho }\sigma ^{\mu \nu }+d_{2}\delta _{\alpha
}^{\mu }\sigma ^{\rho \nu }+d_{3}\delta _{\alpha }^{\nu }\sigma ^{\rho \mu },
\label{sec33a} \\
\left( P^{\rho \mu \nu }\right) _{\alpha \beta \gamma }\left( X,Y,Z\right)
&=&d_{4}\delta _{\lbrack \alpha }^{\rho }\delta _{\beta }^{\mu }\sigma
_{\gamma ]}^{\nu },  \label{sec33b}
\end{eqnarray}%
where $\left( d_{i}\right) _{i=1,2,3,4}$ are arbitrary polynomials in $X$, $Y
$, and $Z$. We remark that (\ref{sec33b}) gives in (\ref{om7}), and thus in $%
S_{1}^{\left( \mathrm{int}\right) }$, a contribution (up to a trivial, $s$%
-exact term) that is already contained in (\ref{sec32a}) since
\begin{eqnarray}
&&\bar{\psi}_{\rho }\left( P^{\rho \mu \nu }\right) _{\alpha \beta \gamma
}\gamma ^{\alpha \beta \gamma }\partial _{\mu }\psi _{\nu }=6d_{4}\bar{\psi}%
_{\mu }\gamma ^{\mu \nu \rho }\partial _{\nu }\psi _{\rho }=  \notag \\
&&s\left( -6\mathrm{i}d_{4}\bar{\psi}_{\mu }\bar{\psi}^{\ast \mu }\right) -6%
\mathrm{i}d_{4}\bar{\psi}_{\mu }\gamma ^{\mu \nu }\psi _{\nu },
\label{sec33c}
\end{eqnarray}%
so we can take, without loss of generality
\begin{equation}
d_{4}=0  \label{e40}
\end{equation}%
in (\ref{sec33b}). Taking into account the last result and inserting (\ref%
{sec33a}) in (\ref{sec33}) and then in (\ref{sec33ab}), we infer that
\begin{eqnarray}
&&V^{\mu \nu }\partial _{\mu }\psi _{\nu }=d_{1}\bar{\psi}_{\rho }\gamma
^{\rho }\partial _{\mu }\psi ^{\mu }+d_{2}\bar{\psi}^{\nu }\gamma ^{\mu
}\partial _{\mu }\psi _{\nu }+d_{3}\bar{\psi}^{\nu }\gamma ^{\mu }\partial
_{\nu }\psi _{\mu }=  \notag \\
&&d_{1}\bar{\psi}_{\rho }\gamma ^{\rho }\partial _{\mu }\psi ^{\mu }+\frac{1%
}{2}\left( d_{2}+d_{3}\right) \bar{\psi}^{\nu }\gamma ^{\mu }\partial _{(\mu
}\psi _{\nu )}+\frac{1}{2}\left( d_{3}-d_{2}\right) \bar{\psi}^{\mu }\gamma
^{\nu }\partial _{\lbrack \mu }\psi _{\nu ]}=  \notag \\
&&d_{1}\bar{\psi}_{\rho }\gamma ^{\rho }\partial _{\mu }\psi ^{\mu }+\frac{1%
}{2}\left( d_{2}+d_{3}\right) \bar{\psi}^{\nu }\gamma ^{\mu }\partial _{(\mu
}\psi _{\nu )}+  \notag \\
&&s\left( -\frac{\mathrm{i}}{4}\left( d_{3}-d_{2}\right) \left( \bar{\psi}%
_{\mu }\gamma ^{\mu \nu }\bar{\psi}_{\nu }^{\ast }-\bar{\psi}_{\mu }\bar{\psi%
}^{\ast \mu }\right) \right)   \notag \\
&&-\frac{\mathrm{i}m}{2}\left( d_{3}-d_{2}\right) \left( \bar{\psi}_{\mu
}\psi ^{\mu }+\bar{\psi}_{\mu }\gamma ^{\mu \nu }\psi _{\nu }\right) .
\label{gama6}
\end{eqnarray}%
Thus, up to an irrelevant, $s$-exact term, $V^{\mu \nu }\partial _{\mu }\psi
_{\nu }$ contains, beside the first two pieces, the last component, which is
a contribution already considered in (\ref{sec32a}). We can thus forget
about it by setting
\begin{equation}
d_{3}-d_{2}=0.  \label{e3minuse2}
\end{equation}%
At this stage, from (\ref{e40}) and (\ref{e3minuse2}) replaced in (\ref%
{sec33a})--(\ref{sec33b}) and the resulting relations further substituted in
(\ref{sec33}), with the help of the representation (\ref{sec33ab}) we
determine the relevant part of $V^{\mu \nu }$ under the form
\begin{equation}
V^{\mu \nu }=d_{1}\left( X,Y,Z\right) \bar{\psi}_{\alpha }\gamma ^{\alpha
}\sigma ^{\mu \nu }+d_{2}\left( X,Y,Z\right) \left( \bar{\psi}^{\nu }\gamma
^{\mu }+\bar{\psi}^{\mu }\gamma ^{\nu }\right) .  \label{finalfmunu}
\end{equation}%
Consequently, we find that $V^{\mu \nu }\partial _{\mu }\psi _{\nu }$ no
longer contains the unwanted (trivial or redundant) contributions, being
precisely given by
\begin{equation}
V^{\mu \nu }\partial _{\mu }\psi _{\nu }=d_{1}\left( X,Y,Z\right) \bar{\psi}%
_{\rho }\gamma ^{\rho }\partial _{\mu }\psi ^{\mu }+d_{2}\left( X,Y,Z\right)
\bar{\psi}^{\nu }\gamma ^{\mu }\partial _{(\mu }\psi _{\nu )}.  \label{red}
\end{equation}%
Based on the relations (\ref{finalfmunu}) and (\ref{red}), we deduce that
the antisymmetric part of $\Pi _{1}^{\mu \nu }$ must vanish
\begin{equation}
\Pi _{1}^{\mu \nu }-\Pi _{1}^{\nu \mu }=0.  \label{qq21}
\end{equation}

As a consequence of this step of the deformation procedure, on the one hand
the results (\ref{sec25}), (\ref{sec32a}), and (\ref{red}) completely
determine the component (\ref{om7}), and hence the cross-coupling part of
the first-order deformation (\ref{om10}) like
\begin{eqnarray}
S_{1}^{\left( \mathrm{int}\right) } &=&\int d^{4}x\left( \psi ^{*\mu }\left(
\partial ^{\nu }\psi _{\mu }\right) \eta _{\nu }+\frac{1}{2}\psi ^{*\mu
}\psi ^{\nu }\partial _{[\mu }\eta _{\nu ]}\right.  \notag \\
&&+\frac{1}{8}\psi ^{*\rho }\gamma ^{\mu \nu }\psi _{\rho }\partial _{[\mu
}\eta _{\nu ]}+\frac{1}{2}\left( \sigma ^{\rho \lambda }\mathcal{L}_{0}^{(%
\mathrm{RS})}-\frac{\mathrm{i}}{2}\bar{\psi}_{\mu }\gamma ^{\mu \nu \rho
}\partial ^{\lambda }\psi _{\nu }\right) h_{\rho \lambda }  \notag \\
&&+\frac{\mathrm{i}}{4}\left( \frac{1}{2}\bar{\psi}^{\mu }\gamma ^{\rho
}\psi ^{\nu }+\sigma ^{\mu \rho }\bar{\psi}^{\nu }\gamma ^{\sigma }\psi
_{\sigma }+\bar{\psi}_{\sigma }\gamma ^{\sigma \rho \mu }\psi ^{\nu }\right)
\partial _{[\mu }h_{\nu ]\rho }  \notag \\
&&\left. +V+d_{1}\bar{\psi}_{\rho }\gamma ^{\rho }\partial _{\mu }\psi ^{\mu
}+d_{2}\bar{\psi}^{\nu }\gamma ^{\mu }\partial _{(\mu }\psi _{\nu )}\right) .
\label{deford1}
\end{eqnarray}
On the other hand, (\ref{sec29}), (\ref{eqpi0}), (\ref{sec32a}), (\ref%
{finalfmunu}), and (\ref{qq21}) offer us the concrete form of $b_{0}^{(%
\mathrm{int})}$ as solution to the equation (\ref{sec13}) like
\begin{eqnarray}
&&b_{0}^{\left( \mathrm{int}\right) }=\frac{1}{8}\mathcal{L}_{0}^{(\mathrm{RS%
})}\left( h^{2}-2h_{\mu \nu }h^{\mu \nu }\right)  \notag \\
&&-\frac{\mathrm{i}}{16}\bar{\psi}^{\mu }\gamma ^{\lambda }\psi ^{\nu
}\left( \left( h_{\rho }^{\lambda }-h\delta _{\rho }^{\lambda }\right)
\partial _{[\mu }h_{\nu ]\lambda }+h_{\nu }^{\sigma }\left( 2\partial _{[\mu
}h_{\sigma ]\lambda }+\partial _{\lambda }h_{\mu \sigma }\right) \right)
\notag \\
&&-\frac{\mathrm{i}}{8}\bar{\psi}^{\mu }\gamma ^{\sigma }\psi _{\sigma
}\left( h\left( \partial _{\mu }h-\partial ^{\nu }h_{\mu \nu }\right)
+h_{\mu }^{\rho }\left( \partial ^{\lambda }h_{\rho \lambda }-\partial
_{\rho }h\right) \right.  \notag \\
&&\left. -2h^{\alpha \beta }\partial _{\mu }h_{\alpha \beta }+\frac{3}{2}%
h^{\rho \lambda }\partial _{\rho }h_{\mu \lambda }+\frac{1}{2}h_{\mu \nu
}\partial _{\rho }h^{\rho \nu }\right)  \notag \\
&&-\frac{\mathrm{i}}{8}\bar{\psi}_{\mu }\gamma ^{\mu \nu \beta }\left(
\partial ^{\alpha }\psi _{\nu }\right) \left( hh_{\alpha \beta }-\frac{3}{2}%
h_{\alpha \sigma }h_{\beta }^{\sigma }\right)  \notag \\
&&-\frac{\mathrm{i}}{16}\bar{\psi}_{\beta }\gamma ^{\beta \mu \alpha
}\psi ^{\nu }\left( \left( h\delta _{\nu }^{\rho }-\frac{1}{2}h_{\nu
}^{\rho }\right) \partial _{[\mu }h_{\alpha ]\rho }+h_{\mu }^{\rho
}\left( 3\partial _{\alpha }h_{\nu \rho }-2\partial _{\rho
}h_{\alpha \nu }\right) \right)
\notag \\
&&+\frac{h}{2}V+\frac{d_{1}}{2}\bar{\psi}_{\rho }\gamma ^{\rho }\left(
h\partial _{\mu }\psi ^{\mu }-\left( \partial _{\mu }\psi _{\nu }\right)
h^{\mu \nu }-\sigma ^{\mu \nu }\psi ^{\sigma }\partial _{[\nu }h_{\sigma
]\mu }-\right.  \notag \\
&&\left. -\frac{1}{4}\sigma ^{\mu \nu }\gamma ^{\alpha \beta }\psi _{\nu
}\partial _{[\alpha }h_{\beta ]\mu }\right) +\frac{d_{2}}{2}\bar{\psi}_{\rho
}\left( h\gamma ^{\mu }\partial _{\mu }\psi ^{\rho }-h^{\mu \nu }\gamma
_{\mu }\partial _{\nu }\psi ^{\rho }\right.  \notag \\
&&\left. -\gamma _{\mu }\psi _{\lambda }\partial ^{[\rho }h^{\lambda ]\mu }-%
\frac{1}{4}\gamma ^{\mu }\gamma ^{\alpha \beta }\psi ^{\rho }\partial
_{[\alpha }h_{\beta ]\mu }\right) +\frac{d_{2}}{2}\left( h\bar{\psi}^{\rho
}\gamma ^{\mu }\partial _{\rho }\psi _{\mu }\right.  \notag \\
&&\left. -h^{\mu \nu }\bar{\psi}_{\mu }\gamma ^{\rho }\partial _{\nu }\psi
_{\rho }-\bar{\psi}^{\mu }\gamma ^{\nu }\psi ^{\rho }\partial _{[\nu
}h_{\rho ]\mu }-\frac{1}{4}\bar{\psi}^{\mu }\gamma ^{\nu }\gamma ^{\alpha
\beta }\psi _{\nu }\partial _{[\alpha }h_{\beta ]\mu }\right) .
\label{sec34}
\end{eqnarray}
At this moment, the components $\left( b_{I}^{\left( \mathrm{int}\right)
}\right) _{I=0,1,2}$ expressed by (\ref{sec27}), (\ref{sec28}), and (\ref%
{sec34}) yield the cross-coupling part of the second-order deformation $%
S_{2}^{\left( \mathrm{int}\right) }=\int d^{4}x\left( b_{0}^{\left( \mathrm{%
int}\right) }+b_{1}^{\left( \mathrm{int}\right) }+b_{2}^{\left( \mathrm{int}%
\right) }\right) $ as
\begin{eqnarray}
S_{2}^{\left( \mathrm{int}\right) } &=&\int d^{4}x\left( \frac{1}{8}\left(
\psi ^{*[\mu }\psi ^{\sigma ]}+\frac{1}{2}\psi ^{*\rho }\gamma ^{\mu \sigma
}\psi _{\rho }\right) h_{\mu }^{\lambda }\partial _{[\sigma }\eta _{\lambda
]}-\frac{1}{2}\psi ^{*\sigma }\left( \partial ^{\mu }\psi _{\sigma }\right)
\eta ^{\nu }h_{\mu \nu }\right.  \notag \\
&&-\frac{1}{4}\left( \psi ^{*\mu }\psi ^{\nu }+\frac{1}{4}\psi ^{*\sigma
}\gamma ^{\mu \nu }\psi _{\sigma }\right) \eta ^{\rho }\partial _{[\mu
}h_{\nu ]\rho }+\frac{1}{8}\mathcal{L}_{0}^{(\mathrm{RS})}\left(
h^{2}-2h_{\mu \nu }h^{\mu \nu }\right)  \notag \\
&&-\frac{\mathrm{i}}{16}\bar{\psi}^{\mu }\gamma ^{\lambda }\psi ^{\nu
}\left( \left( h_{\rho }^{\lambda }-h\delta _{\rho }^{\lambda }\right)
\partial _{[\mu }h_{\nu ]\lambda }+h_{\nu }^{\sigma }\left( 2\partial _{[\mu
}h_{\sigma ]\lambda }+\partial _{\lambda }h_{\mu \sigma }\right) \right)
\notag \\
&&-\frac{\mathrm{i}}{8}\bar{\psi}^{\mu }\gamma ^{\sigma }\psi _{\sigma
}\left( h\left( \partial _{\mu }h-\partial ^{\nu }h_{\mu \nu }\right)
+h_{\mu }^{\rho }\left( \partial ^{\lambda }h_{\rho \lambda }-\partial
_{\rho }h\right) \right.  \notag \\
&&\left. -2h^{\alpha \beta }\partial _{\mu }h_{\alpha \beta }+\frac{3}{2}%
h^{\rho \lambda }\partial _{\rho }h_{\mu \lambda }+\frac{1}{2}h_{\mu \nu
}\partial _{\rho }h^{\rho \nu }\right)  \notag \\
&&-\frac{\mathrm{i}}{8}\bar{\psi}_{\mu }\gamma ^{\mu \nu \beta }\left(
\partial ^{\alpha }\psi _{\nu }\right) \left( hh_{\alpha \beta }-\frac{3}{2}%
h_{\alpha \sigma }h_{\beta }^{\sigma }\right)  \notag \\
&&-\frac{\mathrm{i}}{16}\bar{\psi}_{\beta }\gamma ^{\beta \mu \alpha
}\psi ^{\nu }\left( \left( h\delta _{\nu }^{\rho }-\frac{1}{2}h_{\nu
}^{\rho }\right) \partial _{[\mu }h_{\alpha ]\rho }+h_{\mu }^{\rho
}\left( 3\partial _{\alpha }h_{\nu \rho }-2\partial _{\rho
}h_{\alpha \nu }\right) \right)
\notag \\
&&+\frac{h}{2}V+\frac{d_{1}}{2}\bar{\psi}_{\rho }\gamma ^{\rho }\left(
h\partial _{\mu }\psi ^{\mu }-\left( \partial _{\mu }\psi _{\nu }\right)
h^{\mu \nu }-\sigma ^{\mu \nu }\psi ^{\sigma }\partial _{[\nu }h_{\sigma
]\mu }-\right.  \notag \\
&&\left. -\frac{1}{4}\sigma ^{\mu \nu }\gamma ^{\alpha \beta }\psi _{\nu
}\partial _{[\alpha }h_{\beta ]\mu }\right) +\frac{d_{2}}{2}\bar{\psi}_{\rho
}\left( h\gamma ^{\mu }\partial _{\mu }\psi ^{\rho }-h^{\mu \nu }\gamma
_{\mu }\partial _{\nu }\psi ^{\rho }\right.  \notag \\
&&\left. -\gamma _{\mu }\psi _{\lambda }\partial ^{[\rho }h^{\lambda ]\mu }-%
\frac{1}{4}\gamma ^{\mu }\gamma ^{\alpha \beta }\psi ^{\rho }\partial
_{[\alpha }h_{\beta ]\mu }\right) +\frac{d_{2}}{2}\left( h\bar{\psi}^{\rho
}\gamma ^{\mu }\partial _{\rho }\psi _{\mu }\right.  \notag \\
&&\left. \left. -h^{\mu \nu }\bar{\psi}_{\mu }\gamma ^{\rho }\partial _{\nu
}\psi _{\rho }-\bar{\psi}^{\mu }\gamma ^{\nu }\psi ^{\rho }\partial _{[\nu
}h_{\rho ]\mu }-\frac{1}{4}\bar{\psi}^{\mu }\gamma ^{\nu }\gamma ^{\alpha
\beta }\psi _{\nu }\partial _{[\alpha }h_{\beta ]\mu }\right) \right) .
\label{deford2}
\end{eqnarray}
This ends the second step of the deformation procedure for the Pauli-Fierz
field and the massive Rarita-Schwinger field.

\section{Lagrangian formulation of the interacting theory}

The main aim of this section is to give an appropriate interpretation of the
Lagrangian formulation of the interacting theory obtained in the previous
section from the deformation of the solution to the master equation. In view
of this, we initially prove that the linearized versions of first- and
second-order formulations of spin-two field theory possess isomorphic local
BRST cohomologies. We start from the first-order formulation of spin-two
field theory
\begin{eqnarray}
S\left[ e_{a}^{\;\;\mu },\omega _{\mu ab}\right] &=&-\frac{1}{\lambda }\int
d^{4}x\left( \omega _{\nu }^{\;\;ab}\partial _{\mu }\left( ee_{a}^{\;\;\mu
}e_{b}^{\;\;\nu }\right) -\omega _{\mu }^{\;\;ab}\partial _{\nu }\left(
ee_{a}^{\;\;\mu }e_{b}^{\;\;\nu }\right) \right.  \notag \\
&&\left. +\frac{1}{2}ee_{a}^{\;\;\mu }e_{b}^{\;\;\nu }\left( \omega _{\mu
}^{\;\;ac}\omega _{\nu \;\;\;c}^{\;\;b}-\omega _{\nu }^{\;\;ac}\omega _{\mu
\;\;\;c}^{\;\;b}\right) \right) ,  \label{xx1}
\end{eqnarray}%
where $e_{a}^{\;\;\mu }$ is the vierbein field and $\omega _{\mu ab}$ are
the components of the spin connection, while $e$ is the inverse of the
vierbein determinant
\begin{equation}
e=\left( \det \left( e_{a}^{\;\;\mu }\right) \right) ^{-1}.  \label{xx2}
\end{equation}%
In order to linearize action (\ref{xx1}), we develop the vierbein like
\begin{equation}
e_{a}^{\;\;\mu }=\delta _{a}^{\;\;\mu }-\frac{\lambda }{2}f_{a}^{\;\;\mu
},\quad e=1+\frac{\lambda }{2}f,  \label{xx3}
\end{equation}%
where $f$ is the trace of $f_{a}^{\;\;\mu }$. Consequently, we find that the
linearized form of (\ref{xx1}) reads as (we come back to the notations $\mu $%
, $\nu $, etc. for flat indices)
\begin{eqnarray}
S_{0}^{\prime }\left[ f_{\mu \nu },\omega _{\mu \alpha \beta }\right]
&=&\int d^{4}x\left( \omega _{\alpha }^{\;\;\alpha \mu }\left( \partial
_{\mu }f-\partial ^{\nu }f_{\mu \nu }\right) +\frac{1}{2}\omega ^{\mu \alpha
\beta }\partial _{\left[ \alpha \right. }f_{\left. \beta \right] \mu }\right.
\notag \\
&&\left. -\frac{1}{2}\left( \omega _{\alpha }^{\;\;\alpha \beta }\omega
_{\;\;\lambda \beta }^{\lambda }-\omega ^{\mu \alpha \beta }\omega _{\alpha
\mu \beta }\right) \right) .  \label{xx4}
\end{eqnarray}%
We mention that the field $f_{\mu \nu }$ contains a symmetric, as well as an
antisymmetric part. The above linearized action is invariant under the gauge
transformations
\begin{equation}
\delta _{\epsilon }f_{\mu \nu }=\partial _{\mu }\epsilon _{\nu }-\epsilon
_{\mu \nu },\quad \delta _{\epsilon }\omega _{\mu \alpha \beta }=-\partial
_{\mu }\epsilon _{\alpha \beta ,}  \label{xy1}
\end{equation}%
where the latter gauge parameters are antisymmetric, $\epsilon _{\alpha
\beta }=-\epsilon _{\beta \alpha }$. Eliminating the spin connection
components on their equations of motion (auxiliary fields) from (\ref{xx4})
\begin{equation}
\omega _{\mu \alpha \beta }\left( f\right) =\frac{1}{2}\left( \partial _{%
\left[ \mu \right. }f_{\left. \alpha \right] \beta }-\partial _{\left[ \mu
\right. }f_{\left. \beta \right] \alpha }-\partial _{\left[ \alpha \right.
}f_{\left. \beta \right] \mu }\right) ,  \label{xx5}
\end{equation}%
we obtain the second-order action
\begin{eqnarray}
&&S_{0}^{\prime }\left[ f_{\mu \nu },\omega _{\mu \alpha \beta }\left(
f\right) \right] =S_{0}^{\prime \prime }\left[ f_{\mu \nu }\right] =-\int
d^{4}x\left( \frac{1}{8}\left( \partial ^{\left[ \mu \right. }f^{\left. \nu %
\right] \alpha }\right) \left( \partial _{\left[ \mu \right. }f_{\left. \nu %
\right] \alpha }\right) \right.  \notag \\
&&\left. +\frac{1}{4}\left( \partial ^{\left[ \mu \right. }f^{\left. \nu %
\right] \alpha }\right) \left( \partial _{\left[ \mu \right. }f_{\left.
\alpha \right] \nu }\right) -\frac{1}{2}\left( \partial _{\mu }f-\partial
^{\nu }f_{\mu \nu }\right) \left( \partial ^{\mu }f-\partial _{\alpha
}f^{\mu \alpha }\right) \right) ,  \label{xx6}
\end{eqnarray}%
subject to the gauge invariances
\begin{equation}
\delta _{\epsilon }f_{\mu \nu }=\partial _{\left( \mu \right. }\epsilon
_{\left. \nu \right) }-\epsilon _{\mu \nu }.  \label{xx7}
\end{equation}%
If we decompose $f_{\mu \nu }$ in its symmetric and antisymmetric parts
\begin{equation}
f_{\mu \nu }=h_{\mu \nu }+B_{\mu \nu },\quad h_{\mu \nu }=h_{\nu \mu },\quad
B_{\mu \nu }=-B_{\nu \mu },  \label{xx8}
\end{equation}%
the action (\ref{xx6}) becomes
\begin{eqnarray}
S_{0}^{\prime \prime }\left[ f_{\mu \nu }\right] =S_{0}^{\prime \prime }%
\left[ h_{\mu \nu },B_{\mu \nu }\right] &=&\int d^{4}x\left( -\frac{1}{2}%
\left( \partial _{\mu }h_{\nu \rho }\right) \left( \partial ^{\mu }h^{\nu
\rho }\right) +\left( \partial _{\mu }h^{\mu \rho }\right) \left( \partial
^{\nu }h_{\nu \rho }\right) \right.  \notag \\
&&\left. -\left( \partial _{\mu }h\right) \left( \partial _{\nu }h^{\nu \mu
}\right) +\frac{1}{2}\left( \partial _{\mu }h\right) \left( \partial ^{\mu
}h\right) \right) ,  \label{xx9}
\end{eqnarray}%
while the accompanying gauge transformations are given by
\begin{equation}
\delta _{\epsilon }h_{\mu \nu }=\partial _{\left( \mu \right. }\epsilon
_{\left. \nu \right) },\quad \delta _{\epsilon }B_{\mu \nu }=-\epsilon _{\mu
\nu }.  \label{xx10}
\end{equation}%
It is easy to see that the right-hand side of (219) is nothing but
the Pauli-Fierz action
\begin{equation}
S_{0}^{\prime \prime }\left[ h_{\mu \nu },B_{\mu \nu }\right]
=S_{0}^{\mathrm{PF}}\left[ h_{\mu \nu }\right] . \label{xx11}
\end{equation}%
As we have previously mentioned, we pass from (\ref{xx4})--(\ref{xy1}) to (%
\ref{xx9})--(\ref{xx10}) via the elimination of the auxiliary fields $\omega
_{\mu \alpha \beta }$, such that the general theorems from Section 15 of the
first reference in~\cite{gen1} ensure the isomorphism
\begin{equation}
H\left( s^{\prime }|d\right) \simeq H\left( s^{\prime \prime }|d\right) ,
\label{xx12}
\end{equation}%
with $s^{\prime }$ and $s^{\prime \prime }$ the BRST differentials
corresponding to (\ref{xx4})--(\ref{xy1}) and respectively to (\ref{xx9})--(%
\ref{xx10}). On the other hand, we observe that the field $B_{\mu
\nu }$ does not appear in (\ref{xx9}) and is subject to a shift
gauge symmetry. Thus, in any cohomological class from $H\left(
s^{\prime \prime }|d\right) $ one can take a representative that is
independent of $B_{\mu \nu }$, the shift ghosts as well as of their
antifields. This is because these variables form contractible pairs
that drop out from $H\left( s^{\prime \prime }|d\right) $ (see the
general results from Section 14 of the first reference
in~\cite{gen1}). As a consequence, we have that
\begin{equation}
H\left( s^{\prime \prime }|d\right) \simeq H\left( s|d\right) ,  \label{xx13}
\end{equation}%
where $s$ is the Pauli-Fierz BRST differential. Combining (\ref{xx12}) and (%
\ref{xx13}), we arrive at
\begin{equation}
H\left( s^{\prime }|d\right) \simeq H\left( s^{\prime \prime }|d\right)
\simeq H\left( s|d\right) .  \label{xx14}
\end{equation}%
Because the local BRST cohomology (in ghost number equal to zero and one)
controls the deformation procedure, it results that the last isomorphisms
allow one to pass in a consistent manner from the Pauli-Fierz version to the
first- and second-order ones (in vierbein formulation) during the
deformation procedure.

It is easy to see that one can go from (\ref{xx9})--(\ref{xx10}) to the
Pauli-Fierz version through the partial gauge-fixing $B_{\mu \nu }=0$. This
gauge-fixing is a consequence of the more general gauge-fixing condition~%
\cite{siegelfields}
\begin{equation}
\sigma _{\mu [a}e_{b]}^{\;\;\mu }=0.  \label{xx15}
\end{equation}
In the context of the larger partial gauge-fixing (\ref{xx15}) simple
computation leads to the vierbein fields $e_{a}^{\;\;\mu }$, their inverse $%
e_{\;\;\mu }^{a}$, the inverse of their determinant $e$, and the
components of the spin connection $\omega _{\mu ab}$ up to the
second order in the coupling constant in terms of the Pauli-Fierz
field as
\begin{eqnarray}
e_{a}^{\;\;\mu } &=&\overset{(0)}{e}_{a}^{\;\;\mu }+\lambda \overset{(1)}{e}%
_{a}^{\;\;\mu }+\lambda ^{2}\overset{(2)}{e}_{a}^{\;\;\mu }+\cdots =\delta
_{a}^{\;\;\mu }-\frac{\lambda }{2}h_{a}^{\;\;\mu }+\frac{3\lambda ^{2}}{8}%
h_{a}^{\;\;\rho }h_{\rho }^{\;\;\mu }+\cdots ,  \label{id1} \\
e_{\;\;\mu }^{a} &=&\overset{(0)}{e}_{\;\;\mu }^{a}+\lambda \overset{(1)}{e}%
_{\;\;\mu }^{a}+\lambda ^{2}\overset{(2)}{e}_{\;\;\mu }^{a}+\cdots =\delta
_{\;\;\mu }^{a}+\frac{\lambda }{2}h_{\;\;\mu }^{a}-\frac{\lambda ^{2}}{8}%
h_{\;\;\rho }^{a}h_{\;\;\mu }^{\rho }+\cdots ,  \label{id1b} \\
e &=&\overset{(0)}{e}+\lambda \overset{(1)}{e}+\lambda ^{2}\overset{(2)}{e}%
+\cdots =1+\frac{\lambda }{2}h+\frac{\lambda ^{2}}{8}\left( h^{2}-2h_{\mu
\nu }h^{\mu \nu }\right) +\cdots ,  \label{id2} \\
\omega _{\mu ab} &=&\lambda \overset{\left( 1\right) }{\omega }_{\mu
ab}+\lambda ^{2}\overset{\left( 2\right) }{\omega }_{\mu ab}+\cdots ,
\label{uv2a}
\end{eqnarray}
where
\begin{equation}
\overset{\left( 1\right) }{\omega }_{\mu ab}=-\partial _{[a}h_{b]\mu },
\label{uv4}
\end{equation}
\begin{equation}
\overset{\left( 2\right) }{\omega }_{\mu ab}=-\frac{1}{4}\left(
2h_{c[a}\left( \partial _{b]}h_{\;\;\mu }^{c}\right) -2h_{\left[ a\right.
}^{\;\;\;\nu }\partial _{\nu }h_{\left. b\right] \mu }-\left( \partial _{\mu
}h_{[a}^{\;\;\;\nu }\right) h_{b]\nu }\right) .  \label{uv5}
\end{equation}
Based on the isomorphisms (\ref{xx14}), we can further pass to the analysis
of the deformed theory obtained in the previous sections.

The component of antighost number equal to zero in $S_{1}^{\left( \mathrm{int%
}\right) }$ is precisely the interacting Lagrangian at order one in the
coupling constant $\mathcal{L}_{1}^{(\mathrm{int})}=a_{0}^{\left( \mathrm{int%
}\right) }+a_{0}^{\left( \mathrm{RS}\right) }$%
\begin{eqnarray}
&&\mathcal{L}_{1}^{(\mathrm{int})}=\left[ \frac{1}{4}\bar{\psi}_{\mu }\left(
-\mathrm{i}\gamma ^{\mu \nu \rho }\partial _{\nu }\psi _{\rho }+m\gamma
^{\mu \nu }\psi _{\nu }\right) h\right] +\left[ \frac{\mathrm{i}}{4}\bar{\psi%
}_{\mu }\gamma ^{\mu \nu \rho }\left( \partial ^{\lambda }\psi _{\rho
}\right) h_{\nu \lambda }\right]  \notag \\
&&+\left[ \frac{\mathrm{i}}{4}\bar{\psi}_{\mu }\gamma ^{\mu \nu \rho }\left(
\partial _{\nu }\psi ^{\lambda }\right) h_{\rho \lambda }\right] +\left[
\frac{\mathrm{i}}{8}\left( \bar{\psi}^{\mu }\gamma ^{\lambda }\psi ^{\nu
}-2\sigma ^{\nu \lambda }\bar{\psi}^{\mu }\gamma ^{\rho }\psi _{\rho
}\right) \partial _{\lbrack \mu }h_{\nu ]\lambda }\right]  \notag \\
&&+\left[ -\frac{\mathrm{i}}{8}\left( 2\bar{\psi}_{\mu }\gamma ^{\mu \nu
\rho }\left( \partial _{\nu }\psi ^{\lambda }\right) h_{\rho \lambda }+\bar{%
\psi}_{\rho }\gamma ^{\rho \mu \nu }\psi ^{\lambda }\partial _{\lbrack \mu
}h_{\nu ]\lambda }\right) \right] +\left[ V\right] +\left[ d_{1}\bar{\psi}%
_{\rho }\gamma ^{\rho }\partial _{\mu }\psi ^{\mu }\right]  \notag \\
&&+\left[ d_{2}\bar{\psi}^{(\mu }\gamma ^{\nu )}\partial _{\mu }\psi _{\nu }%
\right] \equiv \overset{(1)}{e}\mathcal{L}_{0}^{(\mathrm{RS})}+\overset{(0)}{%
e}\overset{(1)}{e}_{b}^{\;\;\mu }\overset{(0)}{e}_{c}^{\;\;\nu }\left( -%
\frac{\mathrm{i}}{2}\bar{\psi}_{a}\gamma ^{abc}\overset{(0)}{D}_{\mu }%
\overset{(0)}{\psi }_{\nu }\right)  \notag \\
&&+\overset{(0)}{e}_{b}^{\;\;\mu }\overset{(1)}{e}_{c}^{\;\;\nu }\left( -%
\frac{\mathrm{i}}{2}\bar{\psi}_{a}\gamma ^{abc}\overset{(0)}{D}_{\mu }%
\overset{(0)}{\psi }_{\nu }\right) +\overset{(0)}{e}\overset{(0)}{e}%
_{b}^{\;\;\mu }\overset{(0)}{e}_{c}^{\;\;\nu }\left( -\frac{\mathrm{i}}{2}%
\bar{\psi}_{a}\gamma ^{abc}\overset{(1)}{D}_{\mu }\overset{(0)}{\psi }_{\nu
}\right)  \notag \\
&&+\overset{(0)}{e}\overset{(0)}{e}_{b}^{\;\;\mu }\overset{(0)}{e}%
_{c}^{\;\;\nu }\left( -\frac{\mathrm{i}}{2}\bar{\psi}_{a}\gamma ^{abc}%
\overset{(0)}{D}_{\mu }\overset{(1)}{\psi }_{\nu }\right) +\overset{(0)}{e}%
V+d_{1}\bar{\psi}_{a}\gamma ^{a}\overset{(0)}{D}_{\mu }\left( \overset{(0)}{e%
}\overset{(0)}{\psi }^{\mu }\right)  \notag \\
&&+d_{2}\overset{(0)}{e}\bar{\psi}^{(a}\gamma ^{b)}\overset{(0)}{e}%
_{b}^{\;\;\mu }\overset{(0)}{e}_{c}^{\;\;\nu }\overset{(0)}{D}_{\mu }\left(
\overset{(0)}{e}\overset{(0)}{\psi }_{\nu }\right) ,  \label{gama7}
\end{eqnarray}%
where
\begin{equation}
\overset{(0)}{D}_{\mu }=\partial _{\mu },  \label{xx19}
\end{equation}%
and
\begin{equation}
\overset{(1)}{D}_{\mu }=\frac{1}{8}\overset{\left( 1\right) }{\omega }_{\mu
ab}\gamma ^{ab},  \label{uv1}
\end{equation}%
with $\overset{\left( 1\right) }{\omega }_{\mu ab}$ given in (\ref{uv4}).
Along the same line, the piece of antighost number equal to zero from the
second-order deformation offers us the interacting Lagrangian at order two
in the coupling constant $\mathcal{L}_{2}^{(\mathrm{int})}=b_{0}^{\left(
\mathrm{int}\right) }$%
\begin{eqnarray}
\mathcal{L}_{2}^{(\mathrm{int})} &=&b_{0}^{\left( \mathrm{int}\right) }=%
\left[ \frac{1}{16}\bar{\psi}_{\mu }\left( -\mathrm{i}\gamma ^{\mu \nu \rho
}\partial _{\nu }\psi _{\rho }+m\gamma ^{\mu \nu }\psi _{\nu }\right) \left(
h^{2}-2h_{\alpha \beta }h^{\alpha \beta }\right) \right]  \notag \\
&&+\left[ \frac{\mathrm{i}}{8}\bar{\psi}_{\mu }\left( \gamma ^{\mu \alpha
\nu }\left( \partial ^{\beta }\psi _{\nu }\right) h_{\alpha \beta }+\gamma
^{\mu \nu \rho }\left( \partial _{\nu }\psi ^{\lambda }\right) h_{\rho
\lambda }\right) h\right]  \notag \\
&&+\left[ \frac{\mathrm{i}h}{16}\left( -\bar{\psi}_{\mu }\gamma ^{\mu \nu
\rho }\left( 2\left( \partial _{\nu }\psi ^{\lambda }\right) h_{\rho \lambda
}+\psi ^{\lambda }\partial _{\left[ \nu \right. }h_{\left. \rho \right]
\lambda }\right) \right. \right.  \notag \\
&&\left. \left. +\left( \bar{\psi}^{\alpha }\gamma ^{\rho }\psi ^{\beta
}-2\sigma ^{\beta \rho }\bar{\psi}^{\alpha }\gamma ^{\mu }\psi _{\mu
}\right) \partial _{\lbrack \alpha }h_{\beta ]\rho }\right) \right]  \notag
\\
&&+\left[ -\frac{\mathrm{i}}{8}\bar{\psi}_{\mu }\gamma ^{\mu \nu \rho
}\left( \partial _{\alpha }\psi _{\beta }\right) h_{\nu }^{\alpha }h_{\rho
}^{\beta }\right]  \notag \\
&&+\left[ \frac{\mathrm{i}}{8}\left( \bar{\psi}_{\alpha }\gamma ^{\alpha
\beta \gamma }\left( h_{\beta }^{\mu }\partial _{\mu }\left( h_{\gamma
}^{\sigma }\psi _{\sigma }\right) +h_{\gamma }^{\mu }\partial _{\beta
}\left( h_{\mu }^{\sigma }\psi _{\sigma }\right) \right) \right. \right.
\notag \\
&&\left. \left. -\frac{1}{2}\left( \bar{\psi}^{\mu }\gamma _{\rho }\psi
^{\nu }h^{\rho \sigma }-2\bar{\psi}^{\mu }\gamma ^{\rho }\psi _{\rho }h^{\nu
\sigma }\right) \partial _{\lbrack \mu }h_{\nu ]\sigma }\right) \right]
\notag \\
&&+\left[ \frac{\mathrm{i}}{8}\left( \bar{\psi}_{\alpha }\gamma ^{\alpha
\beta \gamma }\partial _{\beta }\left( h_{\gamma }^{\mu }h_{\mu }^{\sigma
}\psi _{\sigma }\right) -\frac{1}{2}\bar{\psi}^{\mu }\left( \gamma ^{\rho
}\psi _{\rho }\left( 3h_{\mu \lambda }\partial _{\sigma }h^{\lambda \sigma
}\right. \right. \right. \right.  \notag \\
&&\left. +h^{\lambda \sigma }\partial _{\lambda }h_{\mu \sigma }-2h_{\mu
\sigma }\partial ^{\sigma }h-2h^{\alpha \beta }\partial _{\mu }h_{\alpha
\beta }\right)  \notag \\
&&\left. \left. \left. -\gamma ^{\lambda }\psi ^{\nu }\left( 2h_{\rho \mu
}\partial _{\nu }h_{\lambda }^{\rho }-2h_{\mu }^{\rho }\partial _{\rho
}h_{\nu \lambda }-h_{\nu \rho }\partial _{\lambda }h_{\mu }^{\rho }\right)
\right) \right) \right]  \notag \\
&&+\left[ \frac{3\mathrm{i}}{16}\bar{\psi}_{\mu }\gamma ^{\mu \nu \beta
}\left( \partial ^{\alpha }\psi _{\nu }\right) h_{\alpha \sigma }h_{\beta
}^{\sigma }\right] +\left[ -\frac{3\mathrm{i}}{16}\bar{\psi}_{\mu }\gamma
^{\mu \nu \rho }\left( \partial _{\nu }\psi _{\lambda }\right) h_{\rho
\sigma }h^{\sigma \lambda }\right]  \notag \\
&&+\left[ \frac{h}{2}V\right] +\left[ d_{1}\bar{\psi}_{\rho }\gamma ^{\rho
}\partial _{\mu }\left( \frac{h}{2}\psi ^{\mu }\right) \right] +\left[ -%
\frac{d_{1}}{2}\bar{\psi}_{\rho }\gamma ^{\rho }\partial _{\mu }\left( \psi
_{\nu }h^{\mu \nu }\right) \right]  \notag \\
&&+\left[ -\frac{d_{1}}{8}\bar{\psi}_{\rho }\gamma ^{\rho }\gamma ^{\alpha
\beta }\psi ^{\mu }\partial _{\lbrack \alpha }h_{\beta ]\mu }\right] +\left[
\frac{d_{2}}{2}h\bar{\psi}^{(\mu }\gamma ^{\nu )}\partial _{\mu }\psi _{\nu }%
\right]  \notag \\
&&+\left[ -\frac{d_{2}}{2}h_{\alpha }^{\mu }\bar{\psi}^{(\alpha }\gamma
^{\nu )}\partial _{\mu }\psi _{\nu }\right]  \notag \\
&&+\left[ -d_{2}\bar{\psi}^{(\mu }\gamma ^{\nu )}\left( \frac{1}{2}\psi
^{\rho }\partial _{\lbrack \nu }h_{\rho ]\mu }+\frac{1}{8}\gamma ^{\alpha
\beta }\psi _{\nu }\partial _{\lbrack \alpha }h_{\beta ]\mu }\right) \right]
\equiv \left[ \overset{(2)}{e}\mathcal{L}_{0}^{(\mathrm{RS})}\right]  \notag
\\
&&+\left[ \overset{(1)}{e}\left( \overset{(0)}{e}_{b}^{\;\;\mu }\overset{(1)}%
{e}_{c}^{\;\;\nu }+\overset{(1)}{e}_{b}^{\;\;\mu }\overset{(0)}{e}%
_{c}^{\;\;\nu }\right) \left( -\frac{\mathrm{i}}{2}\bar{\psi}_{a}\gamma
^{abc}\overset{(0)}{D}_{\mu }\overset{(0)}{\psi }_{\nu }\right) \right]
\notag \\
&&+\left[ \overset{(1)}{e}\overset{(0)}{e}_{b}^{\;\;\mu }\overset{(0)}{e}%
_{c}^{\;\;\nu }\left( -\frac{\mathrm{i}}{2}\bar{\psi}_{a}\gamma ^{abc}\left(
\overset{(0)}{D}_{\mu }\overset{(1)}{\psi }_{\nu }+\overset{(1)}{D}_{\mu }%
\overset{(0)}{\psi }_{\nu }\right) \right) \right]  \notag \\
&&+\left[ \overset{(0)}{e}\overset{(1)}{e}_{b}^{\;\;\mu }\overset{(1)}{e}%
_{c}^{\;\;\nu }\left( -\frac{\mathrm{i}}{2}\bar{\psi}_{a}\gamma ^{abc}%
\overset{(0)}{D}_{\mu }\overset{(0)}{\psi }_{\nu }\right) \right]  \notag \\
&&+\left[ \overset{(0)}{e}\overset{(1)}{e}_{b}^{\;\;\mu }\overset{(0)}{e}%
_{c}^{\;\;\nu }\left( -\frac{\mathrm{i}}{2}\bar{\psi}_{a}\gamma ^{abc}\left(
\overset{(0)}{D}_{\mu }\overset{(1)}{\psi }_{\nu }+\overset{(1)}{D}_{\mu }%
\overset{(0)}{\psi }_{\nu }\right) \right) \right.  \notag \\
&&\left. +\overset{(0)}{e}\overset{(0)}{e}_{b}^{\;\;\mu }\overset{(1)}{e}%
_{c}^{\;\;\nu }\left( -\frac{\mathrm{i}}{2}\bar{\psi}_{a}\gamma ^{abc}\left(
\overset{(0)}{D}_{\mu }\overset{(1)}{\psi }_{\nu }+\overset{(1)}{D}_{\mu }%
\overset{(0)}{\psi }_{\nu }\right) \right) \right]  \notag \\
&&+\left[ \overset{(0)}{e}\overset{(0)}{e}_{b}^{\;\;\mu }\overset{(0)}{e}%
_{c}^{\;\;\nu }\left( -\frac{\mathrm{i}}{2}\bar{\psi}_{a}\gamma ^{abc}\left(
\overset{(0)}{D}_{\mu }\overset{(2)}{\psi }_{\nu }+\overset{(1)}{D}_{\mu }%
\overset{(1)}{\psi }_{\nu }\right. \right. \right.  \notag \\
&&\left. \left. \left. +\overset{(2)}{D}_{\mu }\overset{(0)}{\psi }_{\nu
}\right) \right) \right] +\left[ \overset{(0)}{e}\overset{(2)}{e}%
_{b}^{\;\;\mu }\overset{(0)}{e}_{c}^{\;\;\nu }\left( -\frac{\mathrm{i}}{2}%
\bar{\psi}_{a}\gamma ^{abc}\overset{(0)}{D}_{\mu }\overset{(0)}{\psi }_{\nu
}\right) \right]  \notag \\
&&+\left[ \overset{(0)}{e}\overset{(0)}{e}_{b}^{\;\;\mu }\overset{(2)}{e}%
_{c}^{\;\;\nu }\left( -\frac{\mathrm{i}}{2}\bar{\psi}_{a}\gamma ^{abc}%
\overset{(0)}{D}_{\mu }\overset{(0)}{\psi }_{\nu }\right) \right] +\left[
\overset{(1)}{e}V\right]  \notag \\
&&+\left[ d_{1}\bar{\psi}_{a}\gamma ^{a}\overset{(0)}{D}_{\mu }\left(
\overset{(1)}{e}\overset{(0)}{\psi }^{\mu }\right) \right] +\left[ d_{1}\bar{%
\psi}_{a}\gamma ^{a}\overset{(0)}{D}_{\mu }\left( \overset{(0)}{e}\overset{%
(1)}{\psi }^{\mu }\right) \right]  \notag \\
&&+\left[ d_{1}\bar{\psi}_{a}\gamma ^{a}\overset{(1)}{D}_{\mu }\left(
\overset{(0)}{e}\overset{(0)}{\psi }^{\mu }\right) \right] +\left[ d_{2}%
\overset{(1)}{e}\overset{(0)}{e}_{a}^{\;\;\mu }\bar{\psi}^{(a}\gamma ^{b)}%
\overset{(0)}{D}_{\mu }\psi _{b}\right]  \notag \\
&&+\left[ d_{2}\overset{(0)}{e}\overset{(1)}{e}_{a}^{\;\;\mu }\bar{\psi}%
^{(a}\gamma ^{b)}\overset{(0)}{D}_{\mu }\psi _{b}\right] +\left[ d_{2}%
\overset{(0)}{e}\overset{(0)}{e}_{a}^{\;\;\mu }\bar{\psi}^{(a}\gamma ^{b)}%
\overset{(1)}{D}_{\mu }\psi _{b}\right] ,  \label{zwq1}
\end{eqnarray}%
where
\begin{equation}
\overset{(2)}{D}_{\mu }=\frac{1}{8}\overset{\left( 2\right) }{\omega }_{\mu
ab}\gamma ^{ab}  \label{uv6}
\end{equation}%
and $\overset{\left( 2\right) }{\omega }_{\mu ab}$ like in (\ref{uv5}). With
the help of (\ref{id1}) and (\ref{id2}) we deduce that $\mathcal{L}_{0}^{(%
\mathrm{RS})}+\lambda \mathcal{L}_{1}^{(\mathrm{int})}+\lambda ^{2}\mathcal{L%
}_{2}^{(\mathrm{int})}+\cdots $ comes from expanding the fully deformed
Lagrangian written in terms of either the original flat Rarita-Schwinger
spinor $\psi _{a}$%
\begin{eqnarray}
&&\mathcal{L}^{\left( \mathrm{int}\right) }=\frac{e}{2}\left( -\mathrm{i}%
\bar{\psi}_{a}e_{b}^{\;\;\nu }e_{c}^{\;\;\rho }\gamma ^{abc}D_{\nu }\left(
e_{\;\;\rho }^{d}\psi _{d}\right) +m\bar{\psi}_{a}\gamma ^{ab}\psi
_{b}\right)  \notag \\
&&+\lambda \left[ eV\left( X,Y,Z\right) +d_{1}\left( X,Y,Z\right) \bar{\psi}%
_{a}\gamma ^{a}D_{\mu }\left( ee_{b}^{\;\;\mu }\psi ^{b}\right) \right.
\notag \\
&&\left. +ed_{2}\left( X,Y,Z\right) e_{a}^{\;\;\mu }\bar{\psi}^{(a}\gamma
^{b)}D_{\mu }\psi _{b}\right]  \label{PFRS5.1b}
\end{eqnarray}%
or the curved Rarita-Schwinger spinor $\psi _{\mu }$%
\begin{eqnarray}
&&\mathcal{L}^{\left( \mathrm{int}\right) }=\frac{e}{2}\left( -\mathrm{i}%
\bar{\psi}_{\mu }e_{a}^{\;\;\mu }e_{b}^{\;\;\nu }e_{c}^{\;\;\rho }\gamma
^{abc}D_{\nu }\psi _{\rho }+m\bar{\psi}_{\mu }e_{a}^{\;\;\mu }\gamma
^{ab}e_{b}^{\;\;\nu }\psi _{\nu }\right)  \notag \\
&&+\lambda \left[ eV\left( X,Y,Z\right) +d_{1}\left( X,Y,Z\right)
e_{a}^{\;\;\nu }\bar{\psi}_{\nu }\gamma ^{a}D_{\mu }\left( e\psi ^{\mu
}\right) \right.  \notag \\
&&\left. +ed_{2}\left( X,Y,Z\right) \left( \bar{\psi}^{\mu }\gamma
^{b}+e_{a}^{\;\;\mu }e_{\;\;\rho }^{b}\bar{\psi}^{\rho }\gamma ^{a}\right)
D_{\mu }\left( e_{b}^{\;\;\nu }\psi _{\nu }\right) \right] .
\label{PFRS5.1a}
\end{eqnarray}%
The notations $D_{\mu }\psi _{a}$ and $D_{\mu }\psi _{\rho }$ denote the
full covariant derivatives of $\psi _{a}$ and respectively of $\psi _{\rho }$%
\begin{eqnarray}
D_{\mu }\psi _{a} &=&\partial _{\mu }\psi _{a}+\frac{1}{2}\omega
_{\mu ab}\psi ^{b}+\frac{1}{8}\gamma ^{bc}\psi _{a}\omega _{\mu bc},
\label{xx19a} \\
D_{\mu }\psi _{\rho } &=&\partial _{\mu }\psi _{\rho }+\frac{1}{8}\omega
_{\mu ab}\gamma ^{ab}\psi _{\rho }.  \label{xx20}
\end{eqnarray}

The pieces linear in the antifields $\psi _{\mu }^{\ast }$ from the deformed
solution to the master equation give us the deformed gauge transformations
for the Rarita-Schwinger fields as
\begin{eqnarray}
&&\delta _{\epsilon }\psi _{\mu }=\lambda \left( \left( \partial ^{\alpha
}\psi _{\mu }\right) \epsilon _{\alpha }+\frac{1}{2}\psi ^{\nu }\partial
_{\lbrack \mu }\epsilon _{\nu ]}+\frac{1}{8}\gamma ^{\alpha \beta }\psi
_{\mu }\partial _{\lbrack \alpha }\epsilon _{\beta ]}\right)   \notag \\
&&+\lambda ^{2}\left( -\frac{1}{2}\left( \partial _{\alpha }\psi _{\mu
}\right) \epsilon _{\beta }h^{\alpha \beta }+\frac{1}{16}\gamma ^{\rho
\lambda }\psi _{\mu }h_{\rho }^{\sigma }\partial _{\lbrack \lambda }\epsilon
_{\sigma ]}\right.   \notag \\
&&+\frac{1}{8}\psi ^{\rho }\left( h_{\mu }^{\lambda }\partial _{\lbrack \rho
}\epsilon _{\lambda ]}-h_{\rho }^{\lambda }\partial _{\lbrack \mu }\epsilon
_{\lambda ]}\right) -\frac{1}{4}\psi ^{\nu }\epsilon ^{\rho }\partial
_{\lbrack \mu }h_{\nu ]\rho }  \notag \\
&&\left. -\frac{1}{16}\gamma ^{\alpha \beta }\psi _{\mu }\epsilon ^{\rho
}\partial _{\lbrack \alpha }h_{\beta ]\rho }\right)   \notag \\
&=&\lambda \overset{(1)}{\delta }_{\epsilon }\psi _{\mu }+\lambda ^{2}%
\overset{(2)}{\delta }_{\epsilon }\psi _{\mu }+\cdots .  \label{gama8}
\end{eqnarray}%
The first two orders of the gauge transformations can be put under
the form
\begin{eqnarray}
\overset{(1)}{\delta }_{\epsilon }\psi _{m} &=&\left( \partial _{\mu
}\psi
_{m}\right) \overset{(0)}{\bar{\epsilon}}^{\mu }+\frac{1}{2}\overset{(0)}{%
\epsilon }_{mn}\psi ^{n}+\frac{1}{4}\gamma ^{ab}\psi _{m}\overset{(0)}%
{\epsilon }_{ab},  \label{uw1} \\
\overset{(2)}{\delta }_{\epsilon }\psi _{m} &=&\left( \partial _{\mu
}\psi
_{m}\right) \overset{(1)}{\bar{\epsilon}}^{\mu }+\frac{1}{2}\overset{(1)}{%
\epsilon }_{mn}\psi ^{n}+\frac{1}{4}\gamma ^{ab}\psi _{m}\overset{(1)}%
{\epsilon }_{ab},  \label{uw2}
\end{eqnarray}%
where we used the notations
\begin{eqnarray}
\overset{(0)}{\bar{\epsilon}}^{\mu } &=&\epsilon ^{\mu }=\epsilon ^{a}\delta
_{a}^{\;\;\mu },\quad \overset{(1)}{\bar{\epsilon}}^{\mu }=-\frac{1}{2}%
\epsilon ^{a}h_{a}^{\;\;\mu },  \label{uv16} \\
\overset{(0)}{\epsilon }_{ab} &=&\frac{1}{2}\partial _{\lbrack a}\epsilon
_{b]},  \label{apx0} \\
\overset{(1)}{\epsilon }_{ab} &=&-\frac{1}{4}\epsilon ^{c}\partial _{\lbrack
a}h_{b]c}+\frac{1}{8}h_{[a}^{c}\partial _{b]}\epsilon _{c}+\frac{1}{8}\left(
\partial _{c}\epsilon _{\lbrack a}\right) h_{b]}^{c}.  \label{apx1}
\end{eqnarray}%
Based on these notations, the gauge transformations of the spinors take the
form
\begin{eqnarray}
&&\delta _{\epsilon }\psi _{m}=\lambda \left( \left( \partial _{\mu }\psi
_{m}\right) \left( \overset{(0)}{\bar{\epsilon}}^{\mu }+\lambda \overset{(1)}%
{\bar{\epsilon}}^{\mu }+\cdots \right) \right.   \notag \\
&&+\left( \overset{(0)}{\epsilon }_{mn}+\lambda \overset{(1)}{\epsilon }%
_{mn}+\cdots \right) \psi ^{n}  \notag \\
&&\left. +\frac{1}{4}\gamma ^{ab}\psi _{m}\left( \overset{(0)}{\epsilon }%
_{ab}+\lambda \overset{(1)}{\epsilon }_{ab}+\cdots \right) \right) .
\label{uw5}
\end{eqnarray}%
The gauge parameters $\overset{(0)}{\epsilon }_{ab}$ and $\overset{(1)}{%
\epsilon }_{ab}$ are precisely the first two terms from the Lorentz
parameters expressed in terms of the flat parameters $\epsilon ^{a}$ via the
partial gauge-fixing (\ref{xx15}). Indeed, (\ref{xx15}) leads to
\begin{equation}
\bar{\delta}_{\epsilon }\sigma _{\mu \lbrack a}e_{b]}^{\;\;\mu }=0,
\label{xx15a}
\end{equation}%
where
\begin{equation}
\bar{\delta}_{\epsilon }e_{a}^{\;\;\mu }=\bar{\epsilon}^{\rho }\partial
_{\rho }e_{a}^{\;\;\mu }-e_{a}^{\;\;\rho }\partial _{\rho }\bar{\epsilon}%
^{\mu }+\epsilon _{a}^{\;\;b}e_{b}^{\;\;\mu }.  \label{id6}
\end{equation}%
Substituting (\ref{id1}) together with the expansions
\begin{equation}
\bar{\epsilon}^{\mu }=\overset{(0)}{\bar{\epsilon}}^{\mu }+\lambda \overset{%
(1)}{\bar{\epsilon}}^{\mu }+\cdots =\left( \delta _{a}^{\;\;\mu }-\frac{%
\lambda }{2}h_{a}^{\;\;\mu }+\cdots \right) \epsilon ^{a}  \label{uv15}
\end{equation}%
and
\begin{equation}
\epsilon _{ab}=\overset{(0)}{\epsilon }_{ab}+\lambda \overset{(1)}{\epsilon }%
_{ab}+\cdots   \label{uv12}
\end{equation}%
in (\ref{xx15a}), we arrive precisely to (\ref{apx0})--(\ref{apx1}).
At this point it is easy to see that the gauge transformations
(\ref{uw5}) come from the perturbative expansion of the full gauge
transformations
\begin{equation}
\delta _{\epsilon }\psi _{m}=\lambda \left( \left( \partial _{\mu
}\psi
_{m}\right) \bar{\epsilon}^{\mu }+\epsilon _{mn}\psi ^{n}+\frac{1%
}{4}\gamma ^{ab}\psi _{m}\epsilon _{ab}\right) .  \label{full}
\end{equation}%
Moreover, based on (\ref{full}) and (\ref{id6}), it is easy to see
that
\begin{equation}
\delta _{\epsilon }\psi ^{\mu }=\lambda \left( \left( \partial _{\sigma
}\psi ^{\mu }\right) \bar{\epsilon}^{\sigma }-\psi ^{\sigma }\partial
_{\sigma }\bar{\epsilon}^{\mu }+\frac{1}{4}\gamma ^{ab}\psi ^{\mu }\epsilon
_{ab}\right) .  \label{full1}
\end{equation}%
In conclusion, under the above mentioned hypotheses we have shown that the
interactions between a massive Rarita-Schwinger field and a spin-two field
are described by the coupled Lagrangian (\ref{PFRS5.1b}) or (\ref{PFRS5.1a}%
), while the gauge transformations of the Rarita-Schwinger spinors are given
by (\ref{full}) or (\ref{full1}). If we require in addition that the
interacting model remains PT-invariant, then the results (\ref{PFRS5.1b})--(%
\ref{PFRS5.1a}) remain valid up to the point that the functions $V$,
$d_{1}$, and $d_{2}$ must depend only on $X$ and $Y$ (and not on
$Z$).

\section{Impossibility of cross-interactions between gravitons in the
presence of the massive Rarita-Schwinger field}

As it has been proved in~\cite{multi}, there are no direct cross-couplings
that can be introduced among a finite number of gravitons and also no
intermediate cross-couplings between different gravitons in the presence of
a scalar field. In this section, under the hypotheses of locality,
smoothness of the interactions in the coupling constant, Poincar\'{e}
invariance, Lorentz covariance, and the preservation of the number of
derivatives on each field, we will prove that there are no intermediate
cross-couplings between different gravitons intermediated by a massive spin-$%
3/2$ field. In order to ensure the stability of the Minkowski vacuum
(absence of negative-energy excitations or of negative-norm states) we
assume in addition that the metric in internal space in positively defined.
It is always possible to bring the internal metric to the form $\delta _{AB}$
by a linear redefinition of the Pauli-Fierz fields. This is the convention
we will work with in the sequel.

In view of this we start from a finite sum of Pauli-Fierz actions and a
massive Rarita-Schwinger action
\begin{eqnarray}
S_{0}^{\mathrm{L}}\left[ h_{\mu \nu }^{A},\psi _{\mu }\right] &=&\int
d^{4}x\left( -\frac{1}{2}\left( \partial _{\mu }h_{\nu \rho }^{A}\right)
\left( \partial ^{\mu }h_{A}^{\nu \rho }\right) +\left( \partial _{\mu
}h_{A}^{\mu \rho }\right) \left( \partial ^{\nu }h_{\nu \rho }^{A}\right)
\right.  \notag \\
&&\left. -\left( \partial _{\mu }h^{A}\right) \left( \partial _{\nu
}h_{A}^{\nu \mu }\right) +\frac{1}{2}\left( \partial _{\mu }h^{A}\right)
\left( \partial ^{\mu }h_{A}\right) \right)  \notag \\
&&+\int d^{4}x\bar{\psi}\left( -\frac{\mathrm{i}}{2}\bar{\psi}_{\mu }\gamma
^{\mu \nu \rho }\partial _{\nu }\psi _{\rho }+\frac{m}{2}\bar{\psi}_{\mu
}\gamma ^{\mu \nu }\psi _{\nu }\right) ,  \label{vu1}
\end{eqnarray}%
where $h_{A}$ denotes the trace of the field $h_{A}^{\mu \nu }$ ($%
h_{A}=\sigma _{\mu \nu }h_{A}^{\mu \nu }$), with $A$ the collection index,
running from $1$ to $n$. The gauge transformations of the action (\ref{vu1})
read as
\begin{equation}
\delta _{\epsilon }h_{\mu \nu }^{A}=\partial _{(\mu }\epsilon _{\nu
)}^{A},\quad \delta _{\epsilon }\psi _{\mu }=0.  \label{5}
\end{equation}%
The BRST complex comprises the fields/ghosts
\begin{equation}
\phi ^{\alpha _{0}}=\left( h_{\mu \nu }^{A},\psi _{\mu }\right) ,\quad \eta
_{\mu }^{A},  \label{vu2}
\end{equation}%
and respectively their antifields
\begin{equation}
\phi _{\alpha _{0}}^{\ast }=\left( h_{A}^{\ast \mu \nu },\psi ^{\ast \mu
}\right) ,\quad \eta _{A}^{\ast \mu }.  \label{vu3}
\end{equation}%
The BRST differential splits in this situation like in (\ref{PFRS4}), while
the actions of $\delta $ and $\gamma $ on the BRST generators are defined by
\begin{eqnarray}
\delta h_{A}^{\ast \mu \nu } &=&2H_{A}^{\mu \nu },\quad \delta \psi ^{\ast
\mu }=m\bar{\psi}_{\lambda }\gamma ^{\lambda \mu }-\mathrm{i}\partial _{\rho
}\bar{\psi}_{\lambda }\gamma ^{\rho \lambda \mu },  \label{12} \\
\delta \eta _{A}^{\ast \mu } &=&-2\partial _{\nu }h_{A}^{\ast \mu \nu },
\label{13} \\
\delta \phi ^{\alpha _{0}} &=&0,\quad \delta \eta _{\mu }^{A}=0,  \label{14}
\\
\gamma \phi _{\alpha _{0}}^{\ast } &=&0,\quad \gamma \eta _{A}^{\ast \mu }=0,
\label{15} \\
\gamma h_{\mu \nu }^{A} &=&\partial _{(\mu }\eta _{\nu )}^{A},\quad \gamma
\psi _{\mu }=0,\quad \gamma \eta _{\mu }^{A}=0,  \label{16}
\end{eqnarray}%
where $H_{A}^{\mu \nu }=K_{A}^{\mu \nu }-\frac{1}{2}\sigma ^{\mu \nu }K_{A}$
is the linearized Einstein tensor for the field $h_{A}^{\mu \nu }$. In this
case the solution to the master equation reads as
\begin{equation}
\bar{S}=S_{0}^{\mathrm{L}}\left[ h_{\mu \nu }^{A},\psi _{\mu }\right] +\int
d^{4}x\left( h_{A}^{\ast \mu \nu }\partial _{(\mu }\eta _{\nu )}^{A}\right) .
\label{18}
\end{equation}

The first-order deformation of the solution to the master equation may be
decomposed in a manner similar to the case of a single graviton
\begin{equation}
\alpha =\alpha ^{\left( \mathrm{PF}\right) }+\alpha ^{\left( \mathrm{int}%
\right) }+\alpha ^{\left( \mathrm{RS}\right) }.  \label{vux}
\end{equation}%
The first-order deformation in the Pauli-Fierz sector, $\alpha ^{\left(
\mathrm{PF}\right) }$, is of the form~\cite{multi}
\begin{equation}
\alpha ^{\left( \mathrm{PF}\right) }=\alpha _{2}^{\left( \mathrm{PF}\right)
}+\alpha _{1}^{\left( \mathrm{PF}\right) }+\alpha _{0}^{\left( \mathrm{PF}%
\right) },  \label{vx1}
\end{equation}%
with
\begin{equation}
\alpha _{2}^{\left( \mathrm{PF}\right) }=\frac{1}{2}f_{BC}^{A}\eta
_{A}^{\ast \mu }\eta ^{B\nu }\partial _{\lbrack \mu }^{\left. {}\right.
}\eta _{\nu ]}^{C}.  \label{vx2}
\end{equation}%
In (\ref{vx2}) all the coefficients $f_{BC}^{A}$ are constant. The condition
that $\alpha _{2}^{\left( \mathrm{PF}\right) }$ indeed produces a consistent
$\alpha _{1}^{\left( \mathrm{PF}\right) }$ implies that these constants must
be symmetric in their lower indices~\cite{multi}\footnote{%
The term (\ref{vx2}) differs from that corresponding to~\cite{multi} through
a $\gamma $-exact term, which does not affect (\ref{vx3}).}
\begin{equation}
f_{BC}^{A}=f_{CB}^{A}.  \label{vx3}
\end{equation}%
With (\ref{vx3}) at hand, we find that
\begin{equation}
\alpha _{1}^{\left( \mathrm{PF}\right) }=f_{BC}^{A}h_{A}^{\ast \mu \rho
}\left( \left( \partial _{\rho }\eta ^{B\nu }\right) h_{\mu \nu }^{C}-\eta
^{B\nu }\partial _{\lbrack \mu }^{\left. {}\right. }h_{\nu ]\rho
}^{C}\right) .  \label{vxx}
\end{equation}%
The requirement that $\alpha _{1}^{\left( \mathrm{PF}\right) }$ leads to a
consistent $\alpha _{0}^{\left( \mathrm{PF}\right) }$ implies that $f_{ABC}$
must be symmetric~\cite{multi}\footnote{%
The piece (\ref{vxx}) differs from that corresponding to~\cite{multi}
through a $\delta $-exact term, which does not change (\ref{3.47}).}
\begin{equation}
f_{ABC}=\frac{1}{3}f_{\left( ABC\right) },  \label{3.47}
\end{equation}%
where, by definition, $f_{ABC}=\delta _{AD}f_{BC}^{D}$. Based on (\ref{3.47}%
), we obtain that the resulting $\alpha _{0}^{\left( \mathrm{PF}\right) }$
reads as in~\cite{multi} (where this component is denoted by $a_{0}$ and $%
f_{ABC}$ by $a_{abc}$).

If one goes along exactly the same line like in the subsection \ref{firstord}%
, we get that $\alpha ^{\left( \mathrm{int}\right) }=\alpha _{1}^{\left(
\mathrm{int}\right) }+\alpha _{0}^{\left( \mathrm{int}\right) }$, where
\begin{eqnarray}
\alpha _{1}^{(\mathrm{int})} &=&k_{A}\psi ^{*\mu }\left( \partial ^{\nu
}\psi _{\mu }\right) \eta _{\nu }^{A}+\frac{k_{A}}{2}\psi ^{*\mu }\psi ^{\nu
}\partial _{[\mu }\eta _{\nu ]}^{A}  \notag \\
&&+\frac{k_{A}}{8}\psi ^{*\rho }\gamma ^{\mu \nu }\psi _{\rho }\partial
_{[\mu }\eta _{\nu ]}^{A},  \label{3.35}
\end{eqnarray}
\begin{eqnarray}
&&\alpha _{0}^{(\mathrm{int})}=\frac{k_{A}}{2}\left( \sigma ^{\rho \lambda }%
\mathcal{L}_{0}^{(\mathrm{RS})}-\frac{\mathrm{i}}{2}\bar{\psi}_{\mu }\gamma
^{\mu \nu \rho }\partial ^{\lambda }\psi _{\nu }\right) h_{\rho \lambda }^{A}
\notag \\
&&+\frac{\mathrm{i}k_{A}}{4}\left( \frac{1}{2}\bar{\psi}^{\mu }\gamma ^{\rho
}\psi ^{\nu }+\sigma ^{\mu \rho }\bar{\psi}^{\nu }\gamma ^{\sigma }\psi
_{\sigma }+\bar{\psi}_{\sigma }\gamma ^{\sigma \rho \mu }\psi ^{\nu }\right)
\partial _{[\mu }h_{\nu ]\rho }^{A},  \label{3.36}
\end{eqnarray}
and $k_{A}$ are some real constants. Meanwhile, we find in a direct manner
that
\begin{equation}
\alpha ^{\left( \mathrm{RS}\right) }=a_{0}^{\left( \mathrm{RS}\right) },
\label{vuw}
\end{equation}
with $a_{0}^{\left( \mathrm{RS}\right) }$ given in (\ref{om7}).

Let us investigate next the consistency of the first-order deformation. If
we perform the notations
\begin{eqnarray}
\hat{S}_{1}^{\left( \mathrm{PF}\right) } &=&\int d^{4}x\alpha ^{\left(
\mathrm{PF}\right) },  \label{vx4} \\
\hat{S}_{1}^{\left( \mathrm{int}\right) } &=&\int d^{4}x\left( \alpha
^{\left( \mathrm{int}\right) }+\alpha ^{\left( \mathrm{RS}\right) }\right) ,
\label{4.1} \\
\hat{S}_{1} &=&\hat{S}_{1}^{\left( \mathrm{PF}\right) }+\hat{S}_{1}^{\left(
\mathrm{int}\right) },  \label{vx5}
\end{eqnarray}
then the equation $\left( \hat{S}_{1},\hat{S}_{1}\right) +2s\hat{S}_{2}=0$
(expressing the consistency of the first-order deformation) equivalently
splits into two independent equations
\begin{eqnarray}
\left( \hat{S}_{1}^{\left( \mathrm{PF}\right) },\hat{S}_{1}^{\left( \mathrm{%
PF}\right) }\right) +2s\hat{S}_{2}^{\left( \mathrm{PF}\right) } &=&0,
\label{vx6} \\
2\left( \hat{S}_{1}^{\left( \mathrm{PF}\right) },\hat{S}_{1}^{\left( \mathrm{%
int}\right) }\right) +\left( \hat{S}_{1}^{\left( \mathrm{int}\right) },\hat{S%
}_{1}^{\left( \mathrm{int}\right) }\right) +2s\hat{S}_{2}^{\left( \mathrm{int%
}\right) } &=&0,  \label{vx7}
\end{eqnarray}
where $\hat{S}_{2}=\hat{S}_{2}^{\left( \mathrm{PF}\right) }+\hat{S}%
_{2}^{\left( \mathrm{int}\right) }$. The equation (\ref{vx6}) requires that
the constants $f_{AB}^{C}$ satisfy the supplementary conditions~\cite{multi}
\begin{equation}
f_{A[B}^{D}f_{C]D}^{E}=0,  \label{4.20}
\end{equation}
so they are the structure constants of a finite-dimensional, commutative,
symmetric, and associative real algebra $\mathcal{A}$. The analysis realized
in~\cite{multi} shows us that such an algebra has a trivial structure (being
expressed like a direct sum of some one-dimensional ideals). So, we obtain
that
\begin{equation}
f_{AB}^{C}=0\quad \mathrm{if}\quad A\neq B.  \label{4.21}
\end{equation}

Let us analyze now the equation (\ref{vx7}). If we denote by $\hat{\Delta}%
^{\left( \mathrm{int}\right) }$ and $\beta ^{\left( \mathrm{int}\right) }$
the non-integrated densities of the functionals $2\left( \hat{S}_{1}^{\left(
\mathrm{PF}\right) },\hat{S}_{1}^{\left( \mathrm{int}\right) }\right)
+\left( \hat{S}_{1}^{\left( \mathrm{int}\right) },\hat{S}_{1}^{\left(
\mathrm{int}\right) }\right) $ and respectively of $\hat{S}_{2}^{\left(
\mathrm{int}\right) }$, then the equation (\ref{vx7}) takes the local form
\begin{equation}
\hat{\Delta}^{\left( \mathrm{int}\right) }=-2s\beta ^{\left( \mathrm{int}%
\right) }+\partial _{\mu }k^{\mu },  \label{4.5}
\end{equation}%
with
\begin{equation}
\mathrm{gh}\left( \hat{\Delta}^{\left( \mathrm{int}\right) }\right) =1,\quad
\mathrm{gh}\left( \beta ^{\left( \mathrm{int}\right) }\right) =0,\quad
\mathrm{gh}\left( k^{\mu }\right) =1.  \label{45a}
\end{equation}%
The computation of $\hat{\Delta}^{\left( \mathrm{int}\right) }$ reveals in
our case the following decomposition along the antighost number
\begin{equation}
\hat{\Delta}^{\left( \mathrm{int}\right) }=\hat{\Delta}_{0}^{\left( \mathrm{%
int}\right) }+\hat{\Delta}_{1}^{\left( \mathrm{int}\right) },\quad \mathrm{%
agh}\left( \hat{\Delta}_{I}^{\left( \mathrm{int}\right) }\right) =I,\quad
I=0,1,  \label{4.6}
\end{equation}%
with
\begin{eqnarray}
&&\hat{\Delta}_{1}^{\left( \mathrm{int}\right) }=\gamma \left( \left( -\frac{%
1}{4}k_{A}f_{BC}^{A}\left( \psi ^{\ast \lbrack \mu }\psi ^{\sigma ]}+\frac{1%
}{2}\psi ^{\ast \rho }\gamma ^{\mu \sigma }\psi _{\rho }\right) \partial
_{\lbrack \sigma }\eta _{\lambda ]}^{B}\sigma ^{\nu \lambda }\right. \right.
\notag \\
&&\left. +\psi ^{\ast \sigma }\left( \partial ^{\mu }\psi _{\sigma }\right)
\eta ^{B\nu }\right) h_{\mu \nu }^{C}  \notag \\
&&\left. +\left( k_{B}k_{C}-\frac{1}{2}k_{A}f_{BC}^{A}\right) \left( \psi
^{\ast \mu }\psi ^{\nu }+\frac{1}{4}\psi ^{\ast \sigma }\gamma ^{\mu \nu
}\psi _{\sigma }\right) \eta ^{B\rho }\partial _{\lbrack \mu }h_{\nu ]\rho
}^{C}\right)   \notag \\
&&+\left( k_{A}f_{BC}^{A}-k_{B}k_{C}\right) \left( \psi ^{\ast \mu }\left(
\partial ^{\nu }\psi _{\mu }\right) \eta ^{B\rho }\partial _{\lbrack \nu
}\eta _{\rho ]}^{C}+\frac{1}{4}\left( \psi ^{\ast \lbrack \mu }\psi ^{\nu
]}\right. \right.   \notag \\
&&\left. \left. +\frac{1}{2}\psi ^{\ast \sigma }\gamma ^{\mu \nu }\psi
_{\sigma }\right) \partial _{\lbrack \mu }\eta _{\rho ]}^{B}\partial
_{\lbrack \nu }\eta _{\lambda ]}^{C}\sigma ^{\rho \lambda }\right) .
\label{4.12a}
\end{eqnarray}%
The concrete form of $\hat{\Delta}_{0}^{\left( \mathrm{int}\right)
}$ is not important in what follows and therefore we will skip it.
Due to the expansion (\ref{4.6}), we have that $\beta ^{\left(
\mathrm{int}\right) }$ and $k^{\mu }$ from (\ref{4.5}) split like
\begin{eqnarray}
\beta ^{\left( \mathrm{int}\right) } &=&\beta _{0}^{\left( \mathrm{int}%
\right) }+\beta _{1}^{\left( \mathrm{int}\right) }+\beta _{2}^{\left(
\mathrm{int}\right) },\quad \mathrm{agh}\left( \beta _{I}^{\left( \mathrm{int%
}\right) }\right) =I,\quad I=0,1,2,  \label{4.7} \\
k^{\mu } &=&k_{0}^{\mu }+k_{1}^{\mu }+k_{2}^{\mu },\quad \mathrm{agh}\left(
k_{I}^{\mu }\right) =I,\quad I=0,1,2.  \label{4.8}
\end{eqnarray}%
By projecting the equation (\ref{4.5}) on the various decreasing values of
the antighost number, we obtain the equivalent tower of equations
\begin{eqnarray}
\gamma \beta _{2}^{\left( \mathrm{int}\right) } &=&\partial _{\mu }\left(
\frac{1}{2}k_{2}^{\mu }\right) ,  \label{4.9a} \\
\hat{\Delta}_{1}^{\left( \mathrm{int}\right) } &=&-2\left( \delta \beta
_{2}^{\left( \mathrm{int}\right) }+\gamma \beta _{1}^{\left( \mathrm{int}%
\right) }\right) +\partial _{\mu }k_{1}^{\mu },  \label{4.9} \\
\hat{\Delta}_{0}^{\left( \mathrm{int}\right) } &=&-2\left( \delta \beta
_{1}^{\left( \mathrm{int}\right) }+\gamma \beta _{0}^{\left( \mathrm{int}%
\right) }\right) +\partial _{\mu }k_{0}^{\mu }.  \label{4.10}
\end{eqnarray}%
By a trivial redefinition, the equation (\ref{4.9a}) can always be replaced
with
\begin{equation}
\gamma \beta _{2}^{\left( \mathrm{int}\right) }=0.  \label{4.10a}
\end{equation}%
Analyzing the expression of $\hat{\Delta}_{1}^{\left( \mathrm{int}\right) }$
in (\ref{4.12a}) we observe that it can be written like in (\ref{4.9}) if
the quantity
\begin{eqnarray}
&&\hat{\chi}=\left( k_{A}f_{BC}^{A}-k_{B}k_{C}\right) \left( \psi ^{\ast \mu
}\left( \partial ^{\nu }\psi _{\mu }\right) \eta ^{B\rho }\partial _{\lbrack
\nu }\eta _{\rho ]}^{C}+\frac{1}{4}\left( \psi ^{\ast \lbrack \mu }\psi
^{\nu ]}\right. \right.   \notag \\
&&\left. \left. +\frac{1}{2}\psi ^{\ast \sigma }\gamma ^{\mu \nu }\psi
_{\sigma }\right) \partial _{\lbrack \mu }\eta _{\rho ]}^{B}\partial
_{\lbrack \nu }\eta _{\lambda ]}^{C}\sigma ^{\rho \lambda }\right)
\label{4.10b}
\end{eqnarray}%
can be put in the form
\begin{equation}
\hat{\chi}=\delta \hat{\varphi}+\gamma \hat{\omega}+\partial _{\mu }j^{\mu }.
\label{4.10c}
\end{equation}%
Assume that (\ref{4.10c}) holds. Then, by applying $\delta $ on this
equation we infer
\begin{equation}
\delta \hat{\chi}=\gamma \left( -\delta \hat{\omega}\right) +\partial _{\mu
}\left( \delta j^{\mu }\right) .  \label{4.10d}
\end{equation}%
On the other hand, if we use the concrete expression (\ref{4.10b}) of $\hat{%
\chi}$, by direct computation we are led to
\begin{eqnarray}
&&\delta \hat{\chi}=\gamma \left( \frac{1}{2}\left(
k_{A}f_{BC}^{A}-k_{B}k_{C}\right) \delta \left( \psi ^{\ast \rho }\psi
_{\rho }\eta _{\nu }^{B}\left( \partial _{\mu }h^{C\mu \nu }-\partial ^{\nu
}h^{C}\right) \right) \right)   \notag \\
&&+\partial ^{\mu }\left( \frac{1}{2}\left(
k_{A}f_{BC}^{A}-k_{B}k_{C}\right) \delta \left( \psi ^{\ast \rho }\psi
_{\rho }\eta ^{B\nu }\partial _{\lbrack \mu }\eta _{\nu ]}^{C}\right)
\right)   \notag \\
&&+\gamma \left( \frac{\mathrm{i}}{4}\left(
k_{A}f_{BC}^{A}-k_{B}k_{C}\right) \left( \left( \bar{\psi}_{\beta }\gamma
^{\alpha \beta \sigma }\left( \partial ^{\mu }\psi _{\sigma }\right)
h_{\alpha }^{B\rho }-\left( \bar{\psi}_{\beta }\gamma ^{\alpha \beta \lbrack
\mu }\psi ^{\nu ]}\right. \right. \right. \right.   \notag \\
&&\left. \left. -\bar{\psi}^{\mu }\gamma ^{\alpha }\psi ^{\nu }-\sigma
^{\alpha \lbrack \mu }\bar{\psi}^{\nu ]}\gamma ^{\sigma }\psi _{\sigma
}\right) \sigma ^{\rho \lambda }\partial _{\lbrack \nu }h_{\lambda ]\alpha
}^{B}\right) \partial _{\lbrack \mu }\eta _{\rho ]}^{C}  \notag \\
&&\left. \left. -2\bar{\psi}_{\beta }\gamma ^{\alpha \beta \mu }\left(
\partial ^{\nu }\psi _{\mu }\right) \eta ^{B\rho }\partial _{\lbrack \nu
}h_{\rho ]\alpha }^{C}\right) \right)   \notag \\
&&+\partial _{\alpha }\left( \frac{\mathrm{i}}{2}\left(
k_{A}f_{BC}^{A}-k_{B}k_{C}\right) \left( \bar{\psi}_{\beta }\gamma ^{\alpha
\beta \sigma }\left( \partial ^{\mu }\psi _{\sigma }\right) \eta ^{B\rho }-%
\frac{1}{4}\left( \bar{\psi}_{\beta }\gamma ^{\alpha \beta \lbrack \mu }\psi
^{\nu ]}\right. \right. \right.   \notag \\
&&\left. \left. \left. -\bar{\psi}^{\mu }\gamma ^{\alpha }\psi ^{\nu
}-\sigma ^{\alpha \lbrack \mu }\bar{\psi}^{\nu ]}\gamma ^{\sigma }\psi
_{\sigma }\right) \sigma ^{\rho \lambda }\partial _{\lbrack \nu }\eta
_{\lambda ]}^{B}\right) \partial _{\lbrack \mu }\eta _{\rho ]}^{C}\right) .
\label{4.10e}
\end{eqnarray}%
The right-hand side of (\ref{4.10e}) can be written like in the right-hand
side of (\ref{4.10d}) if the following conditions are simultaneously
fulfilled
\begin{eqnarray}
&&\frac{\mathrm{i}}{4}\left( k_{A}f_{BC}^{A}-k_{B}k_{C}\right) \left\{ \left[
\bar{\psi}_{\beta }\gamma ^{\alpha \beta \sigma }\left( \partial ^{\mu }\psi
_{\sigma }\right) h_{\alpha }^{\rho }-\left( \bar{\psi}_{\beta }\gamma
^{\alpha \beta \lbrack \mu }\psi ^{\nu ]}\right. \right. \right.   \notag \\
&&\left. \left. -\bar{\psi}^{\mu }\gamma ^{\alpha }\psi ^{\nu }-\sigma
^{\alpha \lbrack \mu }\bar{\psi}^{\nu ]}\gamma ^{\sigma }\psi _{\sigma
}\right) \sigma ^{\rho \lambda }\partial _{\lbrack \nu }h_{\lambda ]\alpha
}^{B}\right] \partial _{\lbrack \mu }\eta _{\rho ]}^{C}  \notag \\
&&\left. -2\bar{\psi}_{\beta }\gamma ^{\alpha \beta \mu }\left( \partial
^{\nu }\psi _{\mu }\right) \eta ^{B\rho }\partial _{\lbrack \nu }h_{\rho
]\alpha }^{C}\right\} =-\delta \hat{\omega}^{\prime },  \label{4.10f}
\end{eqnarray}%
\begin{eqnarray}
&&\frac{\mathrm{i}}{2}\left( k_{A}f_{BC}^{A}-k_{B}k_{C}\right) \left( \bar{%
\psi}_{\beta }\gamma ^{\alpha \beta \sigma }\left( \partial ^{\mu }\psi
_{\sigma }\right) \eta ^{B\rho }-\frac{1}{4}\left( \bar{\psi}_{\beta }\gamma
^{\alpha \beta \lbrack \mu }\psi ^{\nu ]}\right. \right.   \notag \\
&&\left. \left. -\bar{\psi}^{\mu }\gamma ^{\alpha }\psi ^{\nu }-\sigma
^{\alpha \lbrack \mu }\bar{\psi}^{\nu ]}\gamma ^{\sigma }\psi _{\sigma
}\right) \sigma ^{\rho \lambda }\partial _{\lbrack \nu }\eta _{\lambda
]}^{B}\right) \partial _{\lbrack \mu }\eta _{\rho ]}^{C}=\delta j^{\prime
\mu }.  \label{4.10g}
\end{eqnarray}%
However, from the action of $\delta $ on the BRST generators we observe that
none of $h^{A\mu \beta }$, $\partial _{\lbrack \alpha }^{\left. {}\right.
}h_{\beta ]\mu }^{A}$, $\eta _{\beta }^{A}$, or $\partial _{\lbrack \lambda
}^{\left. {}\right. }\eta _{\beta ]}^{A}$ are $\delta $-exact. In
consequence, the relations (\ref{4.10f})--(\ref{4.10g}) hold if the
equations
\begin{equation}
\bar{\psi}_{\beta }\gamma ^{\alpha \beta \sigma }\left( \partial _{\mu }\psi
_{\sigma }\right) =\delta \Omega _{\;\;\mu }^{\alpha },  \label{5.1x}
\end{equation}%
and
\begin{equation}
\bar{\psi}_{\beta }\gamma ^{\alpha \beta \lbrack \mu }\psi ^{\nu ]}-\bar{\psi%
}^{\mu }\gamma ^{\alpha }\psi ^{\nu }-\sigma ^{\alpha \lbrack \mu }\bar{\psi}%
^{\nu ]}\gamma ^{\sigma }\psi _{\sigma }=\delta \Gamma ^{\mu \nu \alpha }
\label{5.2x}
\end{equation}%
take place simultaneously. The last equations are precisely the equations (%
\ref{sec21a}) and respectively (\ref{sec21b}). Due to the fact that they do
not involve (Pauli-Fierz) collection indices, some arguments identical to
those employed in subsection \ref{secorddef} ensure that (\ref{5.1x}) and (%
\ref{5.2x}) cannot be satisfied. As a consequence, $\hat{\chi}$ must vanish,
which further implies that
\begin{equation}
k_{D}f_{AB}^{D}-k_{A}k_{B}=0.  \label{4.12c}
\end{equation}%
Using (\ref{4.12c}) and (\ref{4.21}) we obtain that for $A\neq B$
\begin{equation}
k_{A}k_{B}=0,  \label{zwx}
\end{equation}%
which shows that the Rarita-Schwinger field can couple to only one graviton,
so the assertion from the beginning of this section is finally proved.

\section{Conclusion}

To conclude with, in this paper we have investigated the couplings between a
collection of massless spin-two fields (described in the free limit by a sum
of Pauli-Fierz actions) and a massive Rarita-Schwinger field using the
powerful setting based on local BRST cohomology. Initially, we have shown
that if we decompose the metric like $g_{\mu \nu }=\sigma _{\mu \nu
}+gh_{\mu \nu }$, then we can couple the massive Rarita-Schwinger field to $%
h_{\mu \nu }$ in the space of formal series with the maximum derivative
order equal to one in $h_{\mu \nu }$. The interacting Lagrangian $\mathcal{L}%
^{\left( \mathrm{int}\right) }$ obtained here contains, besides the standard
minimal couplings, also three types of non-minimal couplings, which are not
discussed in the literature, but are nevertheless consistent with the gauge
symmetries of the Lagrangian $\mathcal{L}_{2}+\mathcal{L}^{\left( \mathrm{int%
}\right) }$, where $\mathcal{L}_{2}$ is the full spin-two Lagrangian in the
vierbein formulation. Next, we have proved, under the hypotheses of
locality, smoothness of the interactions in the coupling constant, Poincar%
\'{e} invariance, (background) Lorentz invariance and the preservation of
the number of derivatives on each field, that there are no consistent
cross-interactions among different gravitons in the presence of a massive
Rarita-Schwinger field if the metric in internal space is positively defined.

\section*{Acknowledgment}

Three of the authors (C.B., E.M.C. and S.O.S) are partially supported by the
European Commission FP6 program MRTN-CT-2004-005104 and by the type A grant
305/2004 with the Romanian National Council for Academic Scientific Research
(C.N.C.S.I.S.) and the Romanian Ministry of Education and Research (M.E.C.).

\appendix{}

\section{Main conventions and properties of the $\protect\gamma $-matrices}

Here, we collect the main conventions and properties of the representation
of the $\gamma $-matrices employed in this paper. We work with the charge
conjugation matrix
\begin{equation}
\mathcal{C}=-\gamma _{0}  \label{lala1}
\end{equation}%
and with that representation of the Clifford algebra
\begin{equation}
\gamma _{\mu }\gamma _{\nu }+\gamma _{\nu }\gamma _{\mu }=2\sigma _{\mu \nu }%
\mathbf{1}  \label{lala2}
\end{equation}%
for which all the $\gamma $-matrices are purely imaginary. In addition, $%
\gamma _{0}$ is Hermitian and antisymmetric, while $\left( \gamma
_{i}\right) _{i=\overline{1,3}}$ are anti-Hermitian and symmetric. We take a
basis in the space of spinor matrices of the form
\begin{equation}
\mathbf{1,}\quad \gamma _{\mu },\quad \gamma _{\mu _{1}\mu _{2}},\quad
\gamma _{\mu _{1}\mu _{2}\mu _{3}},\quad \gamma _{\mu _{1}\mu _{2}\mu
_{3}\mu _{4}},  \label{def0}
\end{equation}%
where
\begin{equation}
\gamma _{\mu _{1}\cdots \mu _{k}}=\frac{1}{k!}\sum\limits_{\sigma \in
S_{k}}\left( -\right) ^{\sigma }\gamma _{\mu _{\sigma (1)}}\gamma _{\mu
_{\sigma (2)}}\cdots \gamma _{\mu _{\sigma (k)}}.  \label{def1}
\end{equation}%
In the above definition $S_{k}$ is the set of permutations of $\left\{
1,2,\ldots ,k\right\} $ and $\left( -\right) ^{\sigma }$ denotes the
signature of a given permutation $\sigma $. This means that any $4\times 4$
matrix $M$ with purely spinor indices can be expressed in terms of the
matrices (\ref{def0}) via
\begin{equation}
M=\frac{1}{4}\sum\limits_{k=0}^{4}\left( -\right) ^{k(k-1)/2}\frac{1}{k!}%
\mathrm{Tr}\left( \gamma ^{\mu _{1}\cdots \mu _{k}}M\right) \gamma _{\mu
_{1}\cdots \mu _{k}}.  \label{gama11}
\end{equation}

We list below some Fierz identities that are useful at the construction of
consistent interactions between the Pauli-Fierz field and the massive
Rarita-Schwinger spinor. They provide the products of the various elements
from (\ref{def0}) in terms of their linear combinations
\begin{eqnarray}
\gamma _{\mu \nu }\gamma ^{\rho } &=&-\delta _{\lbrack \mu }^{\rho }\gamma
_{\nu ]}+\gamma _{\mu \nu }^{\;\;\;\;\rho },  \label{fie1} \\
\gamma _{\mu \nu }\gamma ^{\rho \lambda } &=&-\delta _{\lbrack \mu }^{\rho
}\delta _{\nu ]}^{\lambda }\mathbf{1}-\delta _{\lbrack \mu }^{[\rho }\gamma
_{\nu ]}^{\;\;\;\lambda ]}+\gamma _{\mu \nu }^{\;\;\;\;\rho \lambda },
\label{fie2} \\
\gamma _{\mu \nu }\gamma ^{\rho \lambda \sigma } &=&-\delta _{\mu }^{[\rho
}\delta _{\nu }^{\lambda }\gamma ^{\sigma ]}-\delta _{\lbrack \mu }^{[\rho
}\gamma _{\nu ]}^{\;\;\;\lambda \sigma ]},  \label{fie3} \\
\gamma _{\mu \nu }\gamma ^{\rho \lambda \sigma \xi } &=&-\delta _{\mu
}^{[\rho }\delta _{\nu }^{\lambda }\gamma ^{\sigma \xi ]},  \label{fie4} \\
\gamma _{\mu \nu \rho }\gamma ^{\alpha } &=&\delta _{\lbrack \mu }^{\alpha
}\gamma _{\nu \rho ]}+\gamma _{\mu \nu \rho }^{\;\;\;\;\ \ \alpha },
\label{fie5} \\
\gamma _{\mu \nu \rho }\gamma ^{\alpha \beta \gamma } &=&-\delta _{\mu
}^{[\alpha }\delta _{\nu }^{\beta }\delta _{\rho }^{\gamma ]}\mathbf{1}%
-\delta _{\lbrack \mu }^{[\alpha }\delta _{\nu }^{\beta }\gamma _{\rho
]}^{\;\;\gamma ]}.  \label{fie6}
\end{eqnarray}%
Moreover, in the chosen representation of the $\gamma $-matrices the
elements of the basis (\ref{def0}) display the following
symmetry/antisymmetry properties:
\begin{equation}
\gamma _{0}\gamma _{\mu },\quad \gamma _{0}\gamma _{\mu \nu }  \label{sim1}
\end{equation}%
are symmetric and
\begin{equation}
\gamma _{0}\gamma _{\mu \nu \rho },\quad \gamma _{0}\gamma _{\mu \nu \rho
\lambda },\quad \gamma _{0}\gamma _{5}  \label{sim2}
\end{equation}%
are antisymmetric. If we take $\gamma _{5}=\mathrm{i}\gamma _{0}\gamma
_{1}\gamma _{2}\gamma _{3}$ and work with $\varepsilon ^{0123}=-\varepsilon
_{0123}=1$, then
\begin{eqnarray}
\gamma ^{\mu \nu \rho \lambda } &=&\varepsilon ^{\mu \nu \rho \lambda
}\gamma ^{0}\gamma ^{1}\gamma ^{2}\gamma ^{3}=\mathrm{i}\varepsilon ^{\mu
\nu \rho \lambda }\gamma _{5},  \label{gama12} \\
\gamma _{\mu \nu \rho \lambda } &=&-\varepsilon _{\mu \nu \rho \lambda
}\gamma _{0}\gamma _{1}\gamma _{2}\gamma _{3}=\mathrm{i}\varepsilon _{\mu
\nu \rho \lambda }\gamma _{5}.  \label{gama13}
\end{eqnarray}

\section{Proof of some assertions made in the subsection \protect\ref%
{firstord}}

Initially, we show that our statement from footnote \ref{hstar} is indeed
valid. The terms linear in the Pauli-Fierz antifield $h^{\ast \mu \nu }$
that can be in principle added to $a_{1}^{(\mathrm{int})}$ have the generic
form
\begin{equation}
\tilde{a}_{1}^{\left( \mathrm{int}\right) }=h^{\ast \mu \nu }\left( M_{\mu
\nu }^{\rho }\eta _{\rho }+M_{\mu \nu }^{\rho \lambda }\partial _{\lbrack
\rho }\eta _{\lambda ]}\right) \equiv \tilde{a}_{1}^{\prime \left( \mathrm{%
int}\right) }+\tilde{a}_{1}^{\prime \prime \left( \mathrm{int}\right) },
\label{atilde1}
\end{equation}%
where $M_{\mu \nu }^{\rho }$ and $M_{\mu \nu }^{\rho \lambda }$ are bosonic,
real, gauge-invariant functions. Imposing that (\ref{atilde1}) satisfies the
requirements i)--ii) from the subsection \ref{firstord}, then the functions $%
M_{\mu \nu }^{\rho }$ and $M_{\mu \nu }^{\rho \lambda }$ are restricted to
depend at most on the undifferentiated Rarita-Schwinger field. The
consistency equation for $\tilde{a}_{1}^{\left( \mathrm{int}\right) }$ in
antighost number zero
\begin{equation}
\delta \tilde{a}_{1}^{\left( \mathrm{int}\right) }+\gamma \tilde{a}%
_{0}^{\left( \mathrm{int}\right) }=\partial _{\mu }\tilde{j}_{0}^{\left(
\mathrm{int}\right) },  \label{consatilde1}
\end{equation}%
is independent of that for $a_{1}^{(\mathrm{int})}$ of the form (\ref%
{PFRS3.23}) since the former piece produces in $\tilde{a}_{0}^{\left(
\mathrm{int}\right) }$ components quadratic in the Pauli-Fierz field, while
the latter introduces in $a_{0}^{(\mathrm{int})}$ terms linear in $h_{\mu
\nu }$. Moreover, the consistency equation of $\tilde{a}_{1}^{\prime \left(
\mathrm{int}\right) }$\ is independent of that implying $\tilde{a}%
_{1}^{\prime \prime \left( \mathrm{int}\right) }$ due to the different
number of derivatives contained in these two types of terms, so (\ref%
{consatilde1}) is equivalent to the equations
\begin{eqnarray}
\delta \tilde{a}_{1}^{\prime \left( \mathrm{int}\right) }+\gamma \tilde{a}%
_{0}^{\prime \left( \mathrm{int}\right) } &=&\partial _{\mu }\tilde{j}%
_{0}^{\prime \left( \mathrm{int}\right) },  \label{consatildea} \\
\delta \tilde{a}_{1}^{\prime \prime \left( \mathrm{int}\right) }+\gamma
\tilde{a}_{0}^{\prime \prime \left( \mathrm{int}\right) } &=&\partial _{\mu }%
\tilde{j}_{0}^{\prime \prime \left( \mathrm{int}\right) }.
\label{consatildeb}
\end{eqnarray}%
Now, we prove that (\ref{atilde1}) is not consistent in antighost number
zero, i.e., there are no solutions $\tilde{a}_{0}^{\prime \left( \mathrm{int}%
\right) }$ or $\tilde{a}_{0}^{\prime \prime \left( \mathrm{int}\right) }$ to
the equations (\ref{consatildea})--(\ref{consatildeb}). To this end we use
the fact that the linearized Einstein tensor (\ref{PFRS13}) can be written
like
\begin{equation}
H^{\mu \nu }=\partial _{\alpha }\partial _{\beta }\phi ^{\mu \alpha \nu
\beta },  \label{inc1}
\end{equation}%
with
\begin{eqnarray}
\phi ^{\mu \alpha \nu \beta } &=&\frac{1}{2}\left( -h^{\mu \nu }\sigma
^{\alpha \beta }+h^{\alpha \nu }\sigma ^{\mu \beta }+h^{\mu \beta }\sigma
^{\alpha \nu }-h^{\alpha \beta }\sigma ^{\mu \nu }\right.  \notag \\
&&\left. +h\left( \sigma ^{\mu \nu }\sigma ^{\alpha \beta }-\sigma ^{\mu
\beta }\sigma ^{\alpha \nu }\right) \right) .  \label{inc2}
\end{eqnarray}%
By direct computation, we find that
\begin{eqnarray}
&&\delta \tilde{a}_{1}^{\prime \left( \mathrm{int}\right) }=-2\partial
_{\alpha }\partial _{\beta }\phi ^{\mu \alpha \nu \beta }M_{\mu \nu }^{\rho
}\eta _{\rho }=  \notag \\
&&\partial _{\alpha }\left( -2\left( \partial _{\beta }\phi ^{\mu \alpha \nu
\beta }\right) M_{\mu \nu }^{\rho }\eta _{\rho }\right) +\partial _{\beta
}\left( 2\phi ^{\mu \alpha \nu \beta }\partial _{\alpha }\left( M_{\mu \nu
}^{\rho }\eta _{\rho }\right) \right)  \notag \\
&&+\phi ^{\mu \alpha \nu \beta }\partial _{\lbrack \mu }M_{\alpha ]\nu
}^{\rho }\partial _{\lbrack \beta }\eta _{\rho ]}+\frac{1}{2}\phi ^{\mu
\alpha \nu \beta }\partial _{\lbrack \mu }M_{\alpha ][\nu ,\beta ]}^{\rho
}\eta _{\rho }  \notag \\
&&+\gamma \left( \phi ^{\mu \alpha \nu \beta }\left( \partial _{\lbrack \mu
}M_{\alpha ]\nu }^{\rho }h_{\beta \rho }-2M_{\mu \nu }^{\rho }\overset{(1)}{%
\Gamma }_{\rho \alpha \beta }\right) \right)  \notag \\
&&-\left( \gamma \phi ^{\mu \alpha \nu \beta }\right) \left( \partial
_{\lbrack \mu }M_{\alpha ]\nu }^{\rho }h_{\beta \rho }-2M_{\mu \nu }^{\rho }%
\overset{(1)}{\Gamma }_{\rho \alpha \beta }\right) ,  \label{inc3}
\end{eqnarray}%
where
\begin{equation}
\overset{(1)}{\Gamma }_{\rho \alpha \beta }=\frac{1}{2}\left( \partial
_{\alpha }h_{\beta \rho }+\partial _{\beta }h_{\alpha \rho }-\partial _{\rho
}h_{\alpha \beta }\right) .  \label{qq87}
\end{equation}%
Comparing (\ref{inc3}) with (\ref{consatildea}) and observing that the term
in (\ref{inc3}) involving $\left( \gamma \phi ^{\mu \alpha \nu \beta
}\right) $ comprises the symmetric derivatives $\partial _{(\beta }\eta
_{\rho )}$, it follows that this piece, which is constrained to contribute
to a full divergence, can only realize this task together with the part
proportional with $\partial _{\lbrack \mu }M_{\alpha ][\nu ,\beta ]}^{\rho }$%
. Accordingly, the $\gamma $-exactness modulo $d$ of the right-hand side of (%
\ref{inc3}), which is demanded by the equation (\ref{consatildea}), requires
that the functions $M_{\mu \nu }^{\rho }$ are subject to the equations
\begin{equation}
\partial _{\lbrack \mu }M_{\alpha ]\nu }^{\rho }=0,  \label{gama14}
\end{equation}%
possessing the trivial solution
\begin{equation}
M_{\alpha \nu }^{\rho }=0  \label{inc4}
\end{equation}%
since $M_{\alpha \nu }^{\rho }$ are derivative-free (they depend only on the
undifferentiated spinor-vector $\psi _{\mu }$). In an identical manner,
starting with
\begin{eqnarray}
&&\delta \tilde{a}_{1}^{\prime \prime \left( \mathrm{int}\right)
}=-2\partial _{\alpha }\partial _{\beta }\phi ^{\mu \alpha \nu \beta }M_{\mu
\nu }^{\rho \lambda }\partial _{\lbrack \rho }\eta _{\lambda ]}=  \notag \\
&&\partial _{\alpha }\left( -2\left( \partial _{\beta }\phi ^{\mu \alpha \nu
\beta }\right) M_{\mu \nu }^{\rho \lambda }\partial _{\lbrack \rho }\eta
_{\lambda ]}\right) +\partial _{\beta }\left( 2\phi ^{\mu \alpha \nu \beta
}\partial _{\alpha }\left( M_{\mu \nu }^{\rho \lambda }\partial _{\lbrack
\rho }\eta _{\lambda ]}\right) \right)  \notag \\
&&+\frac{1}{2}\phi ^{\mu \alpha \nu \beta }\partial _{\lbrack \mu }M_{\alpha
][\nu ,\beta ]}^{\rho \lambda }\partial _{\lbrack \rho }\eta _{\lambda ]}
\notag \\
&&+\gamma \left( 2\phi ^{\mu \alpha \nu \beta }\left( \partial
_{\lbrack \mu }M_{\alpha ]\nu }^{\rho \lambda }\partial _{\lbrack
\rho }h_{\lambda ]\beta }-M_{\mu \nu }^{\rho }\partial _{\alpha
}\partial _{\lbrack \rho }h_{\lambda
]\beta }\right) \right)  \notag \\
&&-2\left( \gamma \phi ^{\mu \alpha \nu \beta }\right) \left( \partial
_{\lbrack \mu }M_{\alpha ]\nu }^{\rho \lambda }\partial _{\lbrack \rho
}h_{\lambda ]\beta }-M_{\mu \nu }^{\rho }\partial _{\alpha }\partial
_{\lbrack \rho }h_{\lambda ]\beta }\right) ,  \label{inc5}
\end{eqnarray}%
we argue that the functions $M_{\mu \nu }^{\rho \lambda }$ must obey the
equations
\begin{equation}
\partial _{\lbrack \mu }M_{\alpha ][\nu ,\beta ]}^{\rho \lambda }=0,
\label{gama15}
\end{equation}%
which, due to the fact that $M_{\mu \nu }^{\rho \lambda }$ are
derivative-free, possess only the trivial solution
\begin{equation}
M_{\mu \nu }^{\rho \lambda }=0.  \label{inc6}
\end{equation}%
If we substitute the results (\ref{inc4}) and (\ref{inc6}) into (\ref%
{atilde1}), we conclude that there is no term linear in the Pauli-Fierz
antifield $h^{\ast \mu \nu }$ that can be added to $a_{1}^{(\mathrm{int})}$
such as to give a consistent component of antighost number zero in the
first-order deformation of the solution to the master equation.

Finally, we show that we can always make the functions $c_{1}$,
$c_{2}$, and $c_{3}$ from (\ref{PFRS3.23}) vanish via adding some
trivial terms and making some redefinitions of the functions
$\bar{N}_{\;\;\;\;\;\;\mu }^{\rho
\lambda \sigma }$. In view of this, we insert (\ref{ww8}) in (\ref{PFRS3.23}%
), such that the part from $a_{1}^{(\mathrm{int})}$ proportional with $c_{1}$%
, $c_{2}$, or $c_{3}$ reads as
\begin{eqnarray}
T\left( c_{1},c_{2},c_{3}\right) &=&\left[ c_{1}\left( \psi ^{*\lambda
}\gamma ^{\mu }\psi _{\mu }-\frac{1}{2}\psi _{\mu }^{*}\gamma ^{\mu \nu
\lambda }\psi _{\nu }\right) +c_{2}\left( \psi ^{*\mu }\gamma ^{\lambda
}\psi _{\mu }-\psi _{\mu }^{*}\gamma ^{\mu \nu \lambda }\psi _{\nu }\right)
\right.  \notag \\
&&\left. +c_{3}\left( \psi ^{*\mu }\gamma _{\mu }\psi ^{\lambda }-\frac{3}{2}%
\psi _{\mu }^{*}\gamma ^{\mu \nu \lambda }\psi _{\nu }\right) \right] \eta
_{\lambda }.  \label{c1c2c3}
\end{eqnarray}
Based on the second definition in (\ref{PFRS8}) related to the Koszul-Tate
differential and on the Fierz identities from the previous appendix section,
we obtain that
\begin{eqnarray}
&&\delta \left( \psi ^{*\lambda }\gamma _{\mu }\bar{\psi}^{*\mu }\right)
=-4m\psi ^{*\lambda }\gamma ^{\mu }\psi _{\mu }+m\psi ^{*\mu }\gamma
^{\lambda }\psi _{\mu }+m\psi _{\mu }^{*}\gamma ^{\mu \nu \lambda }\psi
_{\nu }  \notag \\
&&+\mathrm{i}\left( 3\psi ^{*\lambda }\gamma ^{\mu \nu }+\psi ^{*[\mu
}\gamma ^{\nu ]\lambda }\right) \partial _{\mu }\psi _{\nu }+\mathrm{i}\psi
_{\mu }^{*}\gamma ^{\mu \nu \rho \lambda }\partial _{\nu }\psi _{\rho },
\label{rely1}
\end{eqnarray}
\begin{eqnarray}
&&\delta \left( \psi _{\mu }^{*}\gamma ^{\lambda }\bar{\psi}^{*\mu }\right)
=-2m\psi ^{*\lambda }\gamma ^{\mu }\psi _{\mu }+2m\psi _{\mu }^{*}\gamma
^{\mu }\psi ^{\lambda }-2m\psi _{\mu }^{*}\gamma ^{\mu \nu \lambda }\psi
_{\nu }  \notag \\
&&+2\mathrm{i}\left( \psi ^{*\lambda }\gamma ^{\mu \nu }\partial _{\mu }\psi
_{\nu }+\psi ^{*\mu }\gamma _{\rho \mu }\partial ^{[\lambda }\psi ^{\rho
]}\right) +2\mathrm{i}\psi _{\mu }^{*}\gamma ^{\mu \nu \rho \lambda
}\partial _{\nu }\psi _{\rho },  \label{rely2}
\end{eqnarray}
\begin{eqnarray}
&&\delta \left( \psi _{\mu }^{*}\gamma ^{\mu \nu \lambda }\bar{\psi}_{\nu
}^{*}\right) =4m\psi _{\mu }^{*}\gamma ^{\mu }\psi ^{\lambda }-4m\psi ^{*\mu
}\gamma ^{\lambda }\psi _{\mu }-2m\psi _{\mu }^{*}\gamma ^{\mu \nu \lambda
}\psi _{\nu }  \notag \\
&&+4\mathrm{i}\psi _{\mu }^{*}\partial ^{[\mu }\psi ^{\lambda ]}+2\mathrm{i}%
\psi ^{*\mu }\gamma ^{\lambda \nu }\partial _{[\mu }\psi _{\nu ]}-2\mathrm{i}%
\psi ^{*\mu }\gamma _{\mu \nu }\partial ^{[\lambda }\psi ^{\nu ]}.
\label{rely3}
\end{eqnarray}
Relying on the above results, we can rewrite the three terms present in (\ref%
{c1c2c3}) in the form
\begin{eqnarray}
&&c_{1}\left( \psi ^{*\lambda }\gamma ^{\mu }\psi _{\mu }-\frac{1}{2}\psi
_{\mu }^{*}\gamma ^{\mu \nu \lambda }\psi _{\nu }\right) \eta _{\lambda }=s%
\left[ \frac{c_{1}}{12m}\left( 4\psi ^{*\rho }\gamma ^{\mu }\bar{\psi}_{\mu
}^{*}-2\psi _{\mu }^{*}\gamma ^{\rho }\bar{\psi}^{*\mu }\right. \right.
\notag \\
&&\left. \left. +\psi _{\mu }^{*}\gamma ^{\mu \nu \rho }\bar{\psi}_{\nu
}^{*}\right) \eta _{\rho }\right] +\frac{\mathrm{i}c_{1}}{3m}\left[ \left(
2\psi ^{*\lambda }\gamma ^{\mu \nu }+\frac{1}{2}\psi ^{*[\mu }\gamma ^{\nu
]\lambda }\right) \partial _{\mu }\psi _{\nu }\right.  \notag \\
&&\left. +\frac{1}{2}\psi ^{*\mu }\gamma _{\mu \rho }\partial ^{[\lambda
}\psi ^{\rho ]}+\psi _{\mu }^{*}\partial ^{[\mu }\psi ^{\lambda ]}+\frac{1}{2%
}\psi _{\mu }^{*}\gamma ^{\mu \nu \rho \lambda }\partial _{\nu }\psi _{\rho }%
\right] \eta _{\lambda },  \label{fa}
\end{eqnarray}
\begin{eqnarray}
&&c_{2}\left( \psi ^{*\mu }\gamma ^{\lambda }\psi _{\mu }-\psi _{\mu
}^{*}\gamma ^{\mu \nu \lambda }\psi _{\nu }\right) \eta _{\lambda }=s\left[
\frac{c_{2}}{3m}\left( \psi ^{*\rho }\gamma ^{\mu }\bar{\psi}_{\mu
}^{*}-2\psi _{\mu }^{*}\gamma ^{\rho }\bar{\psi}^{*\mu }\right. \right.
\notag \\
&&\left. \left. +\psi _{\mu }^{*}\gamma ^{\mu \nu \rho }\bar{\psi}_{\nu
}^{*}\right) \eta _{\rho }\right] +\frac{\mathrm{i}c_{2}}{3m}\left[ -\left(
\psi ^{*\lambda }\gamma ^{\mu \nu }+\psi ^{*[\mu }\gamma ^{\nu ]\lambda
}\right) \partial _{\mu }\psi _{\nu }\right.  \notag \\
&&\left. +2\psi ^{*\mu }\gamma _{\mu \rho }\partial ^{[\lambda }\psi ^{\rho
]}+4\psi _{\mu }^{*}\partial ^{[\mu }\psi ^{\lambda ]}-3\psi _{\mu
}^{*}\gamma ^{\mu \nu \rho \lambda }\partial _{\nu }\psi _{\rho }\right]
\eta _{\lambda },  \label{fb}
\end{eqnarray}
\begin{eqnarray}
&&c_{3}\left( \psi ^{*\mu }\gamma _{\mu }\psi ^{\lambda }-\frac{3}{2}\psi
_{\mu }^{*}\gamma ^{\mu \nu \lambda }\psi _{\nu }\right) \eta _{\lambda }=s%
\left[ \frac{c_{3}}{12m}\left( 4\psi ^{*\rho }\gamma ^{\mu }\bar{\psi}_{\mu
}^{*}-8\psi _{\mu }^{*}\gamma ^{\rho }\bar{\psi}^{*\mu }\right. \right.
\notag \\
&&\left. \left. +\psi _{\mu }^{*}\gamma ^{\mu \nu \rho }\bar{\psi}_{\nu
}^{*}\right) \eta _{\rho }\right] +\frac{\mathrm{i}c_{3}}{12m}\left[ \left(
-4\psi ^{*\lambda }\gamma ^{\mu \nu }+2\psi ^{*[\mu }\gamma ^{\nu ]\lambda
}\right) \partial _{\mu }\psi _{\nu }\right.  \notag \\
&&\left. +14\psi ^{*\mu }\gamma _{\mu \rho }\partial ^{[\lambda }\psi ^{\rho
]}+4\psi _{\mu }^{*}\partial ^{[\mu }\psi ^{\lambda ]}-12\psi _{\mu
}^{*}\gamma ^{\mu \nu \rho \lambda }\partial _{\nu }\psi _{\rho }\right]
\eta _{\lambda }.  \label{fc}
\end{eqnarray}
By adding the relations (\ref{fa})--(\ref{fc}),\ we observe that
$T\left( c_{1},c_{2},c_{3}\right) $ can be made to vanish by adding
some $s$-exact terms to the first-order deformation
$a^{(\mathrm{int})}$ and by appropriately redefining the functions
$\bar{N}_{\;\;\;\;\;\;\mu }^{\rho \lambda \sigma }$.

\end{document}